\documentclass[aps,prd,twocolumn,a4paper,superscriptaddress]{revtex4-1}
\usepackage{graphicx}
\begin{document}

\newcommand{\be}{\begin{equation}}
\newcommand{\ee}{\end{equation}}
\newcommand{\bq}{\begin{eqnarray}}
\newcommand{\eq}{\end{eqnarray}}
\newcommand{\bsq}{\begin{subequations}}
\newcommand{\esq}{\end{subequations}}
\newcommand{\bc}{\begin{center}}
\newcommand{\ec}{\end{center}}

\title{Fine-structure constant constraints on dark energy: II. Extending the parameter space}

\author{C. J. A. P. Martins}
\email{Carlos.Martins@astro.up.pt}
\affiliation{Centro de Astrof\'{\i}sica da Universidade do Porto, Rua das Estrelas, 4150-762 Porto, Portugal}
\affiliation{Instituto de Astrof\'{\i}sica e Ci\^encias do Espa\c co, CAUP, Rua das Estrelas, 4150-762 Porto, Portugal}
\author{A. M. M. Pinho}
\email[]{Ana.Pinho@astro.up.pt}
\affiliation{Centro de Astrof\'{\i}sica da Universidade do Porto, Rua das Estrelas, 4150-762 Porto, Portugal}
\affiliation{Faculdade de Ci\^encias, Universidade do Porto, Rua do Campo Alegre 687, 4169-007 Porto, Portugal}
\author{P. Carreira}
\email[]{up201506301@fc.up.pt}
\affiliation{Agrupamento de Escolas da Guia, Rua Fundadores do Col\'egio, 3105-075 Guia--Pombal, Portugal}
\affiliation{Faculdade de Ci\^encias, Universidade do Porto, Rua do Campo Alegre 687, 4169-007 Porto, Portugal}
\author{A. Gusart}
\email[]{arnau.gusart.v@gmail.com}
\affiliation{Escola CINGLE, Carretera de Montcada 512, 08223 Terrassa, Spain}
\author{J. L\'opez}
\email[]{julietaa98@gmail.com}
\affiliation{Institut Reguissol, Pg. Reguissol s/n, 08460 Sta. Maria de Palautordera, Spain}
\author{C. I. S. A. Rocha}
\email[]{up201505702@fc.up.pt}
\affiliation{Externato Ribadouro, Rua de Santa Catarina 1346, 4000-447 Porto, Portugal}
\affiliation{Faculdade de Ci\^encias, Universidade do Porto, Rua do Campo Alegre 687, 4169-007 Porto, Portugal}

\date{11 November 2015}

\begin{abstract}
Astrophysical tests of the stability of fundamental couplings, such as the fine-structure constant $\alpha$, are a powerful probe of new physics. Recently these measurements, combined with local atomic clock tests and Type Ia supernova and Hubble parameter data, were used to constrain the simplest class of dynamical dark energy models where the same degree of freedom is assumed to provide both the dark energy and (through a dimensionless coupling, $\zeta$, to the electromagnetic sector) the $\alpha$ variation. One caveat of these analyses was that it was based on fiducial models where the dark energy equation of state was described by a single parameter (effectively its present day value, $w_0$). Here we relax this assumption and study broader dark energy model classes, including the Chevallier-Polarski-Linder and Early Dark Energy parametrizations. Even in these extended cases we find that the current data constrains the coupling $\zeta$ at the $10^{-6}$ level and $w_0$ to a few percent (marginalizing over other parameters), thus confirming the robustness of earlier analyses. On the other hand, the additional parameters are typically not well constrained. We also highlight the implications of our results for constraints on violations of the Weak Equivalence Principle and improvements to be expected from forthcoming measurements with high-resolution ultra-stable spectrographs.
\end{abstract}
\maketitle

\section{Introduction}

The nature of dark energy, which is seemingly behind the recent acceleration of the universe \cite{SN1,SN2}, is arguably the most pressing problem of modern physics and cosmology. A cosmological constant remains the simplest available explanation (at least in the sense of requiring the smallest number of additional parameters), though at the cost of very significant fine-tuning problems. Considerable efforts are therefore underway, both to study possible alternative theoretical scenarios and to identify new observational tools that allow for a more detailed characterization of the dark energy properties and may ultimately lead to discriminating tests between competing paradigms.

The most natural alternative explanation for dark energy would involve scalar fields, an example of which is the recently discovered Higgs field \cite{ATLAS,CMS}. If dynamical scalar fields are indeed present and responsible for dark energy, one expects them to couple to the rest of the model, unless a yet-unknown symmetry is postulated to suppress these couplings \cite{Carroll}. In particular, a coupling of the field to the electromagnetic sector will lead to spacetime variations of the fine-structure constant $\alpha$---see \cite{uzanLR,cjmGRG} for recent reviews on this topic. There are some indications of such a variation \cite{Webb}, at the relative level of variation of a few parts per million and in the approximate redshift range $1<z<3$. An ongoing dedicated Large Program at ESO's Very Large Telescope (VLT) is aiming to test them \cite{LP1,LP3}. Regardless of the outcome of these studies (i.e., whether they provide detections of variations or just null results) these measurements have cosmological implications that go beyond the mere fundamental nature of the tests themselves.

Motivated by the imminent availability of more precise measurements, we have started a systematic analysis of the cosmology and fundamental physics constraints implied by these tests. This relies on the minimal and natural assumption that the same dynamical degree of freedom is responsible for the dark energy and the $\alpha$ variations---these are known as Class I models in the classification of \cite{cjmGRG}. In this case any astrophysical or laboratory tests of the stability of $\alpha$ will directly constrain dark energy. The future impact of these methods as a dark energy probe has recently been assessed in some detail \cite{Amendola,Leite,LeiteNEW}, but in \cite{Pinho} we first pointed out how the currently available measurements already provide non-trivial constraints on dynamical dark energy scenarios. In \cite{Pinho2} we further showed, building upon work in \cite{Dvali,Chiba}, how the same datasets lead to indirect constraints on violations of the Weak Equivalence Principle (WEP) that are one order of magnitude stronger than the best currently available direct ones, coming from torsion balance and lunar laser ranging experiments.

Specifically, the main outcome of \cite{Pinho,Pinho2} was that the current data constrains the coupling $\zeta$ of the scalar field to the electromagnetic sector of the theory (to be rigorously defined below) at the $10^{-6}$ level, the present day value of the dark energy equation of state $w_0$ to a few percent, and the E\"{o}tv\"{o}s parameter $\eta$ (parametrizing WEP violations) at the $10^{-14}$ level; all such constraints are at the two-sigma confidence level. One caveat of these earlier studies was that they assumed relatively simple dark energy parametrizations. While in \cite{Pinho} a constant equation of state (that is, $w(z)=w_0$) was assumed, in \cite{Pinho2} we considered two examples of freezing and thawing models (in the classification of \cite{FrTh}) but in both of them the dark energy equation of state, although redshift-dependent, was still parametrized by a single parameter---effectively its present day value, $w_0$.

Given that there are degeneracies between the coupling $\zeta$ and $w_0$ (which are partially broken by the cosmological datasets) one may legitimately ask how robust these constraints are. The main goal of the present work is to answer this question, by extending the analysis to more general---and, arguably, more realistic---dark energy models. Specifically we consider the well-known Chevallier-Polarski-Linder \cite{CPL1,CPL2} (hereafter CPL) and early dark energy \cite{EDE} (hereafter EDE) classes, as well as a parametrization recently discussed by Mukhanov \cite{MKH}. Compared to the models studied in previous works, each of these has one additional free parameter, but this extra parameter plays a different role in each of the parametrizations.

Taken together, these three classes of models provide a reasonable sample of the allowed parameter space. Thus we can study and quantify how the relevant constraints depend on the choice of model (as well as of priors) while preserving some conceptual simplicity. We will show that the marginalized constraints on $w_0$ and $\zeta$ are only very mildly weakened, whereas the constraints on the additional dark energy parameter depend on the model being considered but are typically weaker. (This occurs since the degeneracies between the relevant parameters are quite model-dependent.) Our results therefore confirm the results of the previous, simpler analyses.

\section{Varying $\alpha$ and dark energy}

Dynamical scalar fields in an effective 4D field theory are naturally expected to couple to the rest of the theory, unless a (still unknown) symmetry is postulated to suppress this coupling \cite{Carroll,Dvali,Chiba}. We will assume that this coupling does exist for the dynamical degree of freedom responsible for the dark energy, denoted $\phi$. Specifically the coupling to the electromagnetic sector is due to a gauge kinetic function $B_F(\phi)$
\begin{equation}
{\cal L}_{\phi F} = - \frac{1}{4} B_F(\phi) F_{\mu\nu}F^{\mu\nu}\,.
\end{equation}
This function can be assumed to be linear,
\begin{equation}
B_F(\phi) = 1- \zeta \kappa (\phi-\phi_0)\,,
\end{equation}
(where $\kappa^2=8\pi G$) since, as has been pointed out in \cite{Dvali}, the absence of such a term would require the presence of a $\phi\to-\phi$ symmetry, but such a symmetry must be broken throughout most of the cosmological evolution.

With these assumptions one can explicitly relate the evolution of $\alpha$ to that of dark energy, as in \cite{Erminia1} whose derivation we summarize here. The evolution of $\alpha$ can be written
\begin{equation}
\frac{\Delta \alpha}{\alpha} \equiv \frac{\alpha-\alpha_0}{\alpha_0} =B_F^{-1}(\phi)-1=
\zeta \kappa (\phi-\phi_0) \,,
\end{equation}
and defining the fraction of the dark energy density
\begin{equation}
\Omega_\phi (z) \equiv \frac{\rho_\phi(z)}{\rho_{\rm tot}(z)} \simeq \frac{\rho_\phi(z)}{\rho_\phi(z)+\rho_m(z)} \,,
\end{equation}
where in the last step we have neglected the contribution from radiation (we will be interested in low redshifts, $z<5$, where it is indeed negligible), the evolution of the scalar field can be expressed in terms of the dark energy properties $\Omega_\phi$ and $w_\phi$ as \cite{Nunes}
\begin{equation}
1+w_\phi = \frac{(\kappa\phi')^2}{3 \Omega_\phi} \,,
\end{equation}
with the prime denoting the derivative with respect to the logarithm of the scale factor. Putting the two together we finally obtain
\begin{equation} \label{eq:dalfa}
\frac{\Delta\alpha}{\alpha}(z) =\zeta \int_0^{z}\sqrt{3\Omega_\phi(z')\left[1+w_\phi(z')\right]}\frac{dz'}{1+z'}\,.
\end{equation}
The above relation assumes a canonical scalar field, but the argument can be repeated for phantom fields \cite{Phantom}, leading to 
\begin{equation} \label{eq:dalfa2}
\frac{\Delta\alpha}{\alpha}(z) =-\zeta \int_0^{z}\sqrt{3\Omega_\phi(z')\left|1+w_\phi(z')\right|}\frac{dz'}{1+z'}\,;
\end{equation}
the change of sign stems from the fact that one expects phantom field to roll up the potential rather than down. Note that in these models the evolution of $\alpha$ can be expressed as a function of cosmological parameters plus the coupling $\zeta$, without explicit reference to the putative underlying scalar field.

In these models the proton and neutron masses are also expected to vary, due to the electromagnetic corrections of their masses. One consequence of this fact is that local tests of the Equivalence Principle lead to the conservative general constraint on the dimensionless coupling parameter (see \cite{uzanLR} for an overview)
\begin{equation}
|\zeta_{\rm local}|<10^{-3}\,.\label{localzeta}
\end{equation}
A few-percent constraint on this coupling was also obtained using CMB and large-scale structure data in combination with direct measurements of the expansion rate of the universe \cite{Erminia1}. We will presently discuss how these constraints can be improved.

We note that there is in principle an additional source term driving the evolution of the scalar field, due to a $F^2B_F'$ term. By comparison to the standard (kinetic and potential energy) terms, the contribution of this term is expected to be subdominant, both because its average is zero for a radiation fluid and because the corresponding term for the baryonic density is constrained to be small by the same reasons discussed in the previous paragraph. For these reasons, in what follows we neglect this term, which would lead to spatial (or, more accurately, environmental) dependencies. We nevertheless note that this term can play a role in scenarios where the dominant standard term is suppressed.

The realization that varying fundamental couplings induce violations of the universality of free fall is several decades old, going back at least to the work of Dicke---we refer the reader to \cite{Damour} for a recent thorough discussion. A light scalar field such as we are considering inevitably couples to nucleons due to the $\alpha$ dependence of their masses, and therefore it mediates an isotope-dependent long-range force. This can be quantified through the dimensionless E\"{o}tv\"{o}s parameter $\eta$, which describes the level of violation of the WEP. One can show that for the class of models we are considering the  E\"{o}tv\"{o}s parameter and the dimensionless coupling $\zeta$ are simply related by \cite{Dvali,Chiba,Damour,uzanLR}
\begin{equation} \label{eq:eotvos}
\eta \approx 10^{-3}\zeta^2\,;
\end{equation}
therefore, the constraints on $\zeta$ obtained in \cite{Pinho,Pinho2} lead to the two-sigma indirect bound
\begin{equation} \label{etaboundfix}
\eta<{\rm few}\times10^{-14}\,,
\end{equation}
the exact factor being somewhat model-dependent. In any case this is roughly one order of magnitude stronger than the current direct bounds that will be discussed below. We emphasize that this relation only applies to Class I models. For other models, called Class II in the classification of \cite{cjmGRG}, the constraints are weaker by about a factor of two \cite{Pinho2}.

\begin{figure}[!]
\begin{center}
\includegraphics[width=3in]{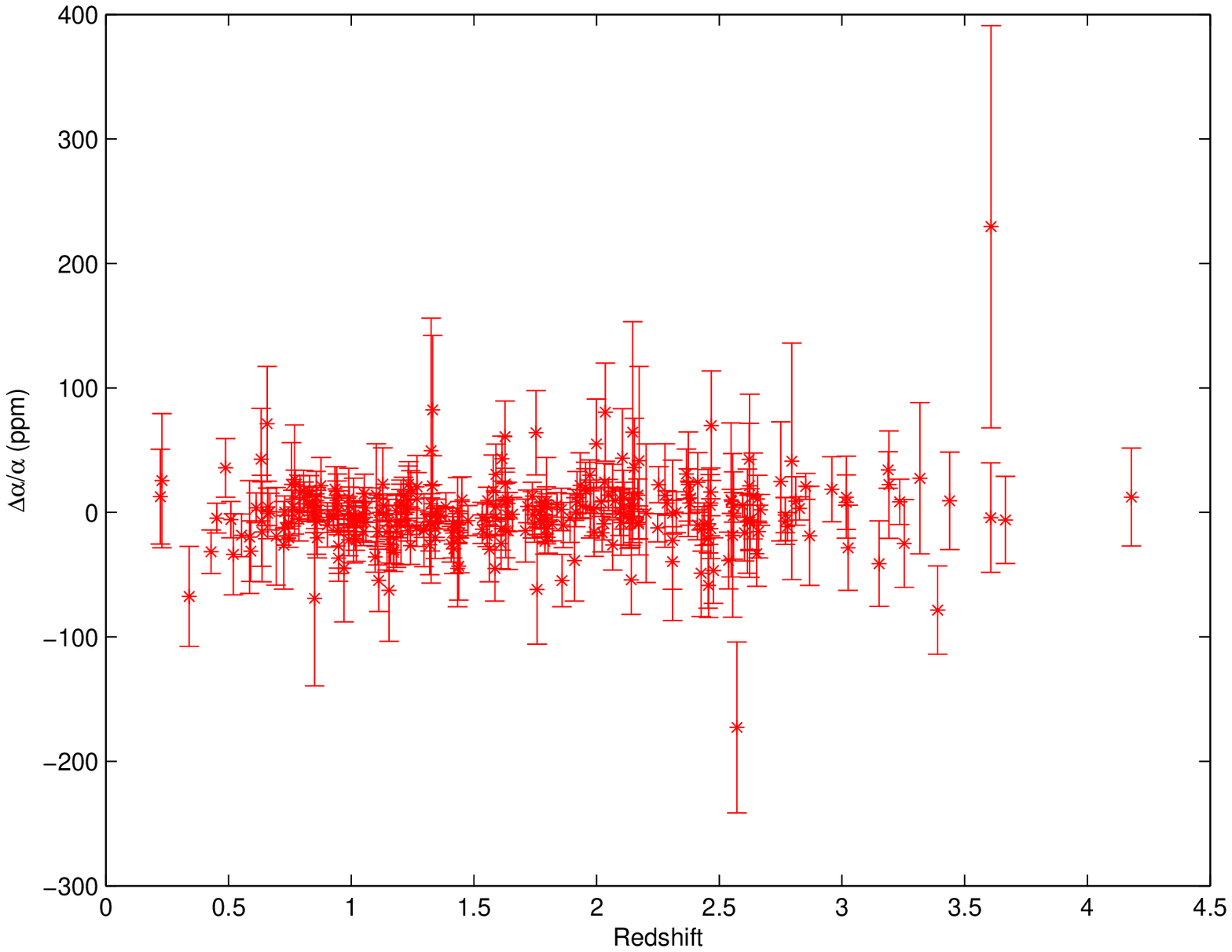}
\includegraphics[width=3in]{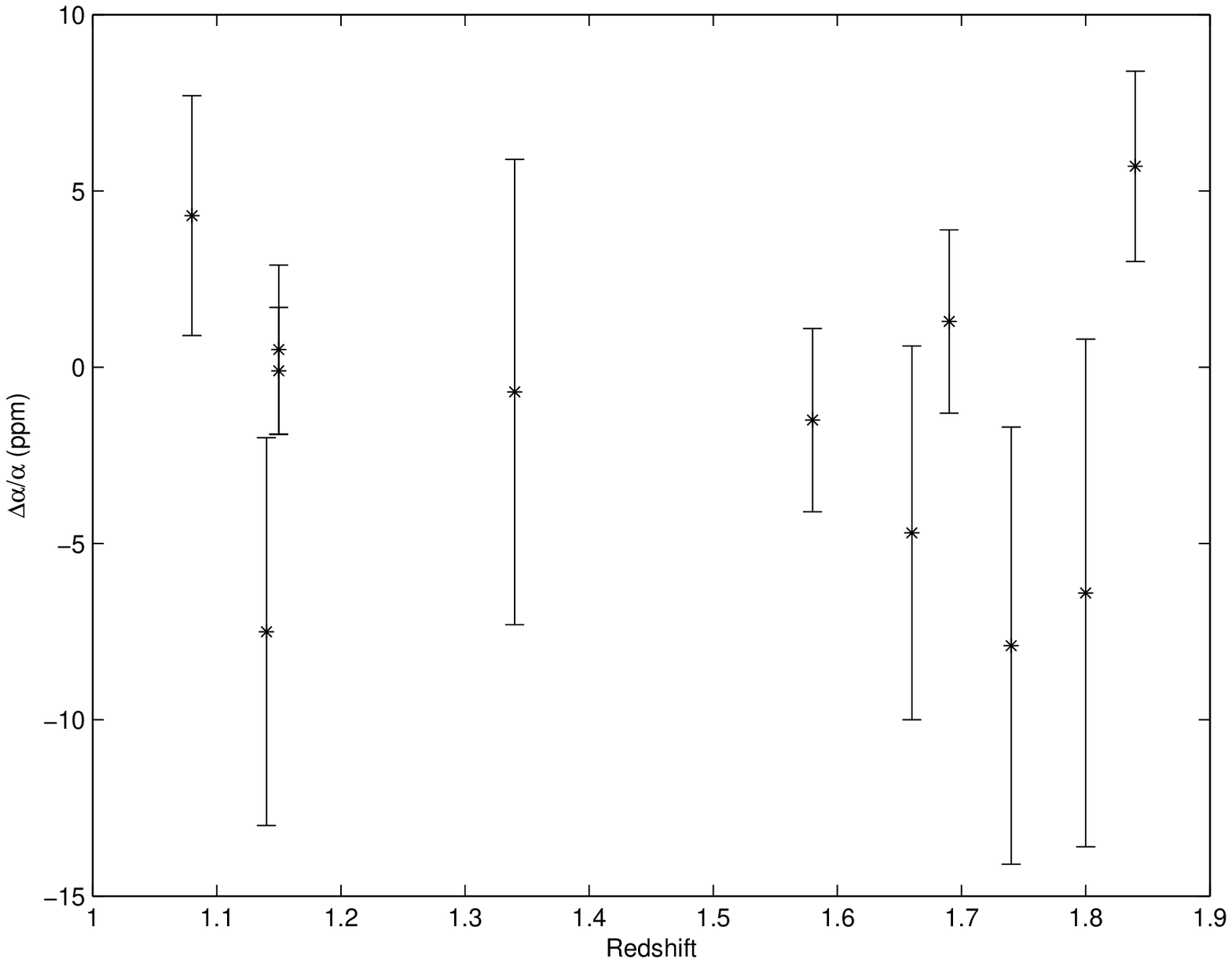}
\end{center}
\caption{\label{fig01}Currently available fine-structure constant measurements, with the relative values $\Delta\alpha/\alpha$ plotted as a function of redshift. The data of \protect\cite{Webb} is shown on the top panel, while the more recent data of Table \protect\ref{table1} is shown on the bottom panel. In both cases the error bars include both statistical and systematic uncertainties, added in quadrature. Note the difference in the vertical scales of both panels.}
\end{figure}

\section{From data to constraints}
\label{sec:data}

We will constrain dynamical dark energy models coupled to the electromagnetic sector, by using the same datasets that were also used in \cite{Pinho,Pinho2}, as follows
\begin{itemize}
\item Cosmological data: we use the Union2.1 dataset of 580 Type Ia supernovas \cite{Union} and the compilation of 28 Hubble parameter measurements from Farooq \& Ratra \cite{Farooq}. These datasets are, to a good approximation, insensitive to the value of the coupling $\zeta$. Strictly speaking a varying $\alpha$ does affect the luminosity of Type Ia supernovas but, as recently shown in \cite{Erminia2}, for parts-per-million level $\alpha$ variations the effect is too small to have an impact on current datasets, and we therefore neglect it in the present analysis. This data constrains the dark energy equation of state, effectively providing us with a prior on it.
\item Laboratory data: we will use the atomic clock constraint on the current drift of $\alpha$ of Rosenband {\it et al.} \cite {Rosenband},
\begin{equation} \label{clocks0}
\frac{\dot\alpha}{\alpha} =(-1.6\pm2.3)\times10^{-17}\,{\rm yr}^{-1}\,.
\end{equation}
which we can also write in a dimensionless form by dividing by the present-day Hubble parameter,
\begin{equation} \label{clocks}
\frac{1}{H_0}\frac{\dot\alpha}{\alpha} =(-2.2\pm3.2)\times10^{-7}\,.
\end{equation}
This is the strongest available laboratory constraint on $\alpha$ only. Other existing laboratory constraints are weaker and also depend on other couplings. (The interested reader can find overviews of atomic clock tests in \cite{Luo,Ferreira,Ferreira2}.) For the models under consideration this translates into
\begin{equation} \label{clocks2}
\frac{1}{H_0}\frac{\dot\alpha}{\alpha} =-\, \Sigma\, \zeta\sqrt{3\Omega_{\phi0}|1+w_0|}\,,
\end{equation}
where $\Sigma$ denotes the sign of $(1+w_0)$, so it is $+1$ for canonical fields and $-1$ for phantom fields.
\item Astrophysical data: we will use both the spectroscopic measurements of $\alpha$ of Webb {\it et al.} \cite{Webb} (a large dataset of 293 archival data measurements) and the smaller but more recent dataset of 11 dedicated measurements listed in Table \ref{table1}. The latter include the early results of the UVES Large Program for Testing Fundamental Physics \cite{LP1,LP3}, which is expected to be the one with a better control of possible systematics. Figure \ref{fig01} depicts both of these datasets.
\end{itemize}

\begin{table}
\centering
\begin{tabular}{|c|c|c|c|c|}
\hline
 Object & z & ${ \Delta\alpha}/{\alpha}$ (ppm) & Spectrograph & Ref. \\ 
\hline\hline
3 sources & 1.08 & $4.3\pm3.4$ & HIRES & \protect\cite{Songaila} \\
\hline
HS1549$+$1919 & 1.14 & $-7.5\pm5.5$ & UVES/HIRES/HDS & \protect\cite{LP3} \\
\hline
HE0515$-$4414 & 1.15 & $-0.1\pm1.8$ & UVES & \protect\cite{alphaMolaro} \\
\hline
HE0515$-$4414 & 1.15 & $0.5\pm2.4$ & HARPS/UVES & \protect\cite{alphaChand} \\
\hline
HS1549$+$1919 & 1.34 & $-0.7\pm6.6$ & UVES/HIRES/HDS & \protect\cite{LP3} \\
\hline
HE0001$-$2340 & 1.58 & $-1.5\pm2.6$ &  UVES & \protect\cite{alphaAgafonova}\\
\hline
HE1104$-$1805A & 1.66 & $-4.7\pm5.3$ & HIRES & \protect\cite{Songaila} \\
\hline
HE2217$-$2818 & 1.69 & $1.3\pm2.6$ &  UVES & \protect\cite{LP1}\\
\hline
HS1946$+$7658 & 1.74 & $-7.9\pm6.2$ & HIRES & \protect\cite{Songaila} \\
\hline
HS1549$+$1919 & 1.80 & $-6.4\pm7.2$ & UVES/HIRES/HDS & \protect\cite{LP3} \\
\hline
Q1101$-$264 & 1.84 & $5.7\pm2.7$ &  UVES & \protect\cite{alphaMolaro}\\
\hline
\end{tabular}
\caption{\label{table1}Recent dedicated measurements of $\alpha$. Listed are, respectively, the object along each line of sight, the redshift of the measurement, the measurement itself (in parts per million), the spectrograph, and the original reference. The first measurement is the weighted average from 8 absorbers in the redshift range $0.73<z<1.53$ along the lines of sight of HE1104-1805A, HS1700+6416 and HS1946+7658, reported in \cite{Songaila} without the values for individual systems. The UVES, HARPS, HIRES and HDS spectrographs are respectively in the VLT, ESO 3.6m, Keck and Subaru telescopes.}
\end{table}

We use these datasets to constrain the dynamical dark energy models which will be described in the following sections. The behavior of $\alpha$ is determined by Eq.(\ref{eq:dalfa}) for canonical equations of state ($w(z)\ge-1$) and Eq.(\ref{eq:dalfa2}) for phantom equations of state ($w(z)<-1$). While in \cite{Pinho,Pinho2} we studied models whose equations of state were parametrized by a single parameter (its present day value, $w_0$) here we relax this assumption and study more general models.

For comparison, we also list here the available direct constraints on the dimensionless E\"{o}tv\"{o}s parameter, quantifying violations to the Weak Equivalence Principle. These stem from torsion balance tests, leading to \cite{Torsion}
\begin{equation}\label{boundetaT}
\eta=(-0.7\pm1.3)\times10^{-13}\,,
\end{equation}
while from lunar laser ranging one obtains \cite{Lunar}
\begin{equation}\label{boundetaL}
\eta=(-0.8\pm1.2)\times10^{-13}\,.
\end{equation}
Both of these are quoted with their one-sigma uncertainties.

Our main interest is in obtaining constraints on $\zeta$ and the dark energy parameters. For this reason we will fix the Hubble parameter to be $H_0=70$ km/s/Mpc and the matter density to be $\Omega_{m0}=0.3$, and further assume a flat universe, so $\Omega_{\phi0}=0.7$. This choice of cosmological parameters is fully consistent with the supernova and Hubble parameter data we use.

Moreover, in \cite{Pinho,Pinho2} we have explicitly verified that allowing $H_0$, $\Omega_m$ or the curvature parameter to vary (within observationally reasonable ranges) and marginalizing over them does not significantly change our results. This should be intuitively clear: a parts-per-million variation of $\alpha$ cannot noticeably affect these cosmological parameters. It is clear that the critical cosmological parameters here are the ones describing the dark energy equation of state, as in Class I models they will be correlated with $\zeta$---cf. Eqs.(\ref{eq:dalfa}--\ref{eq:dalfa2}). We therefore consider 3D grids of $\zeta$, $w_0$ and the additional (model-dependent) parameter, and use standard maximum likelihood techniques to compare the models and the data. Flat priors on the relevant parameters will be used, unless otherwise stated.

\section{Standard dark energy}

We will start by studying canonical dark energy models in which the fraction of the dark energy density $\Omega_\phi (z)$ tends to zero at high redshift. Note that in this case the relative variation of $\alpha$ will tend to a constant in the same limit. We will study the standard CPL parametrization but will also aim to gain some insight on the degree of model-dependence of these results by considering an alternative parametrization.

\subsection{CPL parametrization}

\begin{figure*}[!]
\centering
\includegraphics[width=3in]{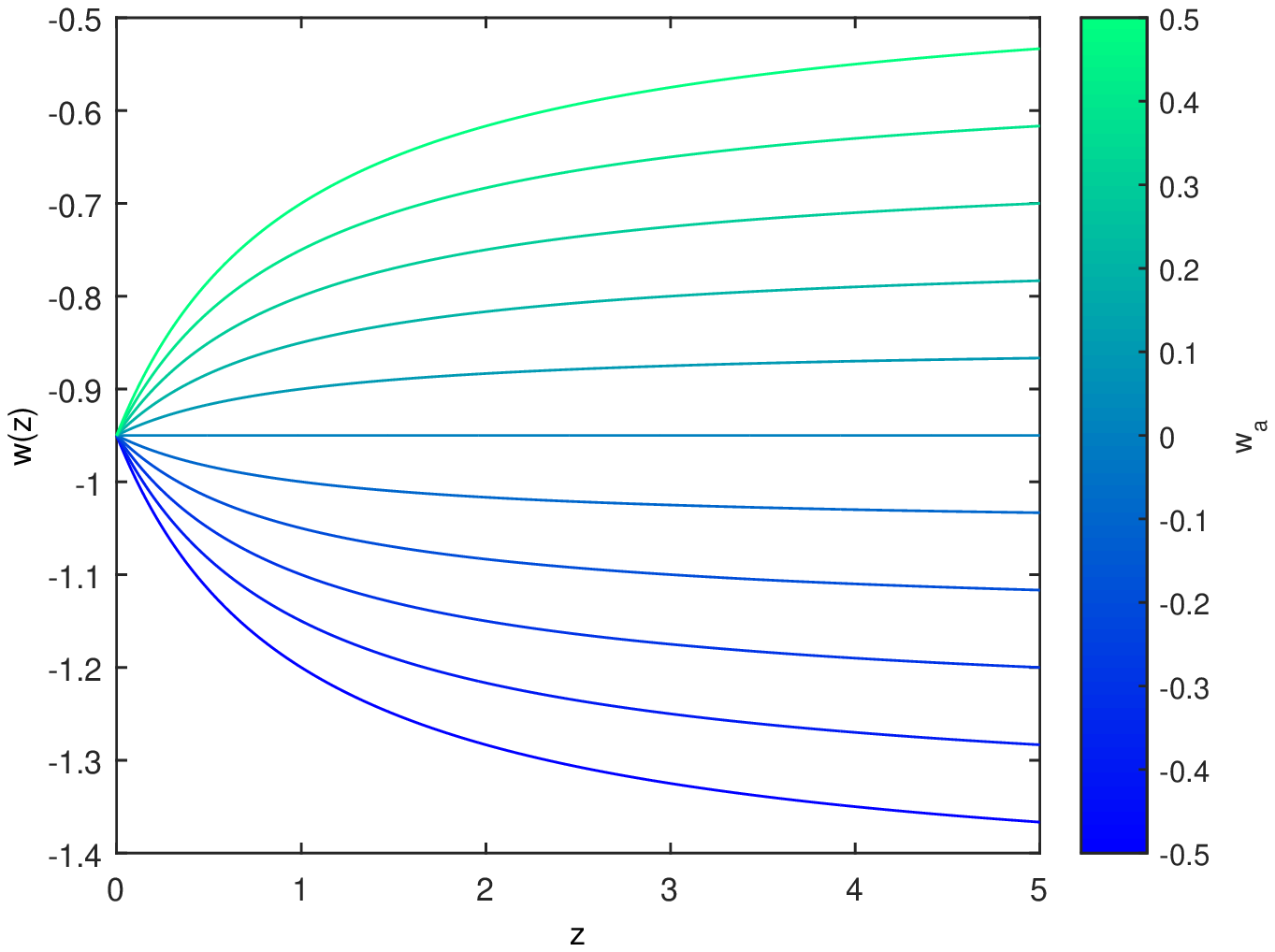}
\includegraphics[width=3in]{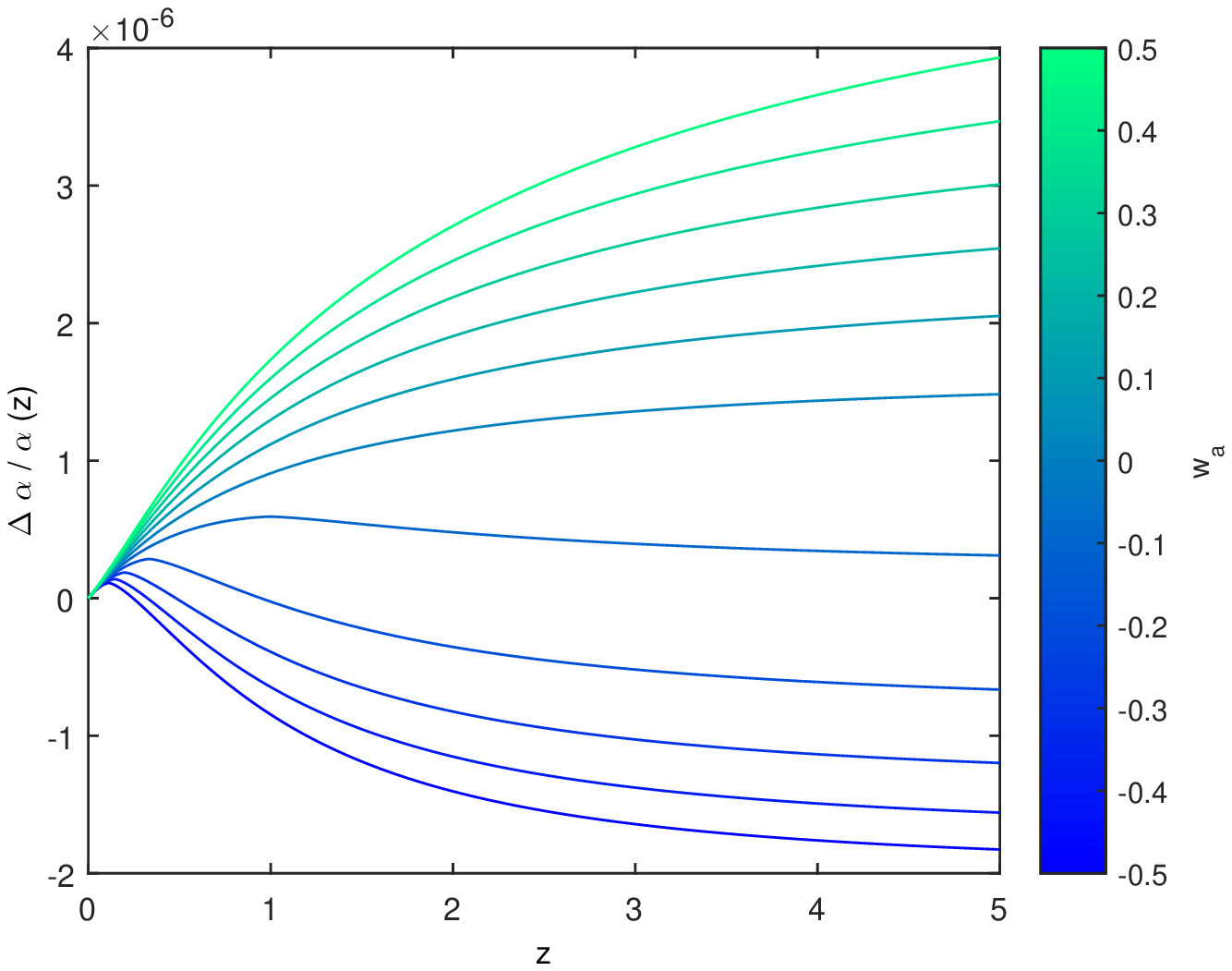}
\includegraphics[width=3in]{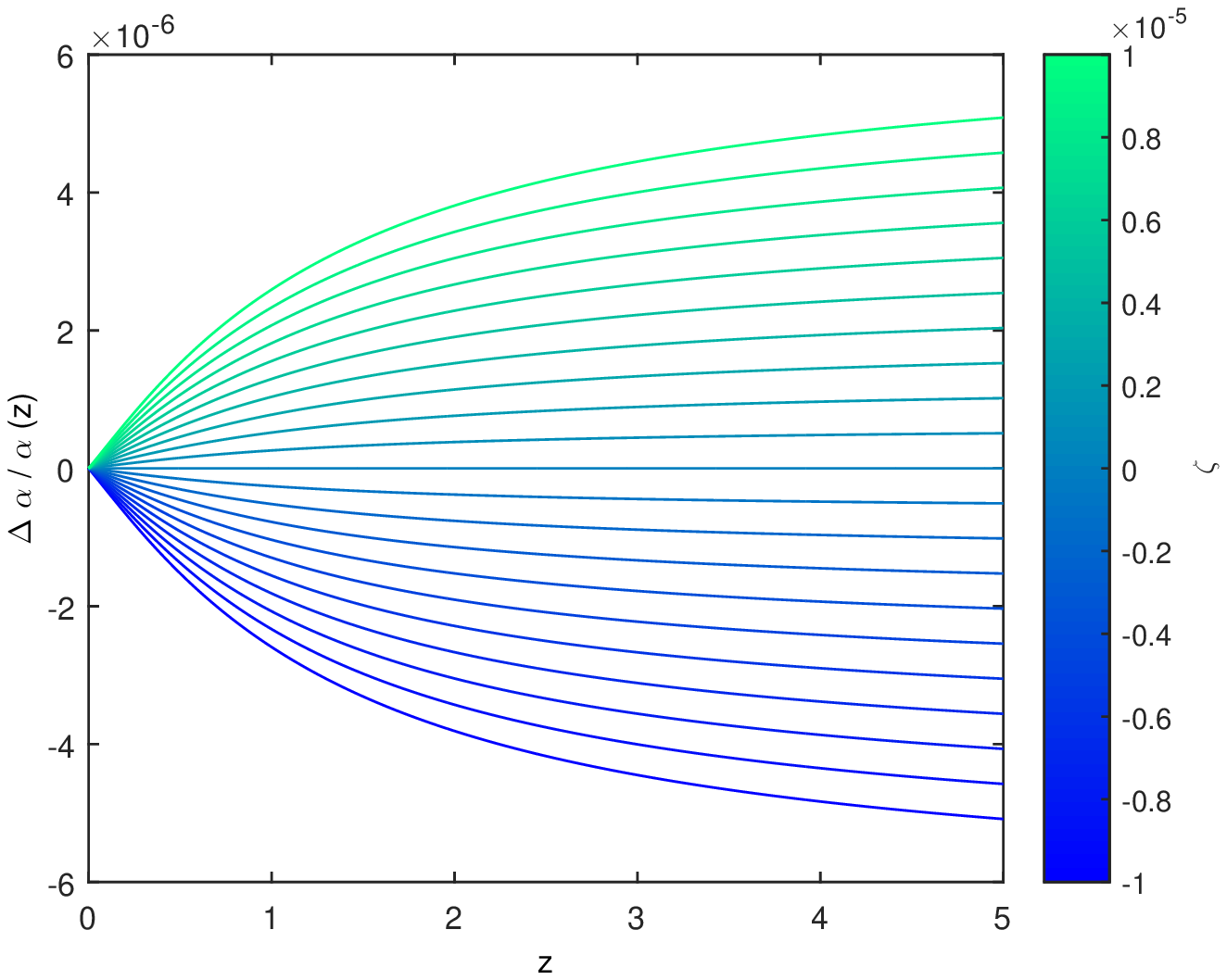}
\includegraphics[width=3in]{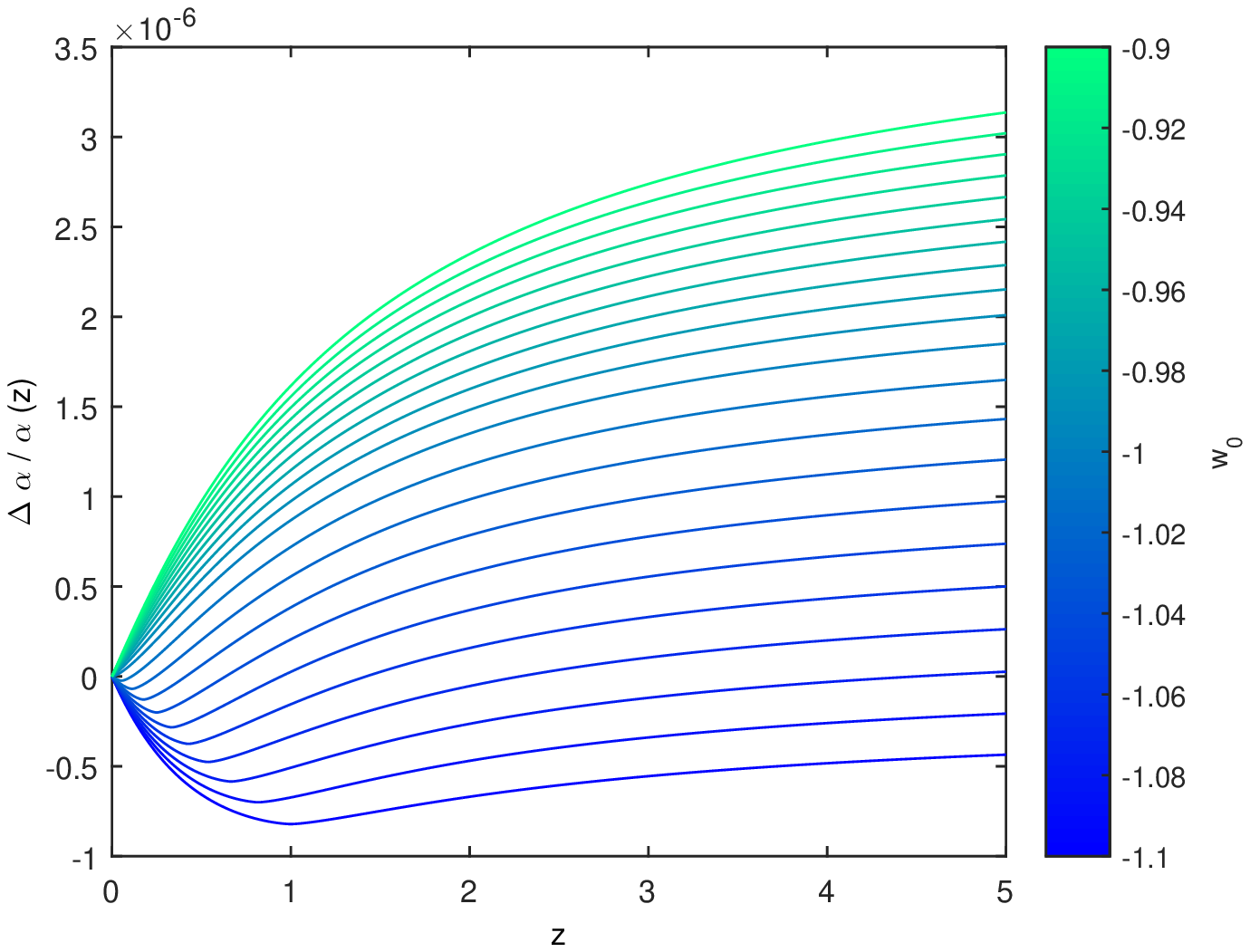}
\caption{\label{figCPL}Redshift dependence of relevant parameters in the CPL model. {\bf Top left}: $w(z)$ with $w_0=-0.95$ and $-0.5\le w_a\le 0.5$; {\bf Top right}: $\Delta\alpha/\alpha(z)$ with $w_0=-0.95$, $\zeta = 5\times10^{-6}$ and $-0.5\le w_a\le 0.5$; {\bf Bottom left}: $\Delta\alpha/\alpha(z)$ with $w_0=-0.95$, $w_a=0.2$ and $-1\times10^{-5}\le \zeta \le+1\times10^{-5}$; {\bf Bottom right}: $\Delta\alpha/\alpha(z)$ with $w_a=0.2$, $\zeta = 5\times10^{-6}$, and $-1.1\le w_0\le -0.9$.}
\end{figure*}

In the Chevallier-Polarski-Linder \cite{CPL1,CPL2} parametrization the dark energy equation of state is written as
\begin{equation} \label{cpl}
w_{\rm CPL}(z)=w_0+w_a \frac{z}{1+z}\,,
\end{equation}
where $w_0$ is its present value and $w_a$ is the coefficient of the time-dependent term. The redshift dependence of this parametrization is not intended to mimic a particular model for dark energy, but rather to allow to probe possible deviations from the $\Lambda$CDM standard paradigm without the assumption of any underlying theory. Nevertheless, we can assume that also this kind of dark energy is produced by a scalar field, coupled to the electromagnetic sector. In this model the fraction of energy density provided by the scalar field is easily found to be
\begin{equation}
\Omega_{\rm CPL}(z)=\frac{1-\Omega_{\rm m}}{1-\Omega_{\rm m}+\Omega_{\rm m}(1+z)^{-3(w_0+w_a)}e^{(3w_az/1+z)}}\,.
\end{equation}
where $\Omega_{\rm m}$ is the present time matter density and we have also assumed a flat universe. Figure \ref{figCPL} illustrates the behavior of $w(z)$ and $\Delta\alpha/\alpha(z)$ in this model for realistic parameter choices, compatible with our cosmological datasets and the recent Planck collaboration results \cite{Planck} (which are also used to choose priors for $w_0$ and $w_a$).

\begin{figure}[!]
\centering
\includegraphics[width=3in]{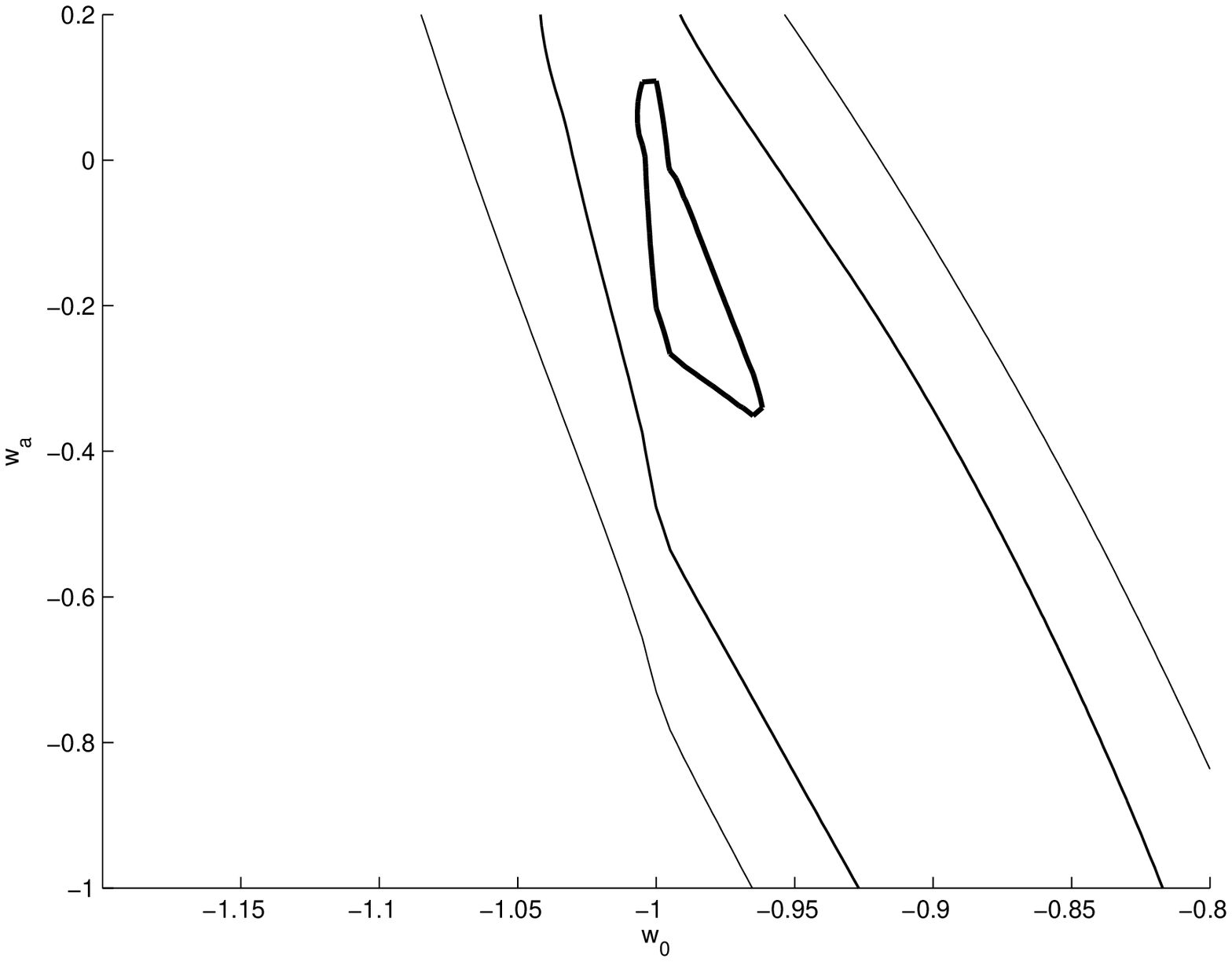}
\includegraphics[width=3in]{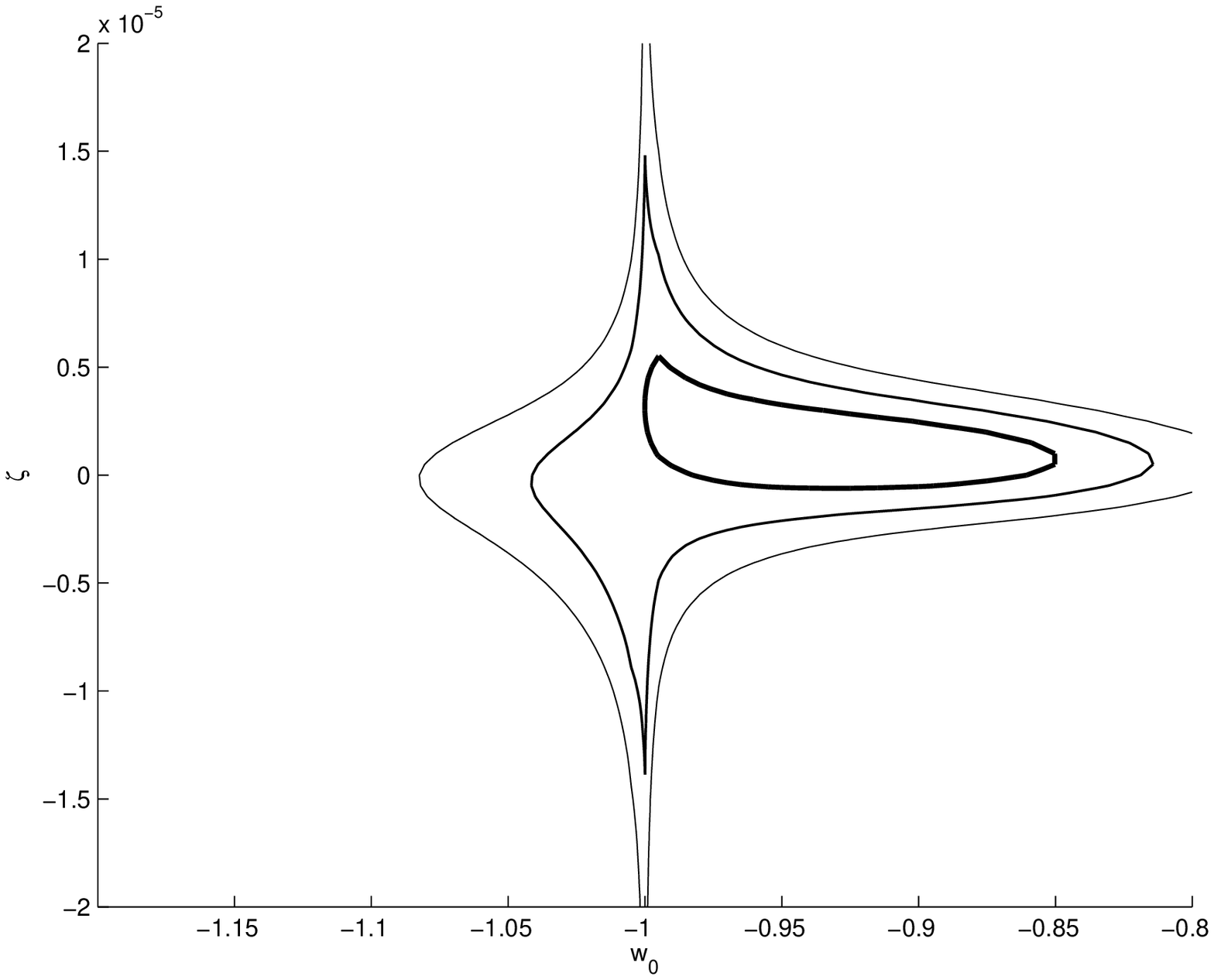}
\includegraphics[width=3in]{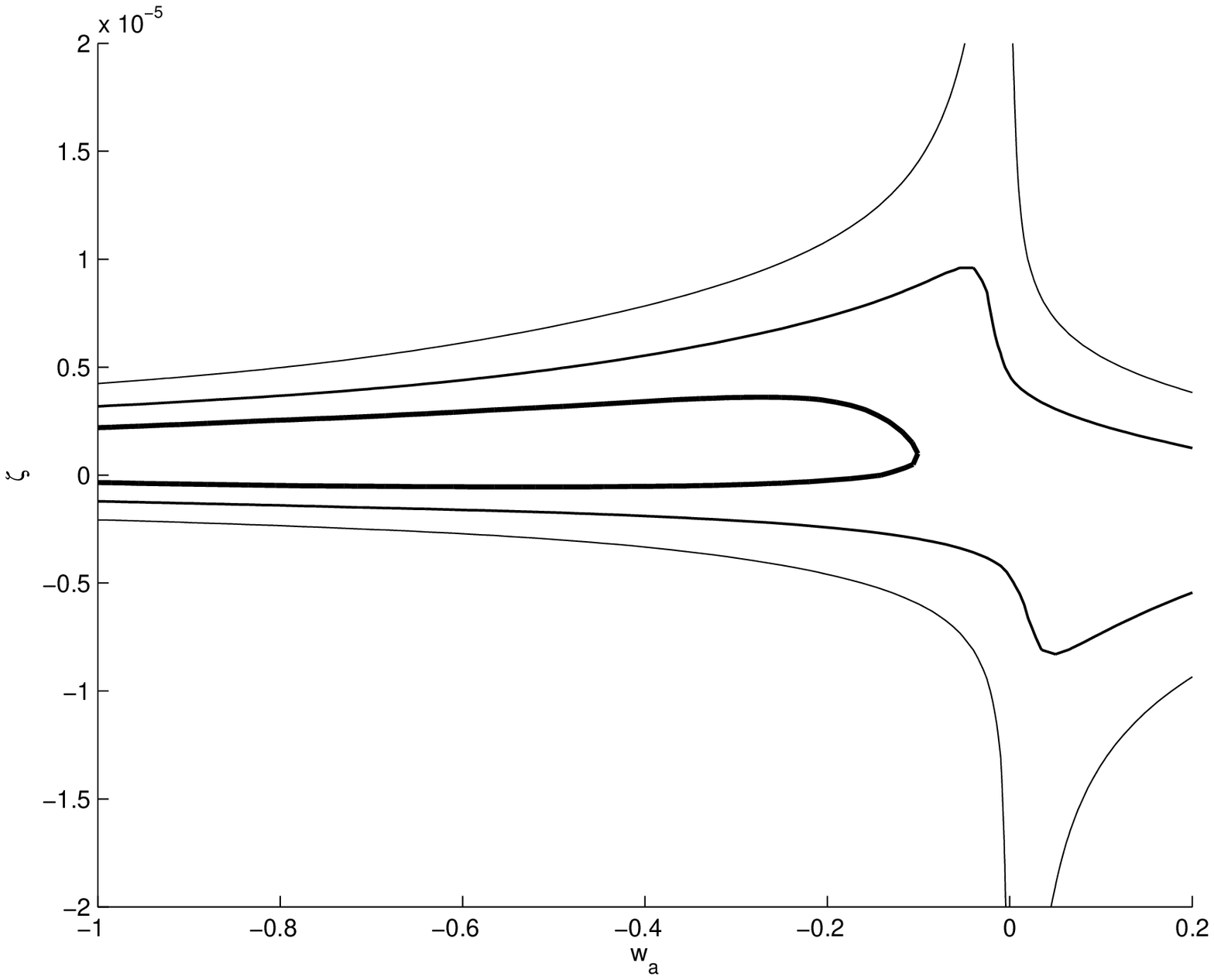}
\caption{\label{figCPL2d}2D constraints on the CPL parametrization from the full (cosmological plus atomic clock plus astrophysical) datasets described in the main text, in the $w_0$-$w_a$ (top panel), $w_0$-$\zeta$ (middle panel) and $w_a$-$\zeta$ (bottom panel) planes, with the remaining parameter marginalized. One, two and three sigma contours are shown in all cases.}
\end{figure}
\begin{figure}[!]
\centering
\includegraphics[width=3in]{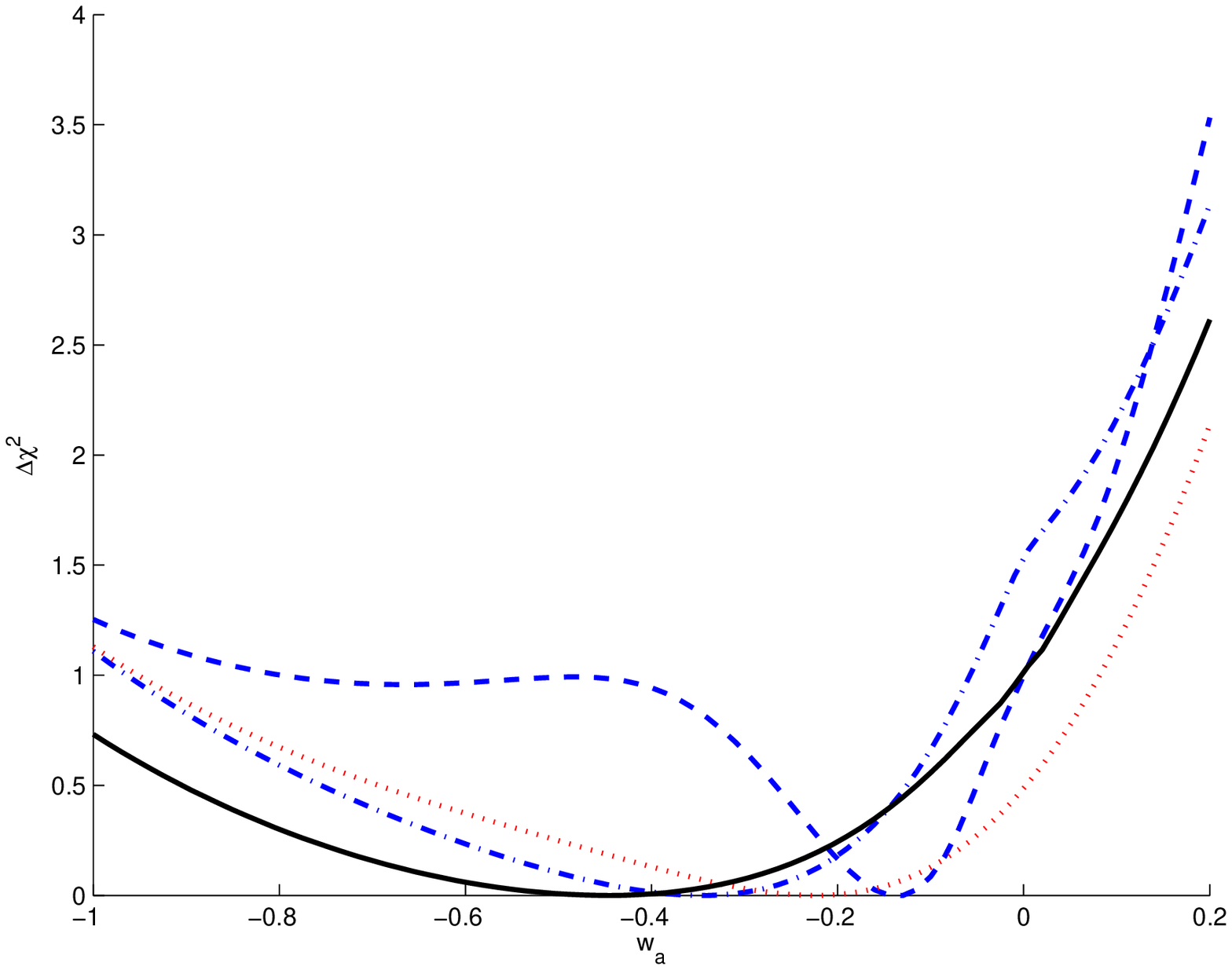}
\includegraphics[width=3in]{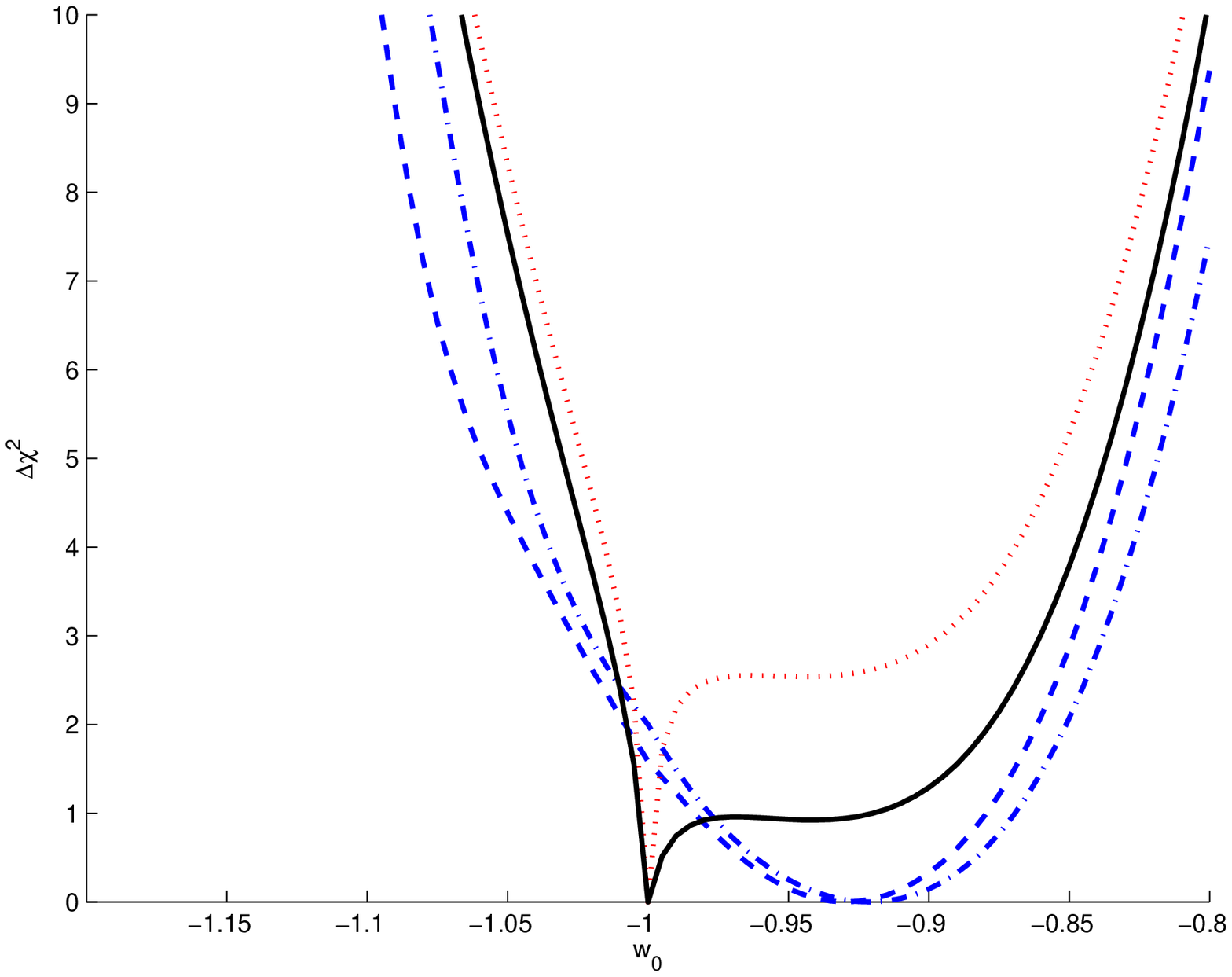}
\includegraphics[width=3in]{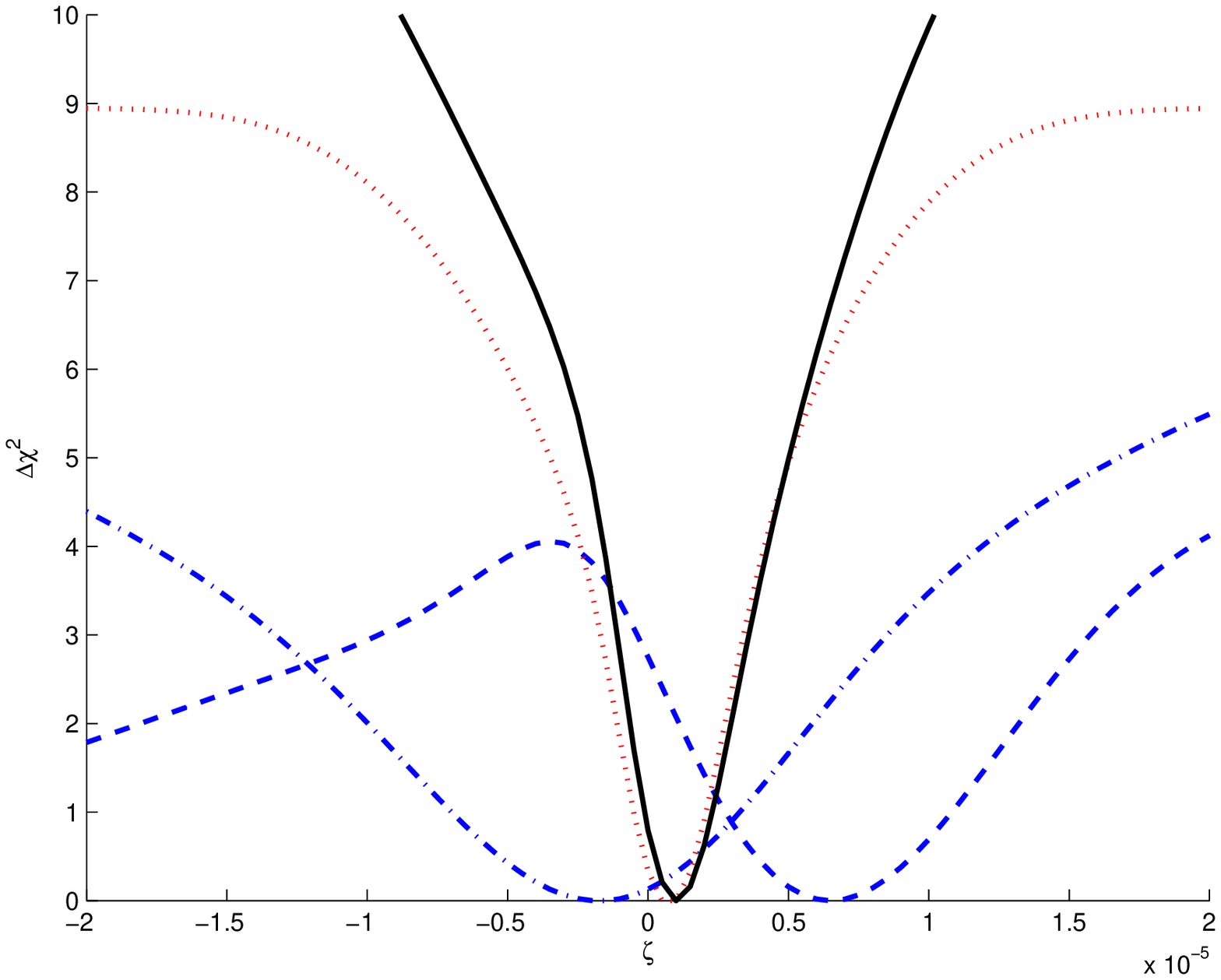}
\caption{\label{figCPL1d}1D marginalized constraints on $w_a$ (top panel), $w_0$ (middle panel) and $\zeta$ (bottom panel) assuming the CPL parametrization. Different lines correspond to different datasets: constraints from cosmological plus Webb \protect\textit{et al.} are shown by the dashed blue lines. cosmological plus dedicated $\alpha$ measurements in dash-dotted blue, cosmological plus atomic clocks in dotted red, and the full sample in solid black lines. In all cases the vertical axis depicts $\Delta\chi^2=\chi^2-\chi^2_{\rm min}$.}
\end{figure}

Figure \ref{figCPL2d} shows the 2D marginalized constraints from our full (i.e., cosmological plus atomic clock plus astrophysical) datasets in the three relevant planes, with the remaining parameter marginalized. One, two and three sigma contours are shown in all cases. Several degeneracies are clearly visible which among other things imply that no significant constraints can be obtained on $w_a$. However, this is not the case for the other two parameters. As explained in previous work \cite{Pinho,Pinho2}, the cosmological datasets effectively provide us with priors on the dark energy behavior close to the present day, partially breaking otherwise unavoidable degeneracies with $\zeta$ and thereby enabling substantive constraints on it.

Figure \ref{figCPL1d} shows the 1D marginalized likelihoods for each of the three parameters, for the full dataset we use as well as for several choices of sub-sets. Specifically one may note the qualitatively different behavior of the Webb \textit{et al.} and dedicated $\alpha$ measurements: the former is not consistent with the null result for $\alpha$ \cite{Webb}, and we correspondingly find a one sigma preference for a non-zero coupling $\zeta$. However, this data is compatible with the null result at the two sigma level. On the other hand, the Table \protect\ref{table1} data is fully compatible with the null result. Finally, the local atomic clock measurement \cite{Rosenband} is more constraining than the astrophysical measurements.

From these we obtain a very weak 1D marginalized constraint on $w_a$
\begin{equation} \label{cplwa}
w_a<0\qquad {\rm (68.3\% C.L.)} \,,
\end{equation}
while that for $w_0$ is stronger
\begin{equation} \label{cplw0}
w_0=-1.00^{+0.15}_{-0.02}\qquad {\rm (95.4\% C.L.)} \,
\end{equation}
and that for the coupling even more so
\begin{equation} \label{cplzeta1}
\zeta=(1\pm3)\times10^{-6}\qquad {\rm (95.4\% C.L.)} \,
\end{equation}
\begin{equation} \label{cplzeta2}
\zeta=(1\pm8)\times10^{-6}\qquad {\rm (99.7\% C.L.)} \,.
\end{equation}
Finally for the E\"{o}tv\"{o}s parameter we obtain
\begin{equation} \label{cpleta}
\eta<1.6\times10^{-14}\qquad {\rm (95.4\% C.L.)} \,
\end{equation}
Compared to earlier results \cite{Pinho,Pinho2} the constraint on $w_0$ becomes weaker (due to the additional freedom provided by the largely unconstrained $w_a$) while that on $\zeta$ (and consequently that on $\eta$) become correspondingly stronger. This is to be expected since $\zeta$ is correlated with the dark energy equation of state parameters: with the equation of state allowed to be further away from a cosmological constant, larger variations of $\alpha$ also become possible, and the existing $\alpha$ measurements therefore impose a tighter constraint on $\zeta$. This effect was also noticed in the case of the forecasts discussed in \cite{Erminia2}.

\subsection{Mukhanov parametrization}

It is interesting to assess the model-dependence of the above constraints on $w_0$ and $\zeta$, and a simple way to do so is to repeat the analysis for a different parametrization of the dark energy equation of state. We will do this through a parametrization recently discussed by Mukhanov in \cite{MKH}. This was introduced in an inflationary context, but it can be trivially applied for the case of the recent acceleration of the universe.

In this parametrization (which we will refer to as MKH) the dark energy equation of state is
\begin{equation} \label{mkheos}
w_{\rm MKH}(z)=-1+\frac{1+w_0}{\left[1+\ln{(1+z)}\right]^\beta}\,,
\end{equation}
where $w_0$ is its present day value and the slope $\beta$ controls the overall redshift dependence. Specifically $\beta<0$ corresponds to freezing models, $\beta=0$ to a constant equation of state and $\beta>0$ to thawing models, in the classification of \cite{FrTh}. This corresponds to the following behavior of the dark energy density
\begin{eqnarray}
\frac{\rho_{\rm MKH}(z)}{\rho_0} &=& \exp\left[3\frac{1+w_0}{1-\beta}\left([1+\ln(1+z)]^{1-\beta}-1\right) \right]\,, \beta\neq1 \\
\frac{\rho_{\rm MKH}(z)}{\rho_0} &=& \left[1+\ln{(1+z)}\right]^{3(1+w_0)} \,, \beta=1
\label{rhomkh}
\end{eqnarray}
and it is easy to verify that this has the correct behavior in the appropriate limits.

\begin{figure}[!]
\centering
\includegraphics[width=3in]{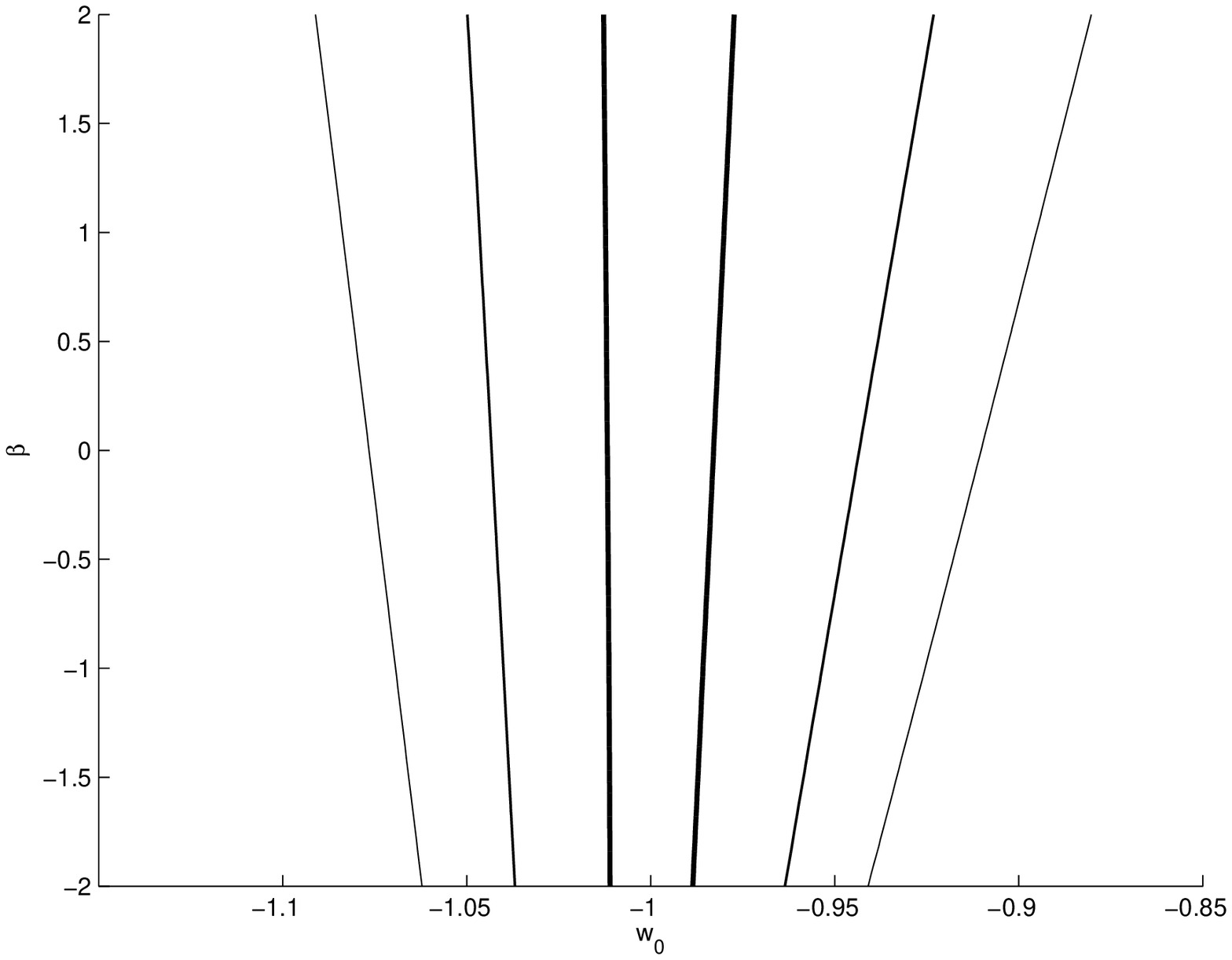}
\includegraphics[width=3in]{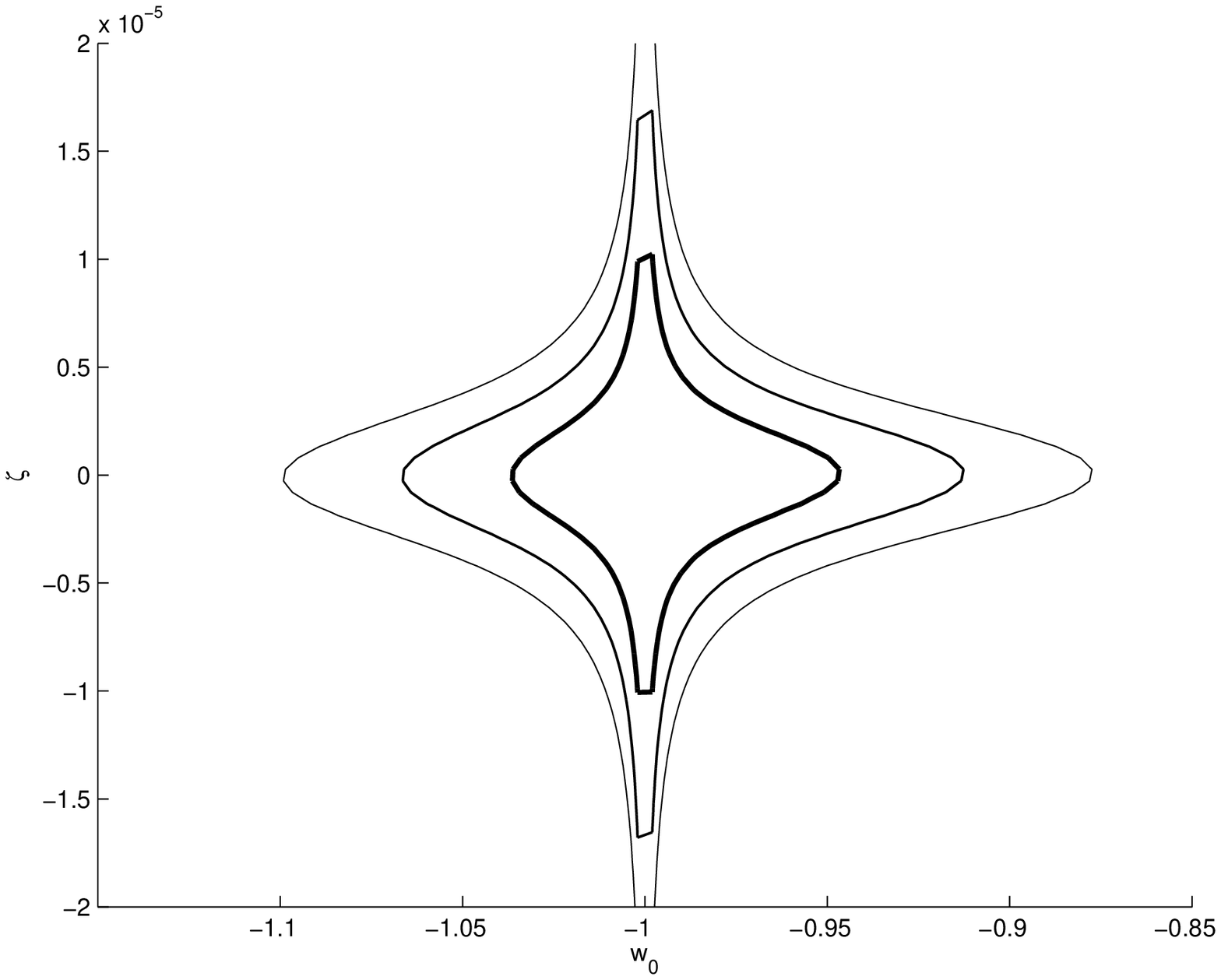}
\includegraphics[width=3in]{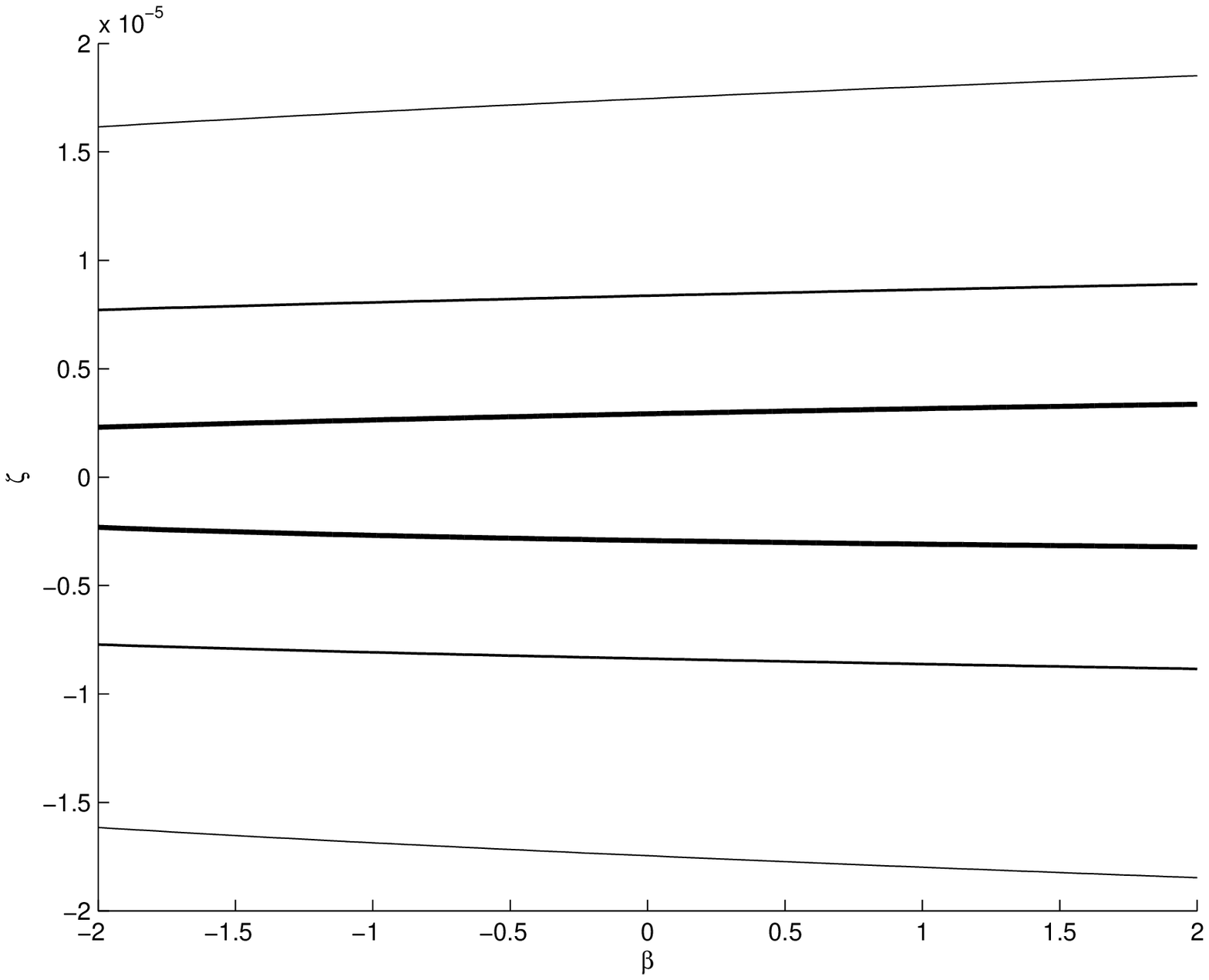}
\caption{\label{figMKH2d}2D constraints on the MKH parametrization from the full (cosmological plus atomic clock plus astrophysical) datasets described in the main text, in the $w_0$-$\beta$ (top panel), $w_0$-$\zeta$ (middle panel) and $\beta$-$\zeta$ (bottom panel) planes, with the remaining parameter marginalized. One, two and three sigma contours are shown in all cases.}
\end{figure}
\begin{figure}[!]
\centering
\includegraphics[width=3in]{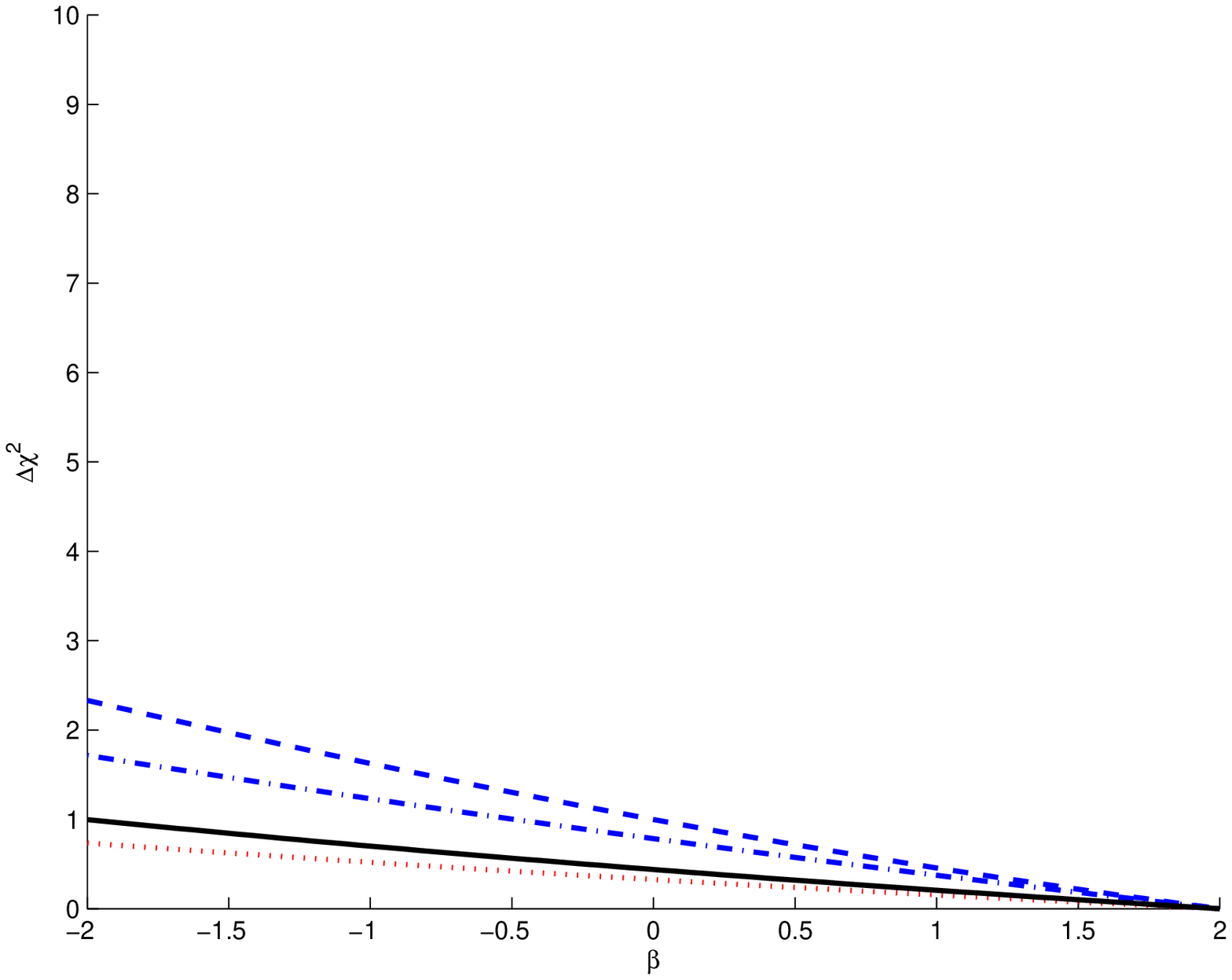}
\includegraphics[width=3in]{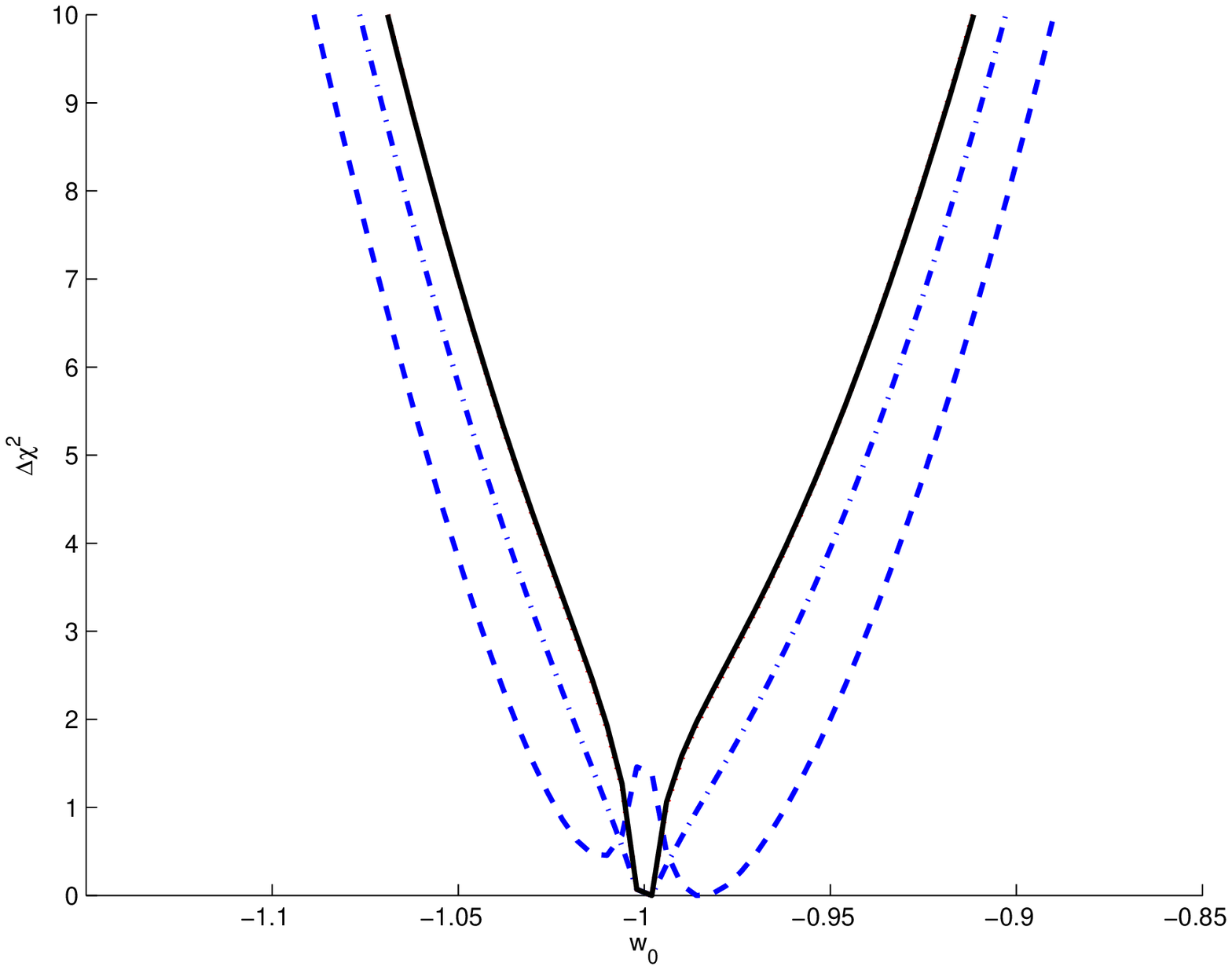}
\includegraphics[width=3in]{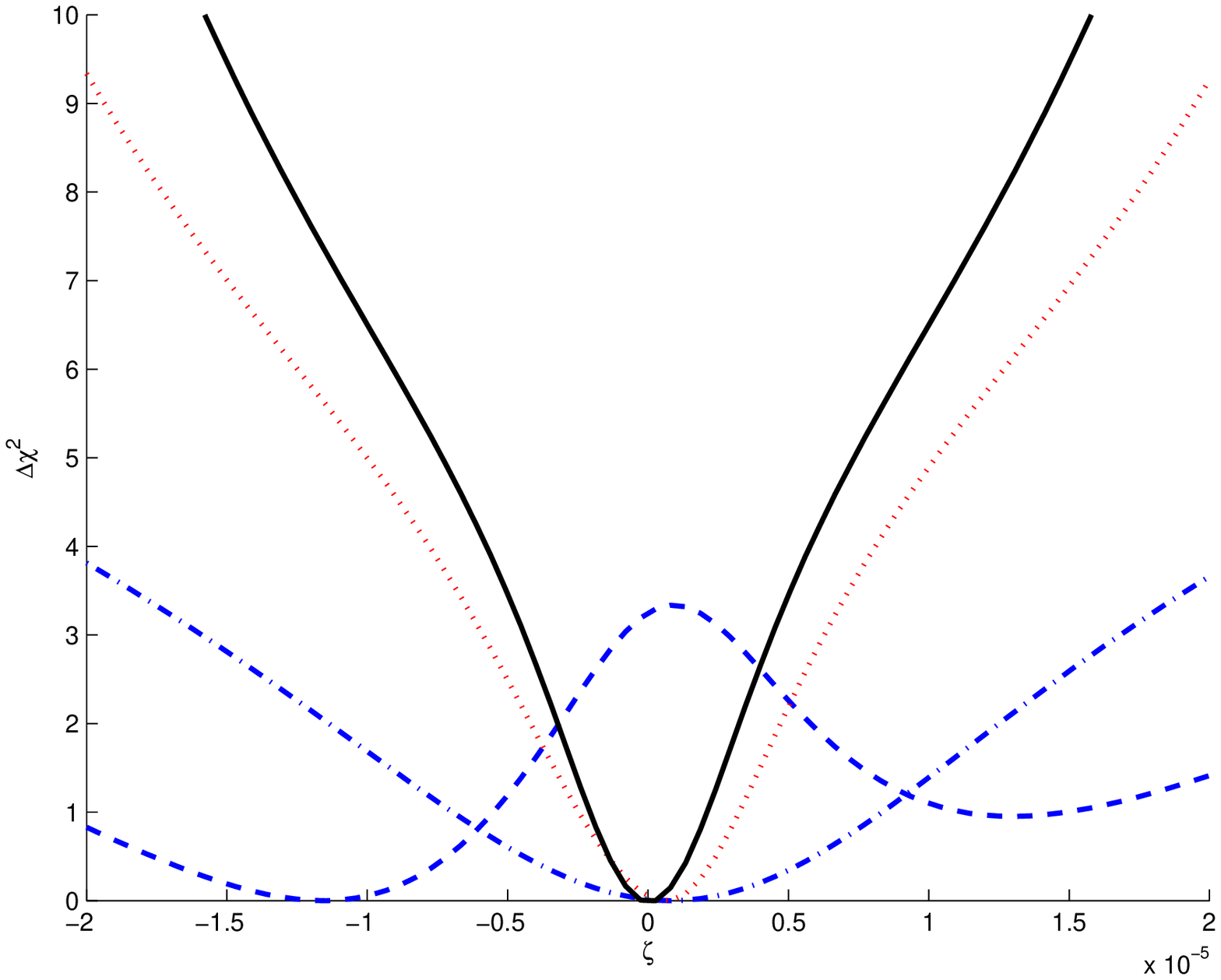}
\caption{\label{figMKH1d}1D marginalized constraints on $\beta$ (top panel), $w_0$ (middle panel) and $\zeta$ (bottom panel) assuming the MKH parametrization. Different lines correspond to different datasets: constraints from cosmological plus Webb \protect\textit{et al.} are shown by the dashed blue lines. cosmological plus dedicated $\alpha$ measurements in dash-dotted blue, cosmological plus atomic clocks in dotted red, and the full sample in solid black lines. In all cases the vertical axis depicts $\Delta\chi^2=\chi^2-\chi^2_{\rm min}$.}
\end{figure}

The analysis can now be repeated for this model, and the results are summarized in Figs. \ref{figMKH2d} and \ref{figMKH1d}, which can be compared to Figs.  \ref{figCPL2d} and \ref{figCPL1d}. Again there is no significant constraint on the slope $\beta$, although freezing models (with $\beta<0$) are comparatively more constrained than thawing ones (with $\beta>0$). Physically the reason for this is clear: for a given value of $w_0$, a freezing model leads to a larger variation of $\alpha$ than an thawing one, and is therefore more tightly constrained by the datasets we are considering.

In this case the 1D marginalized constraint on $w_0$ is
\begin{equation} \label{mkhw02}
w_0=-1.00^{+0.04}_{-0.03}\qquad {\rm (95.4\% C.L.)} 
\end{equation}
\begin{equation} \label{mkhw03}
w_0=-1.00^{+0.08}_{-0.07}\qquad {\rm (99.7\% C.L.)} \,
\end{equation}
while for the coupling we now find
\begin{equation} \label{mkhzeta}
\zeta=(0\pm6)\times10^{-6}\qquad {\rm (95.4\% C.L.)} \,
\end{equation}
leading to
\begin{equation} \label{mkheta}
\eta<3.6\times10^{-14}\qquad {\rm (95.4\% C.L.)} \,.
\end{equation}

In these case, and as compared to the CPL case, we get stronger constraints on $w_0$ and correspondingly weaker ones on $\zeta$. These constraints are comparable to those of the simpler models studied in \cite{Pinho,Pinho2}.

\section{Early dark energy}

We now study the Early Dark Energy (EDE) class of models \cite{EDE}. In this case the dark energy density fraction is
\begin{equation}
\Omega_{\rm EDE}(z) = \frac{1-\Omega_m - \Omega_e \left[1- (1+z)^{3 w_0}\right] }{1-\Omega_m + \Omega_m (1+z)^{-3w_0}} + \Omega_e \left[1- (1+z)^{3 w_0}\right] \label{edeomega} 
\end{equation}
while the dark energy equation of state is
\begin{equation}
w_{\rm EDE}(z)=-\frac{1}{3[1-\Omega_{\rm EDE}]} \frac{d\ln\Omega_{\rm EDE}}{d\ln a} + \frac{a_{eq}}{3(a + a_{eq})}\,;
\label{eq:edew}
\end{equation}
here $a_{eq}$ is the scale factor. The energy density $\Omega_{\rm EDE}(z)$ has a scaling behavior evolving with time and approaching a finite constant $\Omega_e$ in the past, rather than approaching zero as was the case for the models in the previous section. A flat universe is also assumed.

\begin{figure*}[!]
\centering
\includegraphics[width=3in]{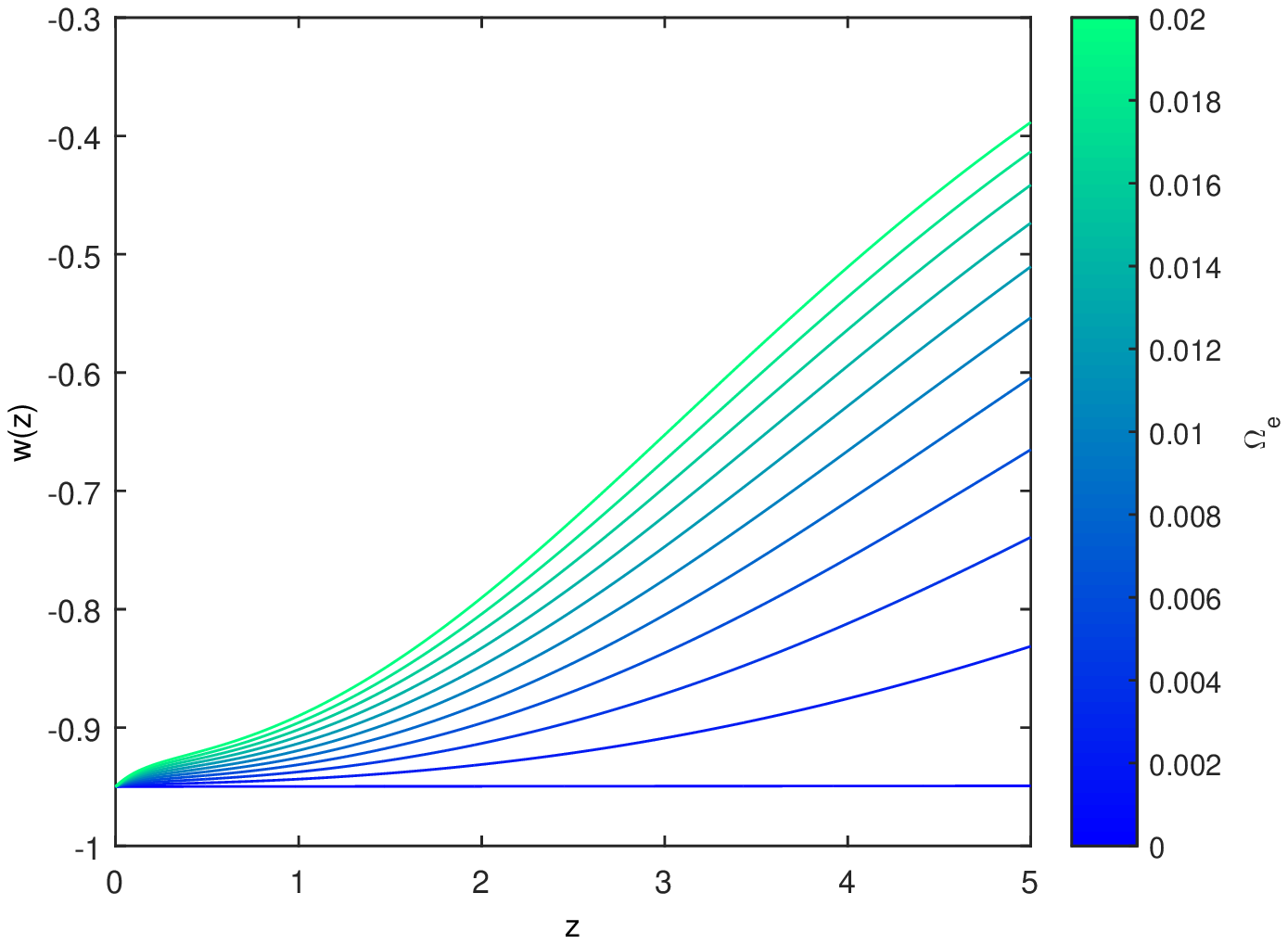}
\includegraphics[width=3in]{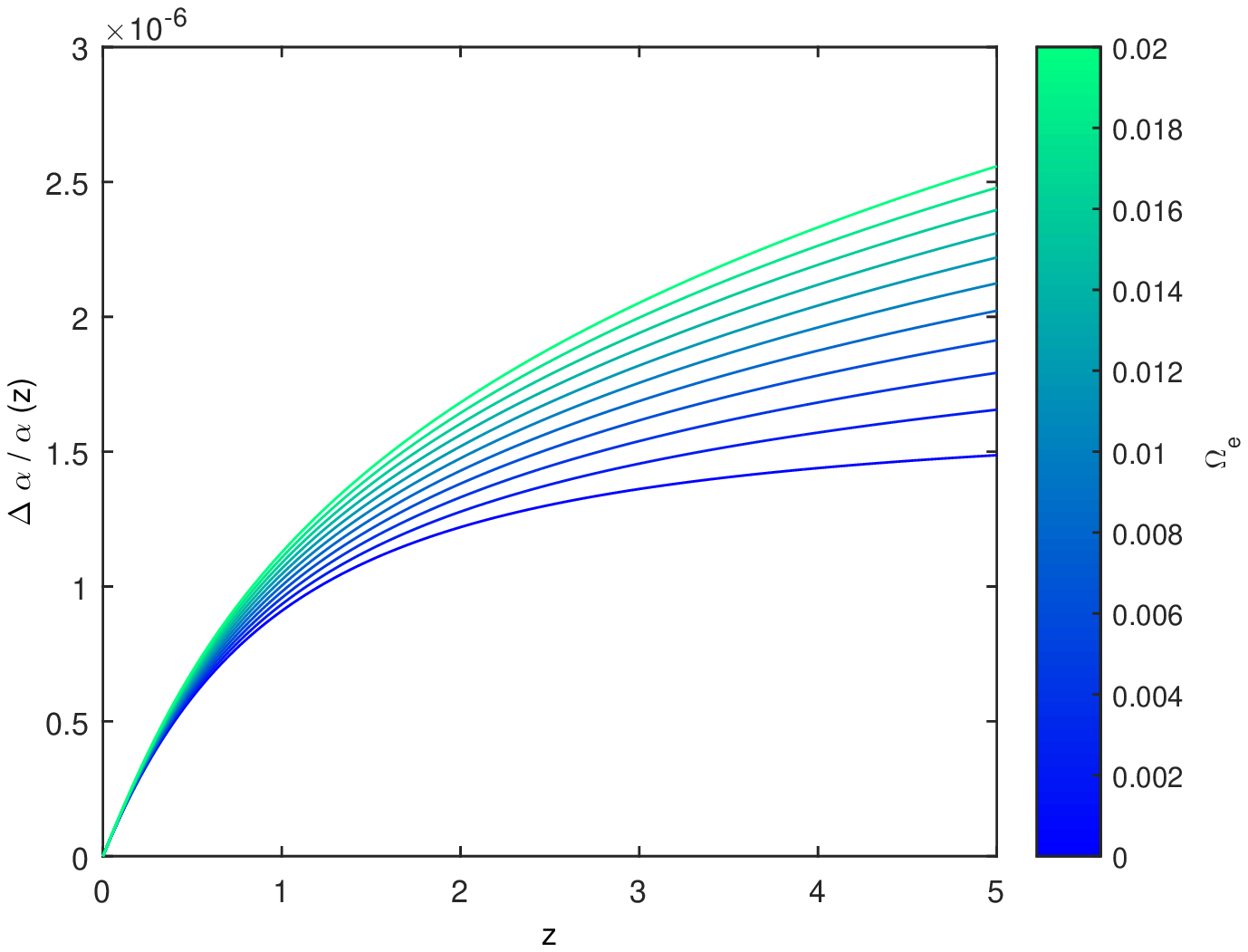}
\includegraphics[width=3in]{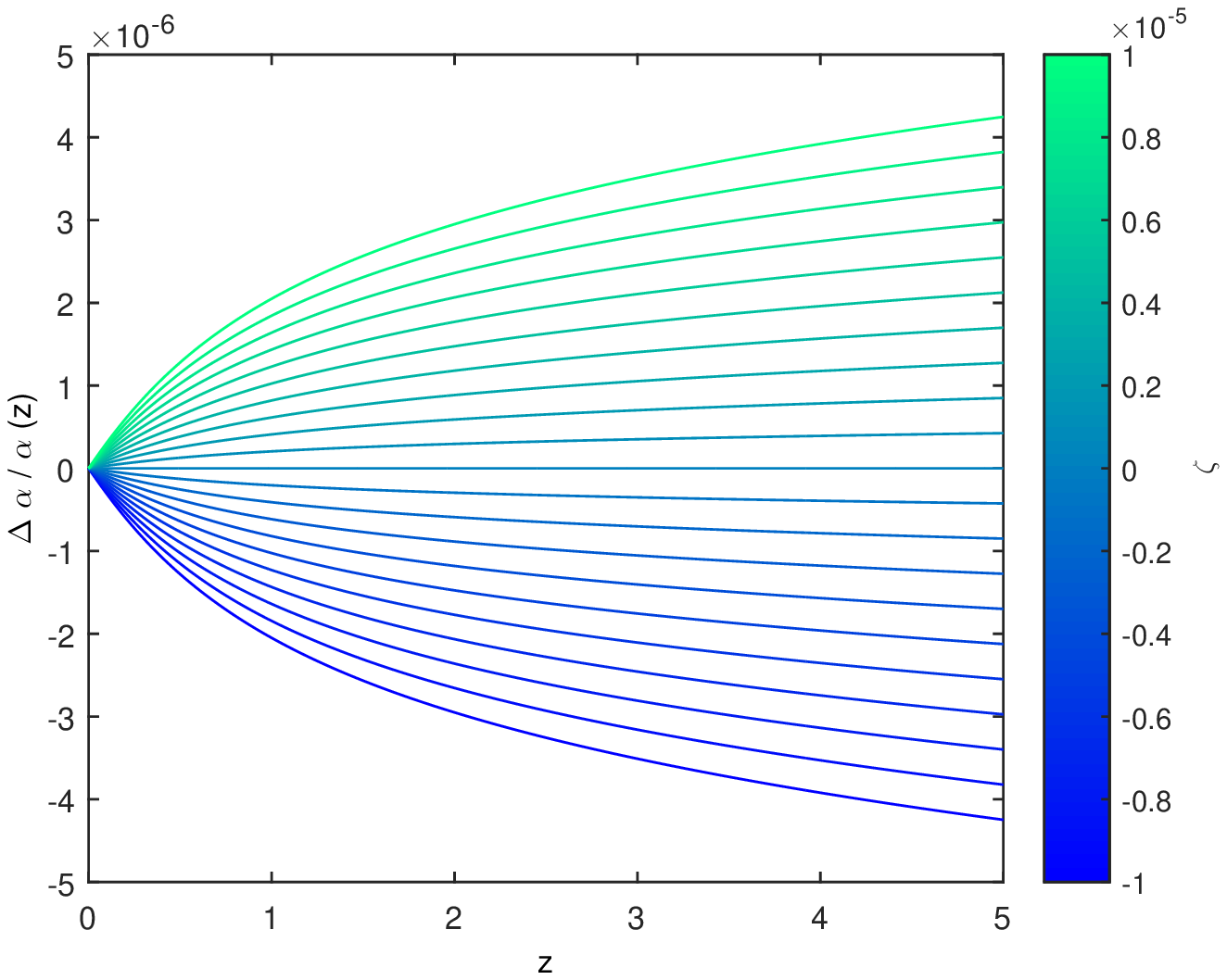}
\includegraphics[width=3in]{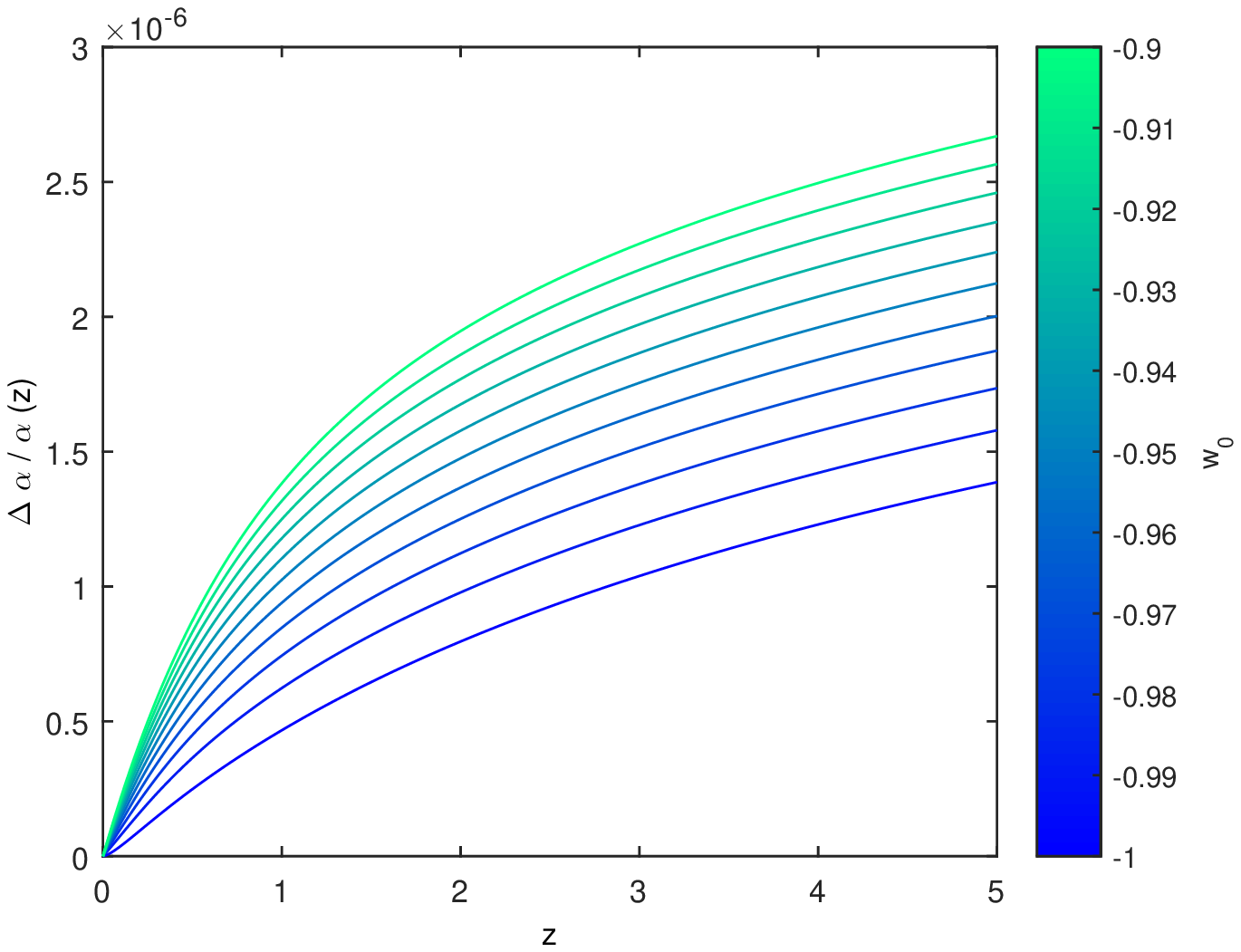}
\caption{\label{figEDE}Redshift dependence of relevant parameters in the EDE model. {\bf Top left}: $w(z)$ with $w_0=-0.95$ and $0.00\le\Omega_e\le 0.02$; {\bf Top right}: $\Delta\alpha/\alpha(z)$ with $w_0=-0.95$, $\zeta = 5\times10^{-6}$ and $0.00\le\Omega_e\le 0.02$; {\bf Bottom left}: $\Delta\alpha/\alpha(z)$ with $w_0=-0.95$, $\Omega_e=0.01$ and $-1\times10^{-5}\le \zeta \le+1\times10^{-5}$; {\bf Bottom right}: $\Delta\alpha/\alpha(z)$ with $\Omega_e=0.01$, $\zeta = 5\times10^{-6}$, and $-1.1\le w_0\le -0.9$.}
\end{figure*}

The present day value of the equation of state is $w_0$, and the equation of state follows the behavior of the dominant component at each cosmic time, with $w_{\rm EDE}\approx1/3$ during radiation domination, and $w_{\rm EDE}\approx0$ during matter domination. Even though this is a phenomenological parametrization, we will again  assume that this kind of dark energy is the result of an underlying scalar field, which couples to the electromagnetic sector. Figure \ref{figEDE} illustrates the behavior of $w(z)$ and $\Delta\alpha/\alpha(z)$ in this model for realistic parameter choices, compatible with the recent Planck collaboration results \cite{Planck}; in particular we use $z_{eq}=3371$.

\subsection{Flat prior}

\begin{figure}[!]
\centering
\includegraphics[width=3in]{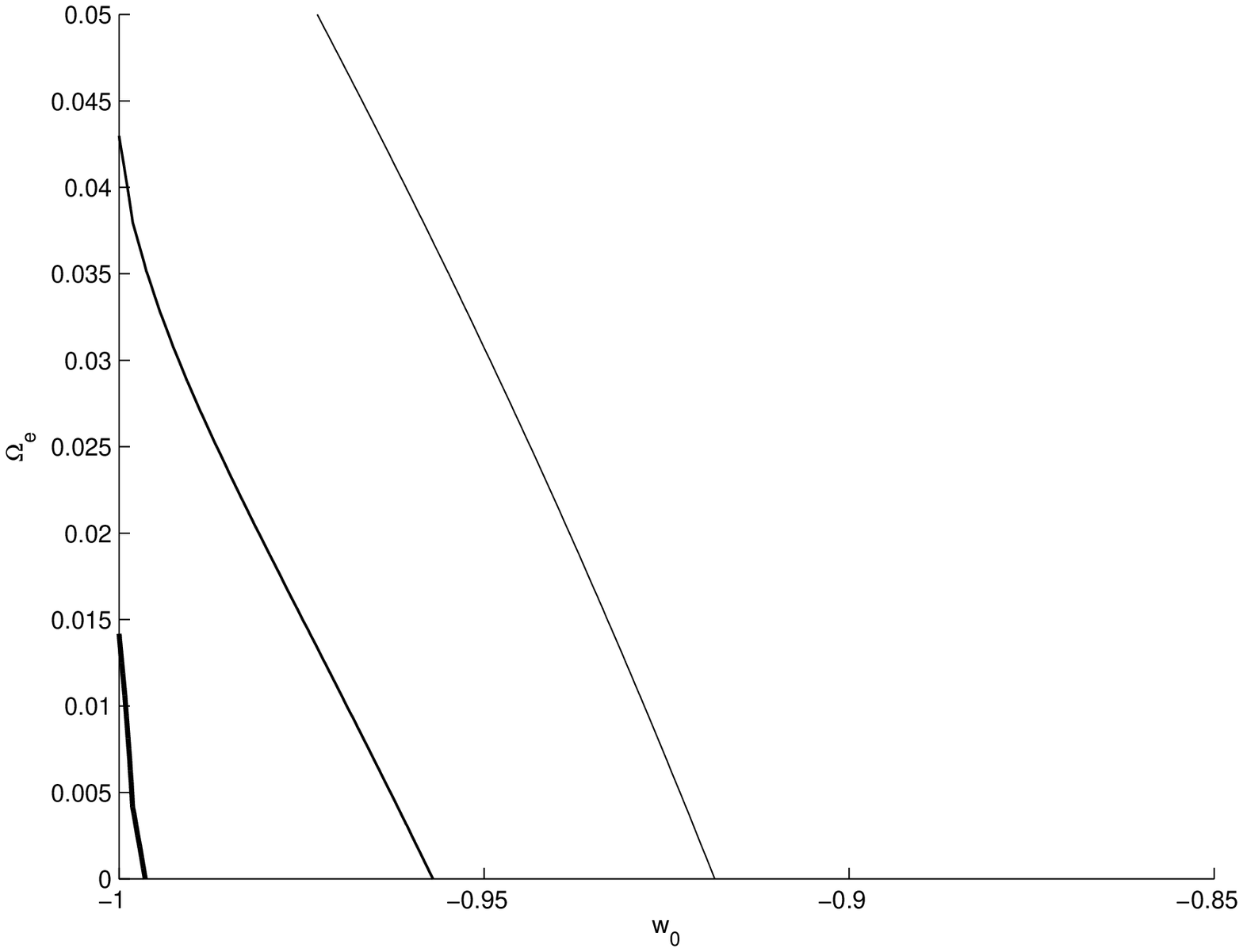}
\includegraphics[width=3in]{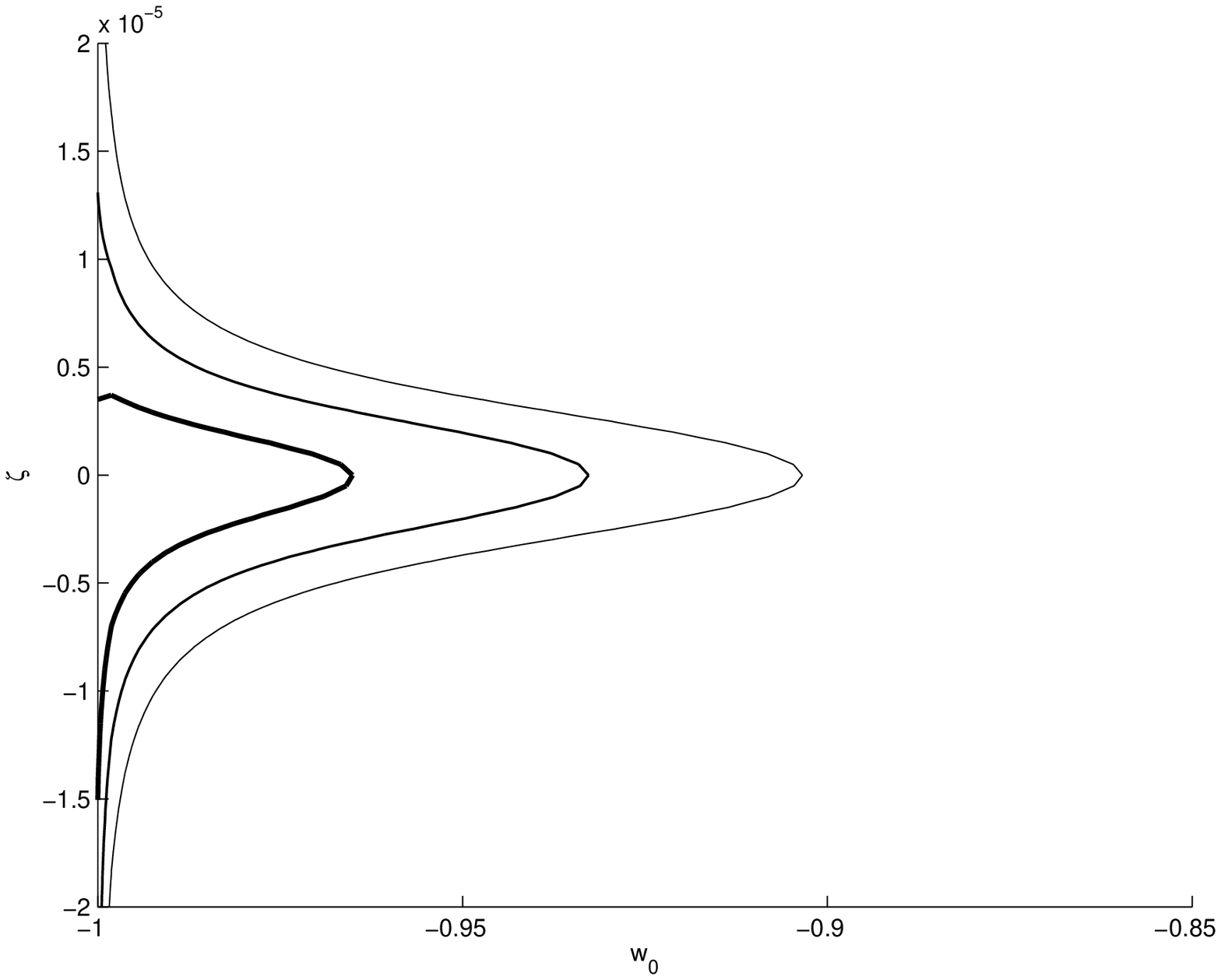}
\includegraphics[width=3in]{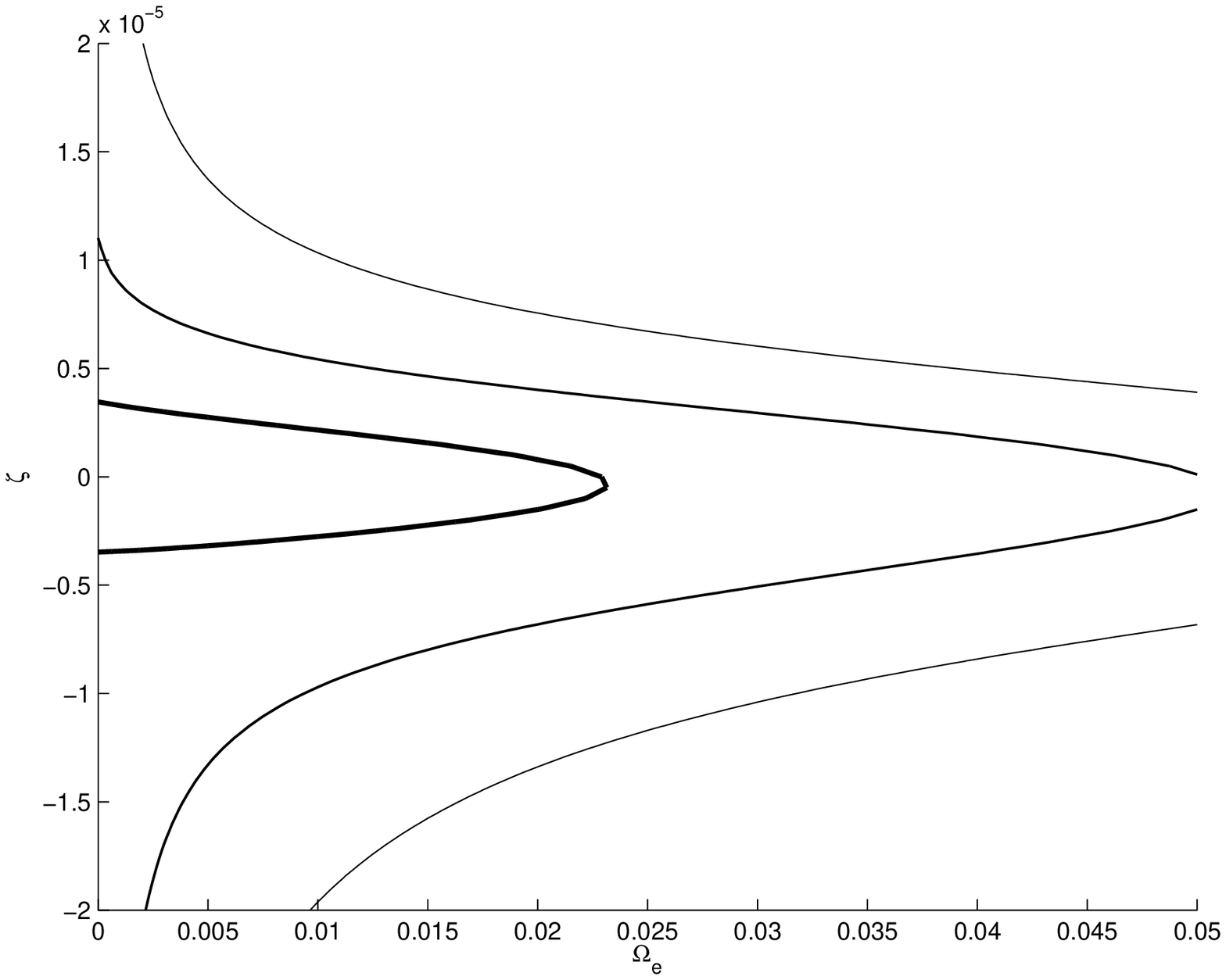}
\caption{\label{figEDE2d}2D constraints on the EDE model with a flat prior on $w_0$, from the full (cosmological plus atomic clock plus astrophysical) datasets described in the main text, in the $w_0$-$\Omega_e$ (top panel), $w_0$-$\zeta$ (middle panel) and $\Omega_e$-$\zeta$ (bottom panel) planes, with the remaining parameter marginalized. One, two and three sigma contours are shown in all cases.}
\end{figure}

\begin{figure}[!]
\centering
\includegraphics[width=3in]{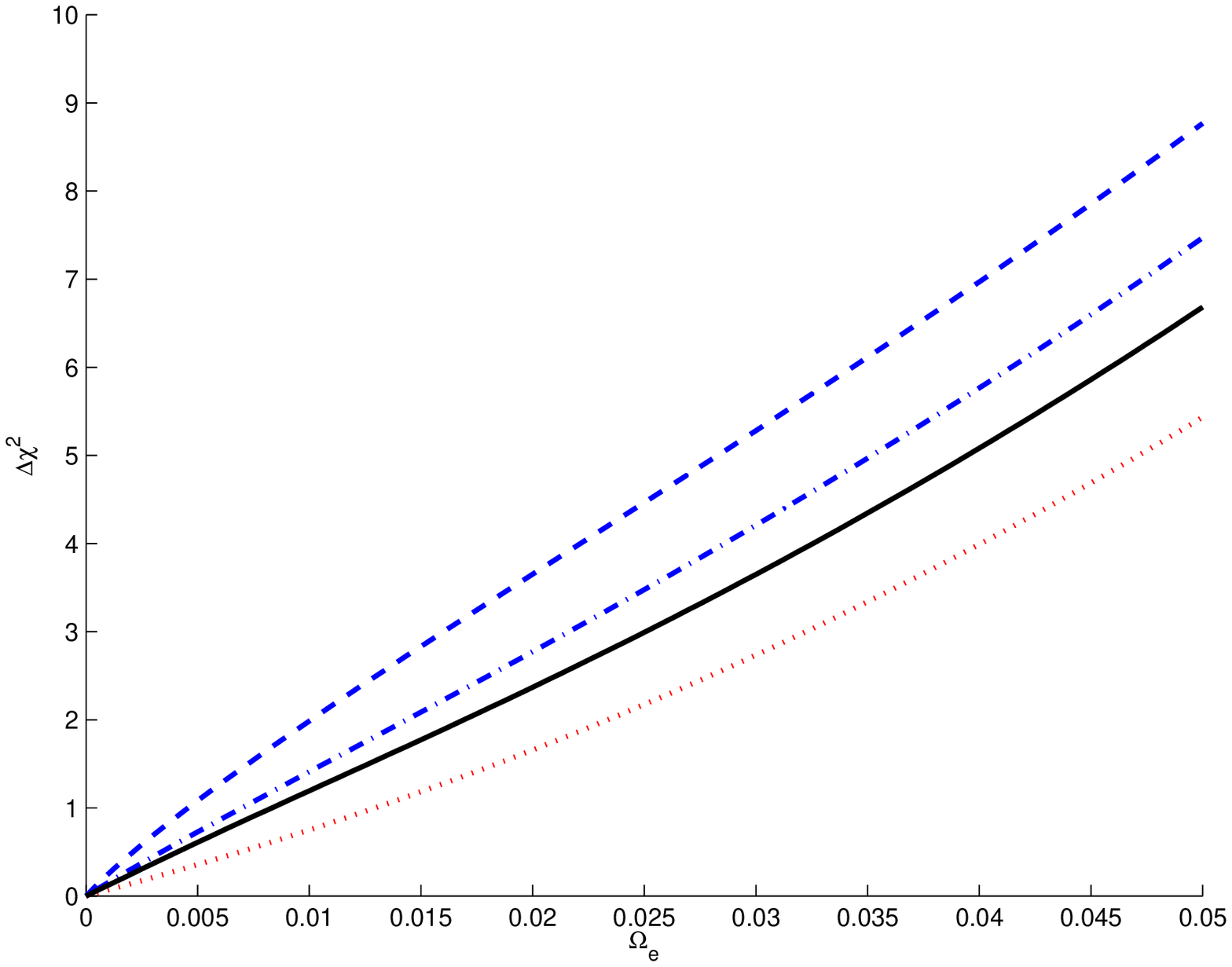}
\includegraphics[width=3in]{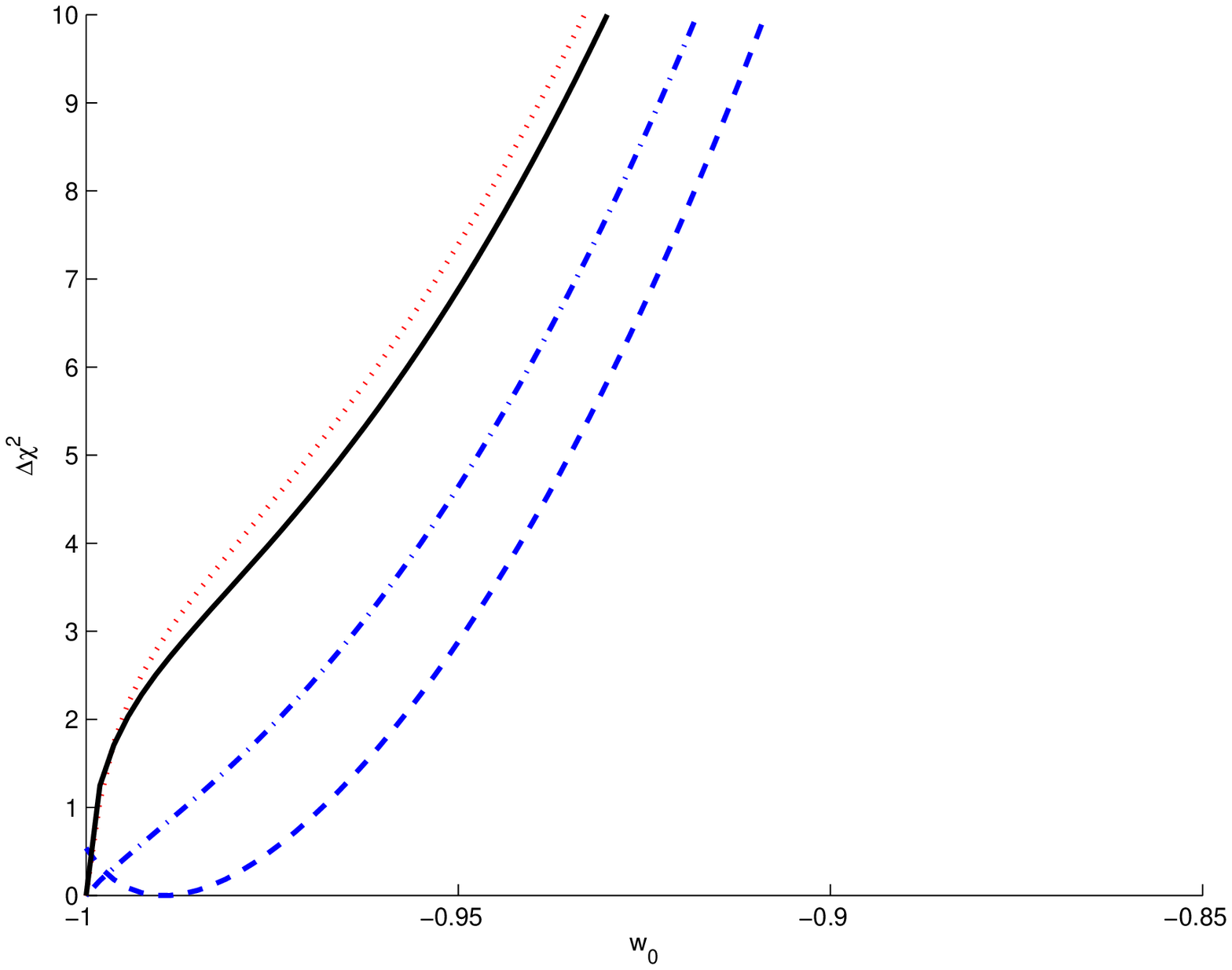}
\includegraphics[width=3in]{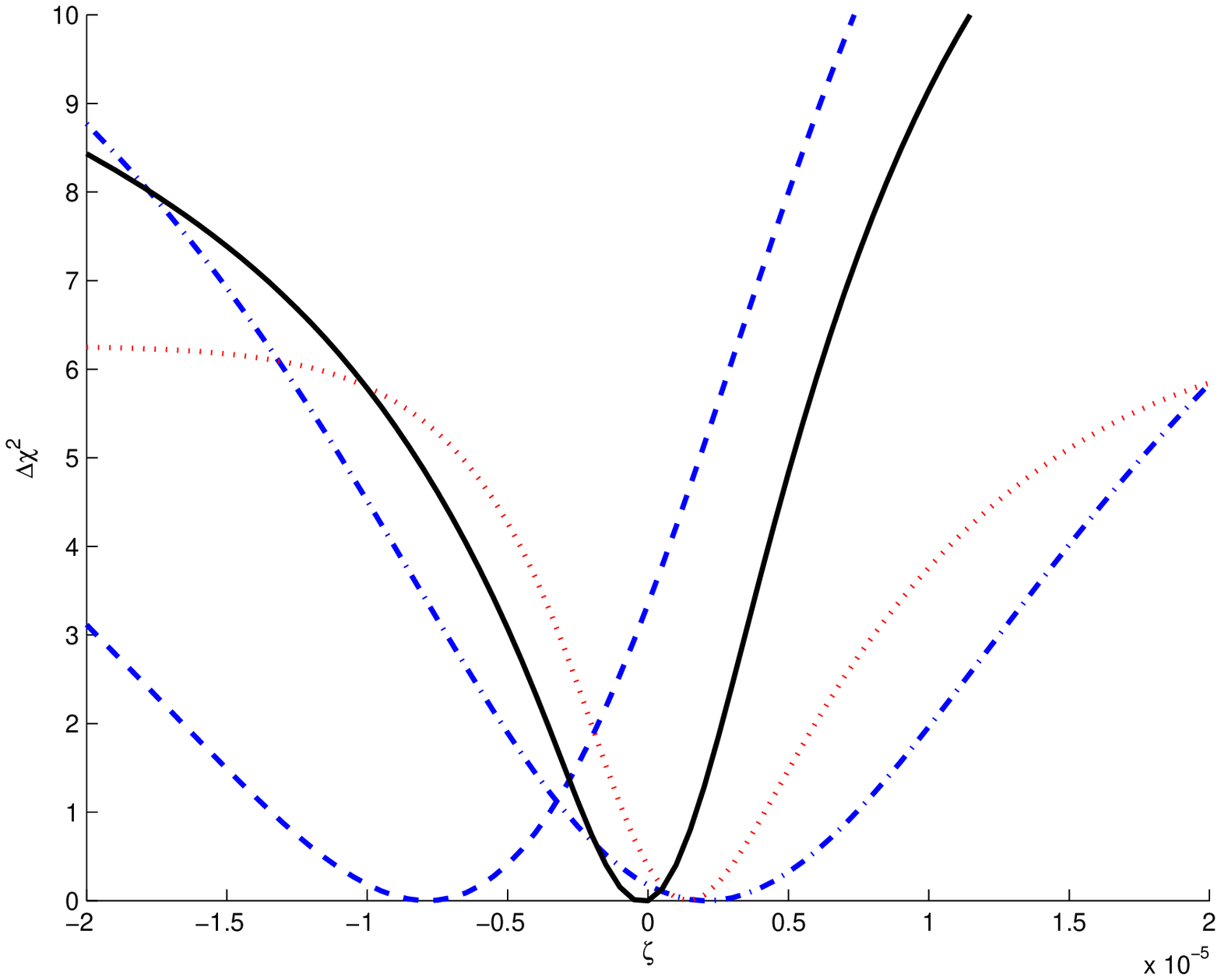}
\caption{\label{figEDE1d}1D marginalized constraints on $\Omega_e$ (top panel), $w_0$ (middle panel) and $\zeta$ (bottom panel) for the EDE model with a flat prior on $w_0$. Different lines correspond to different datasets: constraints from cosmological plus Webb \protect\textit{et al.} are shown by the dashed blue lines. cosmological plus dedicated $\alpha$ measurements in dash-dotted blue, cosmological plus atomic clocks in dotted red, and the full sample in solid black lines. In all cases the vertical axis depicts $\Delta\chi^2=\chi^2-\chi^2_{\rm min}$.}
\end{figure}

We start by studying the EDE model using a flat prior on $w_0$ and further assuming that $w_0\ge-1$. We then carry out an analysis similar to that of the previous section, the results of which can be seen in Figs. \ref{figEDE2d} and \ref{figEDE1d}.


In this case the correlation between $\zeta$ and $w_0$ is also clear, although it is partially broken by the cosmological data. The same happens with the early dark energy density, for which we can also obtain non-trivial constraints. Here we obtain the following 1D marginalized constraints
\begin{equation} \label{edeome}
\Omega_e<0.033\qquad {\rm (95.4\% C.L.)} \,,
\end{equation}
which is about a factor of 3 weaker than the standard one without allowing for possible $\alpha$ variations. On the other hand, for $w_0$ we obtain
\begin{equation} \label{edew02}
w_0<-0.97\qquad {\rm (95.4\% C.L.)}
\end{equation}
\begin{equation} \label{edew03}
w_0<-0.93\qquad {\rm (99.7\% C.L.)} \,
\end{equation}
and finally for the coupling we obtain
\begin{equation} \label{edezeta}
\zeta=(-1\pm5)\times10^{-6}\qquad {\rm (95.4\% C.L.)} \,
\end{equation}
leading to
\begin{equation} \label{edeeta}
\eta<3.6\times10^{-14}\qquad {\rm (95.4\% C.L.)} \,.
\end{equation}

Here, by comparison to the CPL case, the slightly stronger constraints on the dark energy sector imply slightly weaker constraints on the coupling $\zeta$.

\subsection{Logarithmic prior}

\begin{figure}[!]
\centering
\includegraphics[width=3in]{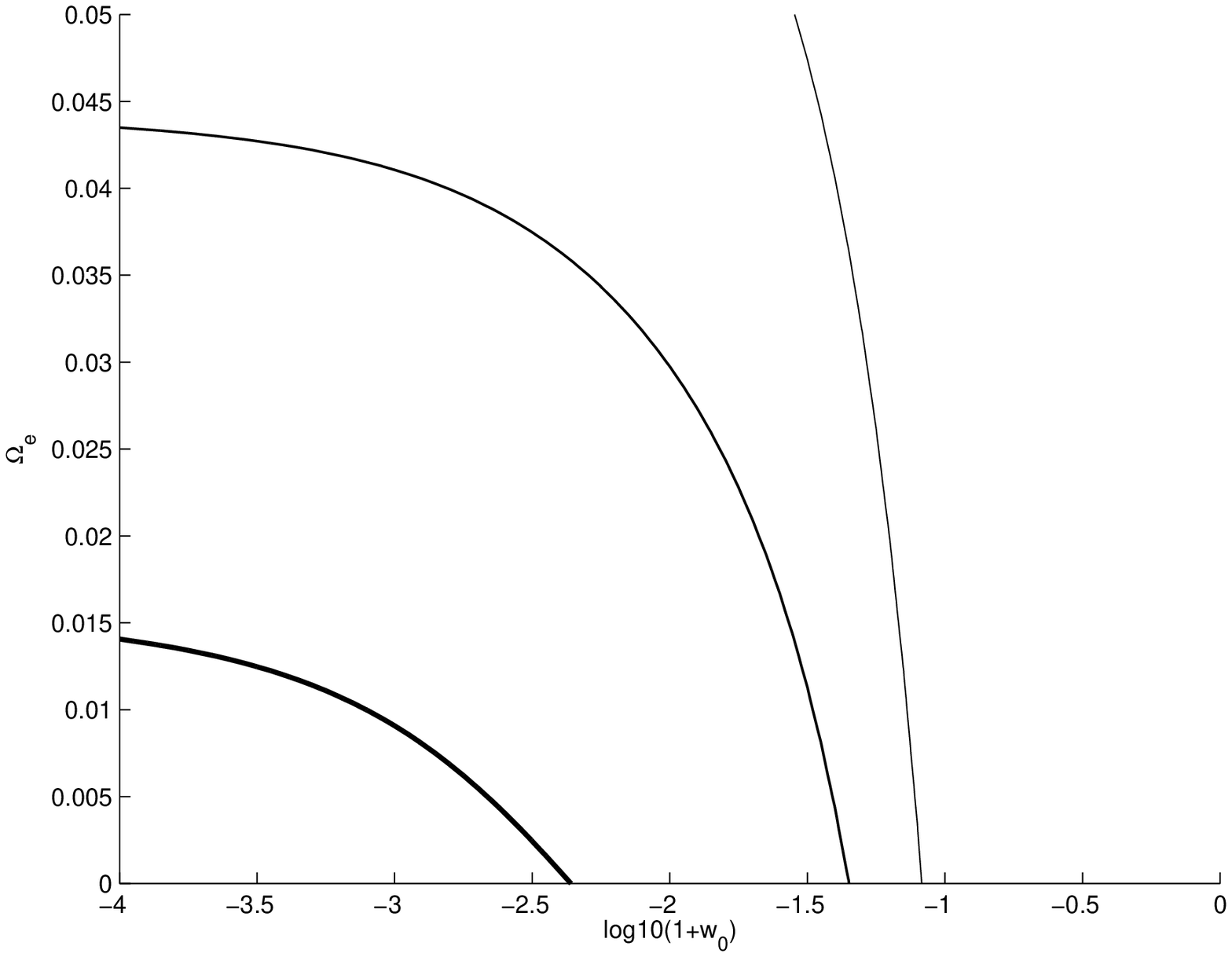}
\includegraphics[width=3in]{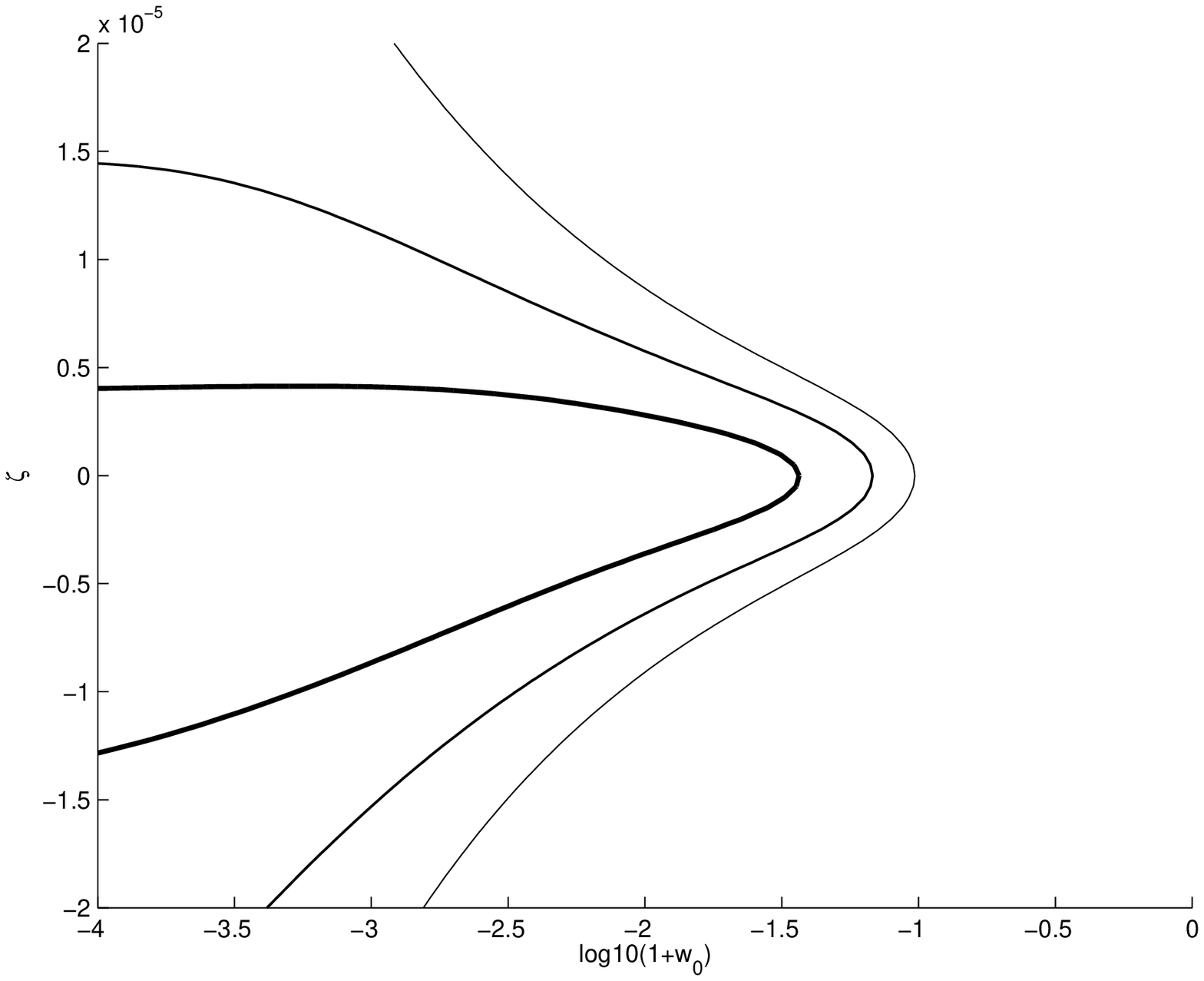}
\includegraphics[width=3in]{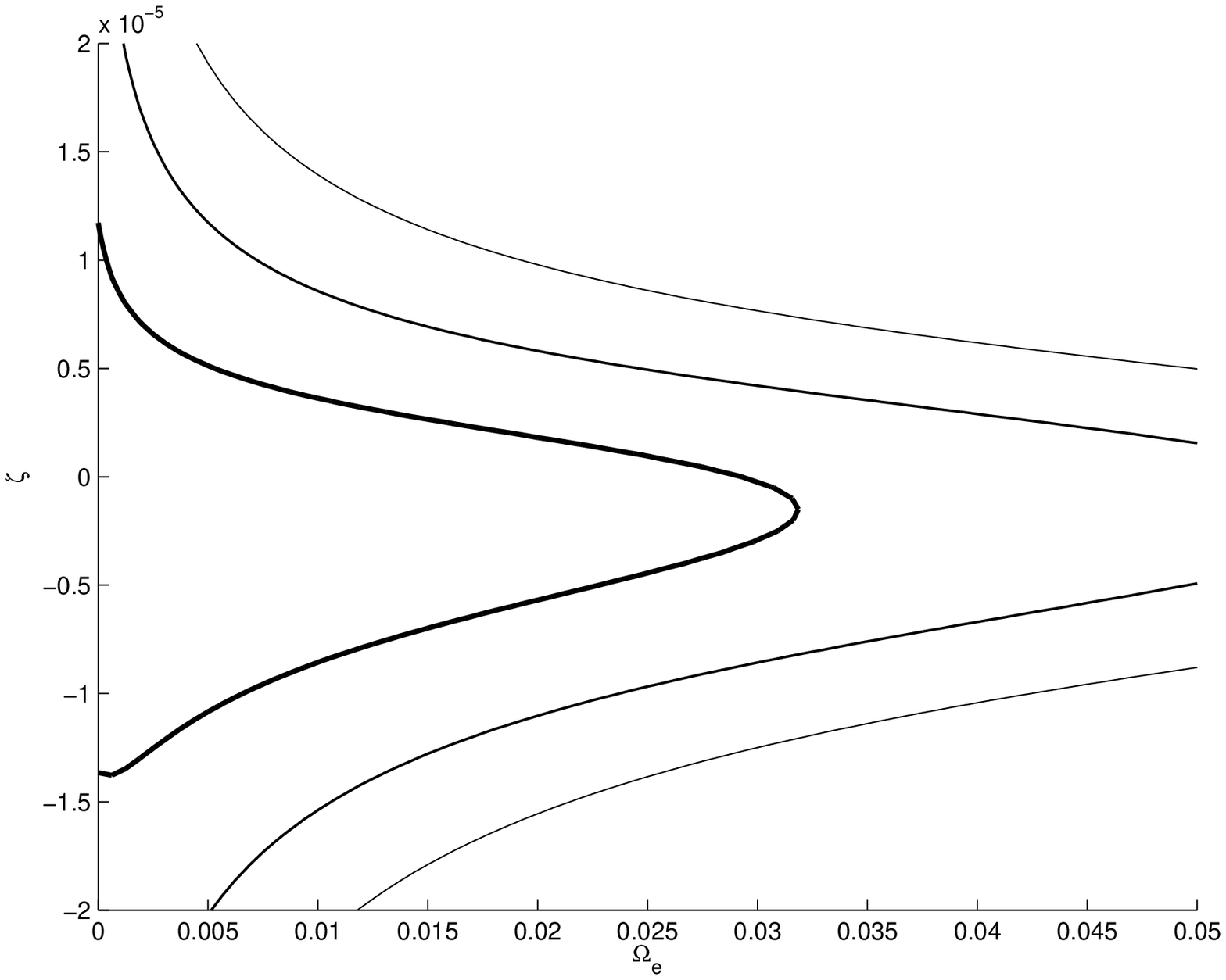}
\caption{\label{figLOG2d}2D constraints on the EDE model with a logarithmic prior on $w_0$, from the full (cosmological plus atomic clock plus astrophysical) datasets described in the main text, in the $w_0$-$\Omega_e$ (top panel), $w_0$-$\zeta$ (middle panel) and $\Omega_e$-$\zeta$ (bottom panel) planes, with the remaining parameter marginalized. One, two and three sigma contours are shown in all cases.}
\end{figure}

\begin{figure}[!]
\centering
\includegraphics[width=3in]{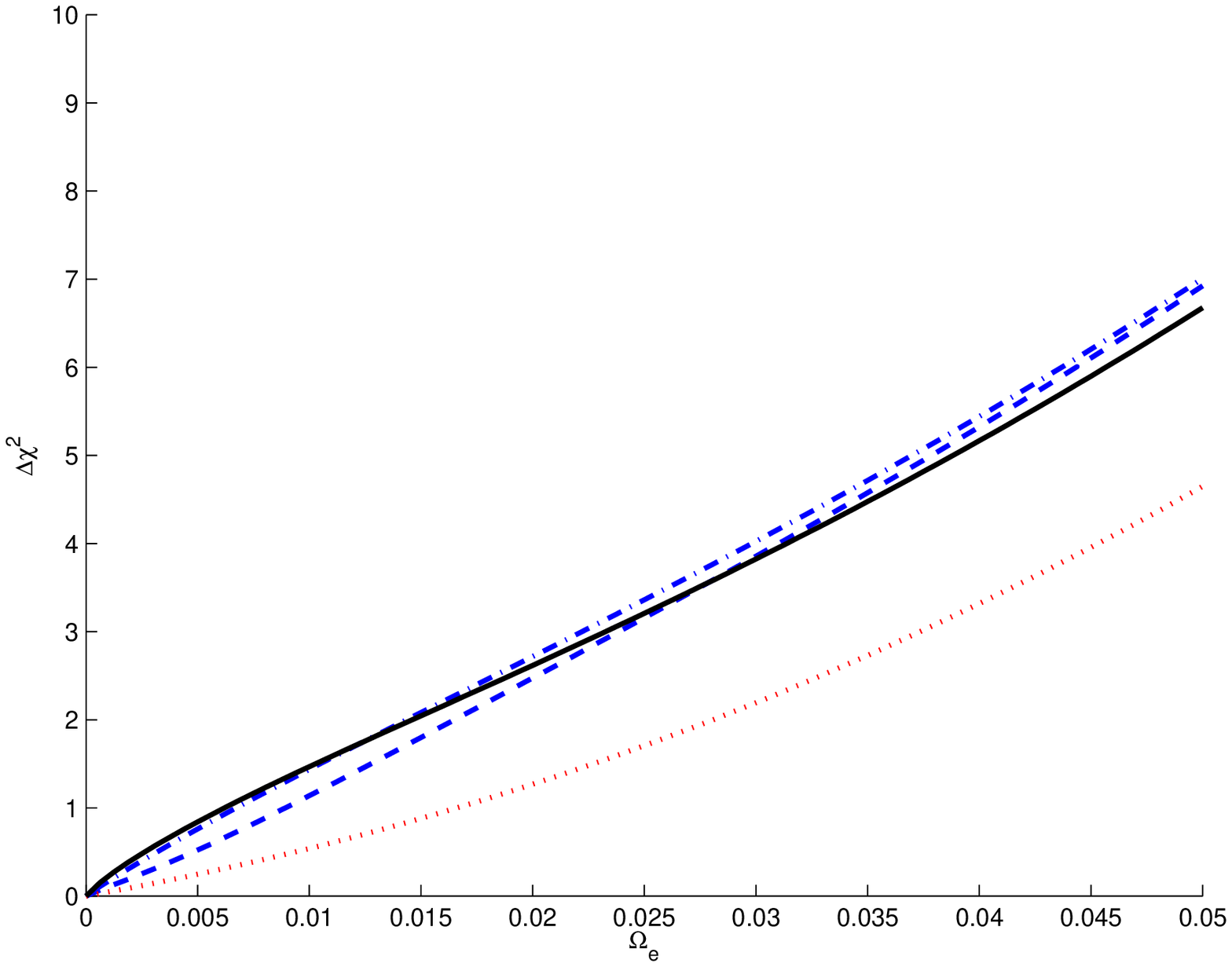}
\includegraphics[width=3in]{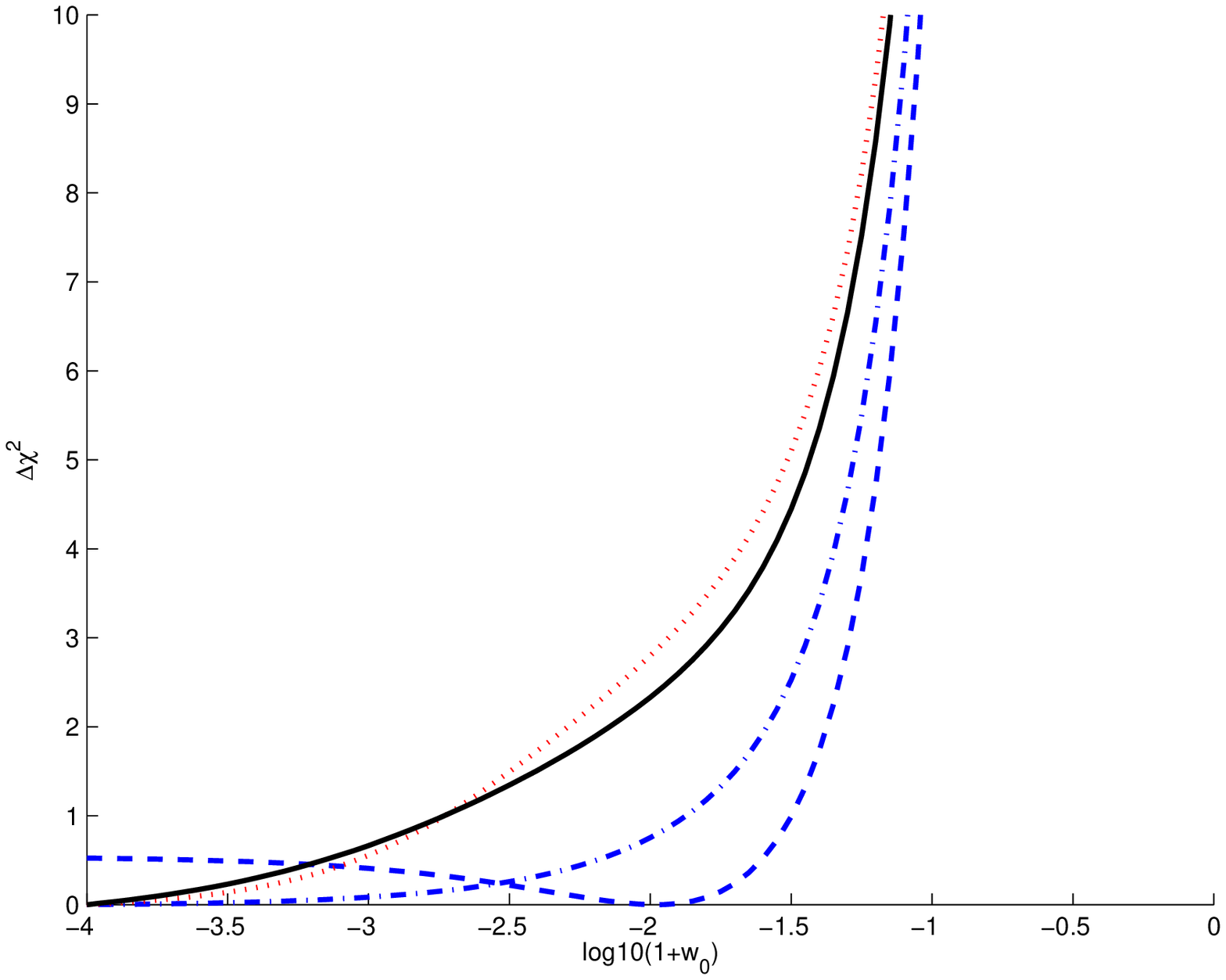}
\includegraphics[width=3in]{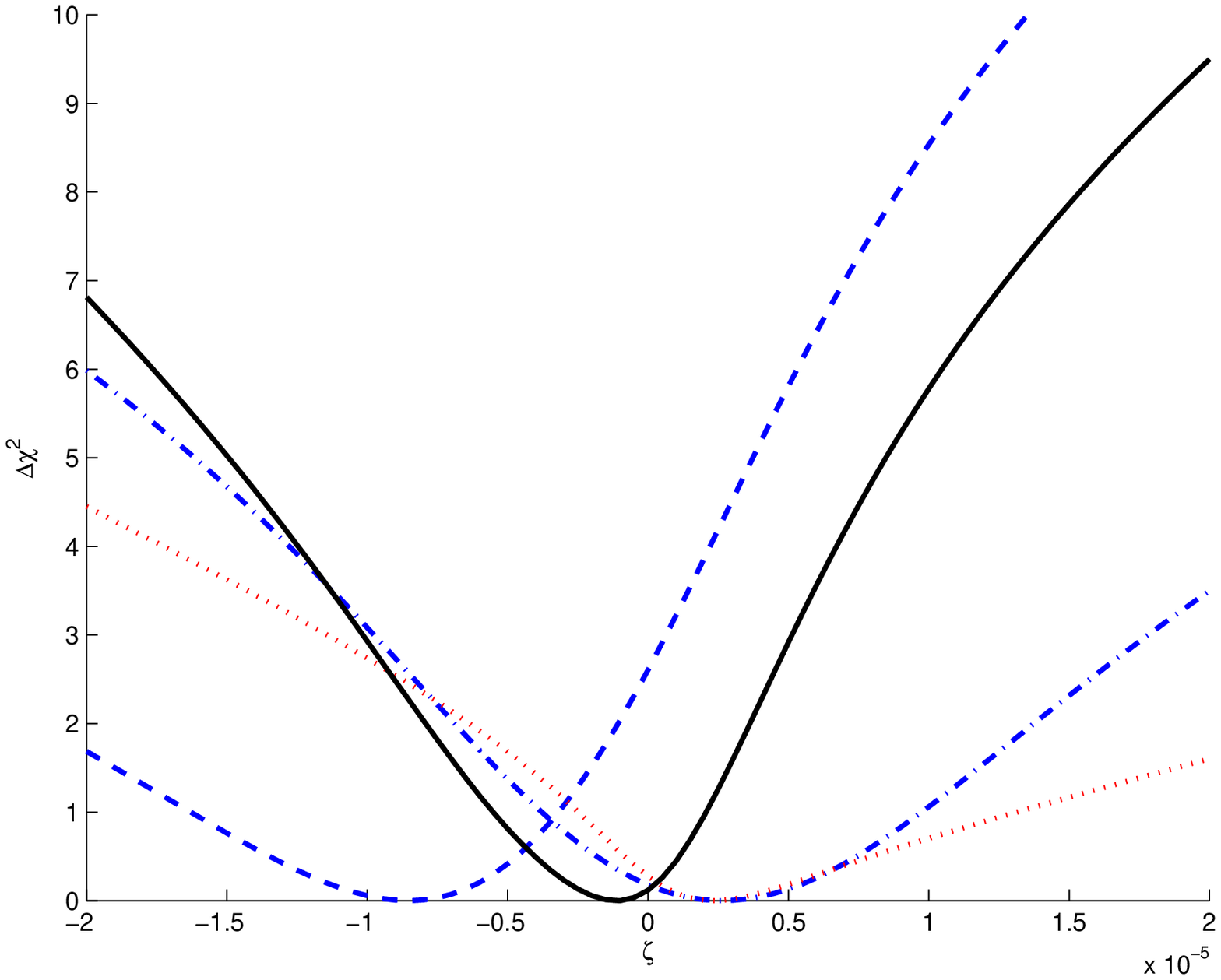}
\caption{\label{figLOG1d}1D marginalized constraints on $\Omega_e$ (top panel), $w_0$ (middle panel) and $\zeta$ (bottom panel) for the EDE model with a logarithmic prior on $w_0$. Different lines correspond to different datasets: constraints from cosmological plus Webb \protect\textit{et al.} are shown by the dashed blue lines. cosmological plus dedicated $\alpha$ measurements in dash-dotted blue, cosmological plus atomic clocks in dotted red, and the full sample in solid black lines. In all cases the vertical axis depicts $\Delta\chi^2=\chi^2-\chi^2_{\rm min}$.}
\end{figure}

We finally study the impact of our choices of priors, specifically by assessing the impact of using a logarithmic (rather than flat) prior on $w_0$. The results can now be seen in Figs. \ref{figLOG2d} and \ref{figLOG1d}.


The 1D marginalized constraints now become
\begin{equation} \label{logome}
\Omega_e<0.030\qquad {\rm (95.4\% C.L.)} \,,
\end{equation}
which is about ten percent stronger than the flat prior case. For $w_0$ the constraints are unchanged,
\begin{equation} \label{logw02}
w_0<-0.97\qquad {\rm (95.4\% C.L.)}
\end{equation}
\begin{equation} \label{logw03}
w_0<-0.93\qquad {\rm (99.7\% C.L.)} \,
\end{equation}
and finally the constraint on the coupling becomes weaker as well as asymmetric
\begin{equation} \label{logzeta}
\zeta=(-1^{+8}_{-11})\times10^{-6}\qquad {\rm (95.4\% C.L.)} \,
\end{equation}
leading to
\begin{equation} \label{logeta}
\eta<14.4\times10^{-14}\qquad {\rm (95.4\% C.L.)} \,;
\end{equation}
note that even in this case this constraint is still marginally stronger than the current direct bounds.

\section{Conclusions and outlook}

We have used a combination of astrophysical spectroscopy and local laboratory tests of the stability of the fine-structure constant $\alpha$, complemented by background cosmological datesets, to constrain several broad class of dynamical dark energy models where the same degree of freedom is responsible for both the dark energy and a variation of $\alpha$. In these models, which are more general than the ones previously studied \cite{Pinho,Pinho2}, the redshift dependence of $\alpha$ is a function both of a fundamental physics parameter (the dimensionless coupling $\zeta$ of the scalar field to the electromagnetic sector) and background 'dark cosmology' parameters, including the present day dark energy equation of state $w_0$.

Our analysis confirms that, despite some mild dependence on the underlying model and the choice of priors, the combination of cosmological, astrophysical and atomic clock data leads to tight and competitive constraints on the dimensionless coupling $\zeta$ of the scalar field to the electromagnetic sector as well as on $w_0$. Current data is consistent with the standard $\Lambda$CDM paradigm where $w_0=-1$ and $\zeta=0$.

Presently these constraints are dominated by the atomic clock tests \cite{Rosenband}, which are only sensitive to the dark energy equation of state today. This is one of the reasons why additional parameters such as $w_a$ in the CPL parametrization and $\beta$ in the MKH parametrization are weakly constrained by the present data. Improvements in astrophysical measurements will allow significantly stronger constraints on a larger parameter space. Forthcoming high-resolution ultra-stable spectrographs such as ESPRESSO and ELT-HIRES will be ideal for this task. A roadmap for these studies is outlined in \cite{cjmGRG}, and more detailed forecasts of the future impact of these measurements may be found in \cite{Leite,LeiteNEW}.

Importantly, in the classes of models we have studied the dynamical degree of freedom responsible for the dark energy and the $\alpha$ variation inevitably couples to nucleons (through the $\alpha$ dependence of their masses) and leads to violations of the Weak Equivalence Principle. Our bounds on the coupling $\zeta$ can therefore be used to obtain indirect bounds on the E\"{o}tv\"{o}s parameter $\eta$. Despite the aforementioned model dependence, these bounds are stronger than the current direct ones, typically by as much as one order of magnitude. Broadly speaking they are at the $\eta\sim$few$\times10^{-14}$ level.

We note that the forthcoming MICROSCOPE mission, currently scheduled for launch in April 2016, should reach $\eta\sim 10^{-15}$ sensitivity \cite{MICROSCOPE}. Should this measure a value of $\eta$ larger than that in our bounds, this would rule out the Class I models we have studied here, and specifically the physically crucial assumption of the coupling of the dynamical dark energy field to the electromagnetic sector. Alternatively, a MICROSCOPE detection of a large $\eta$ would imply that the measurements of $\alpha$ on which they rely are incorrect and dominated by unaccounted systematics.

For the same reason, the next generation of high-resolution ultra-stable spectrographs will also provide significantly tighter constraints on $\eta$. Specifically, for Class I models and on the basis of the forecasts of \cite{Leite,LeiteNEW}, we may conservatively expect a sensitivity of $\eta\sim$few$\times10^{-16}$ for ESPRESSO and $\eta\sim 10^{-18}$ for ELT-HIRES. The latter is comparable to the expected sensitivity of the proposed STEP satellite \cite{STEP}.

Thus the next decade will bring forth tests of these cornerstone principles with unprecedented accuracy. Null results from STEP and the E-ELT would then imply that any putative coupling of light scalar fields to the standard model would need to be unnaturally small, which in turn would mean that either WEP violating fields do not exist at all in nature or that these couplings are suppressed by some currently unknown mechanism. This would be an astrophysical analog of the strong CP problem in Quantum Chromodynamics \cite{Peccei}. In any case, our analysis shows that astrophysical tests of the stability of fundamental couplings are a crucial probe of fundamental physics and cosmology.

\begin{acknowledgments}

We are grateful to Ana Catarina Leite and David Corre for many helpful discussions on the subject of this work. This work was done in the context of project PTDC/FIS/111725/2009 (FCT, Portugal). CJM is also supported by an FCT Research Professorship, contract reference IF/00064/2012, funded by FCT/MCTES (Portugal) and POPH/FSE (EC). A.G. and J.L. acknowledge financial support from Programa Joves i Ci\`encia, funded by Fundaci\'o Catalunya-La Pedrera.

\end{acknowledgments}

\bibliography{paper3}

\begin{thebibliography}{43}%
\makeatletter
\providecommand \@ifxundefined [1]{%
 \@ifx{#1\undefined}
}%
\providecommand \@ifnum [1]{%
 \ifnum #1\expandafter \@firstoftwo
 \else \expandafter \@secondoftwo
 \fi
}%
\providecommand \@ifx [1]{%
 \ifx #1\expandafter \@firstoftwo
 \else \expandafter \@secondoftwo
 \fi
}%
\providecommand \natexlab [1]{#1}%
\providecommand \enquote  [1]{``#1''}%
\providecommand \bibnamefont  [1]{#1}%
\providecommand \bibfnamefont [1]{#1}%
\providecommand \citenamefont [1]{#1}%
\providecommand \href@noop [0]{\@secondoftwo}%
\providecommand \href [0]{\begingroup \@sanitize@url \@href}%
\providecommand \@href[1]{\@@startlink{#1}\@@href}%
\providecommand \@@href[1]{\endgroup#1\@@endlink}%
\providecommand \@sanitize@url [0]{\catcode `\\12\catcode `\$12\catcode
  `\&12\catcode `\#12\catcode `\^12\catcode `\_12\catcode `\%12\relax}%
\providecommand \@@startlink[1]{}%
\providecommand \@@endlink[0]{}%
\providecommand \url  [0]{\begingroup\@sanitize@url \@url }%
\providecommand \@url [1]{\endgroup\@href {#1}{\urlprefix }}%
\providecommand \urlprefix  [0]{URL }%
\providecommand \Eprint [0]{\href }%
\providecommand \doibase [0]{http://dx.doi.org/}%
\providecommand \selectlanguage [0]{\@gobble}%
\providecommand \bibinfo  [0]{\@secondoftwo}%
\providecommand \bibfield  [0]{\@secondoftwo}%
\providecommand \translation [1]{[#1]}%
\providecommand \BibitemOpen [0]{}%
\providecommand \bibitemStop [0]{}%
\providecommand \bibitemNoStop [0]{.\EOS\space}%
\providecommand \EOS [0]{\spacefactor3000\relax}%
\providecommand \BibitemShut  [1]{\csname bibitem#1\endcsname}%
\let\auto@bib@innerbib\@empty
\bibitem [{\citenamefont {Riess}\ \emph {et~al.}(1998)\citenamefont {Riess}
  \emph {et~al.}}]{SN1}%
  \BibitemOpen
  \bibfield  {author} {\bibinfo {author} {\bibfnamefont {A.~G.}\ \bibnamefont
  {Riess}} \emph {et~al.} (\bibinfo {collaboration} {Supernova Search Team}),\
  }\href {\doibase 10.1086/300499} {\bibfield  {journal} {\bibinfo  {journal}
  {Astron.J.}\ }\textbf {\bibinfo {volume} {116}},\ \bibinfo {pages} {1009}
  (\bibinfo {year} {1998})},\ \Eprint {http://arxiv.org/abs/astro-ph/9805201}
  {arXiv:astro-ph/9805201 [astro-ph]} \BibitemShut {NoStop}%
\bibitem [{\citenamefont {Perlmutter}\ \emph {et~al.}(1999)\citenamefont
  {Perlmutter} \emph {et~al.}}]{SN2}%
  \BibitemOpen
  \bibfield  {author} {\bibinfo {author} {\bibfnamefont {S.}~\bibnamefont
  {Perlmutter}} \emph {et~al.} (\bibinfo {collaboration} {Supernova Cosmology
  Project}),\ }\href {\doibase 10.1086/307221} {\bibfield  {journal} {\bibinfo
  {journal} {Astrophys.J.}\ }\textbf {\bibinfo {volume} {517}},\ \bibinfo
  {pages} {565} (\bibinfo {year} {1999})},\ \Eprint
  {http://arxiv.org/abs/astro-ph/9812133} {arXiv:astro-ph/9812133 [astro-ph]}
  \BibitemShut {NoStop}%
\bibitem [{\citenamefont {Aad}\ \emph {et~al.}(2012)\citenamefont {Aad} \emph
  {et~al.}}]{ATLAS}%
  \BibitemOpen
  \bibfield  {author} {\bibinfo {author} {\bibfnamefont {G.}~\bibnamefont
  {Aad}} \emph {et~al.} (\bibinfo {collaboration} {ATLAS Collaboration}),\
  }\href {\doibase 10.1016/j.physletb.2012.08.020} {\bibfield  {journal}
  {\bibinfo  {journal} {Phys.Lett.}\ }\textbf {\bibinfo {volume} {B716}},\
  \bibinfo {pages} {1} (\bibinfo {year} {2012})},\ \Eprint
  {http://arxiv.org/abs/1207.7214} {arXiv:1207.7214 [hep-ex]} \BibitemShut
  {NoStop}%
\bibitem [{\citenamefont {Chatrchyan}\ \emph {et~al.}(2012)\citenamefont
  {Chatrchyan} \emph {et~al.}}]{CMS}%
  \BibitemOpen
  \bibfield  {author} {\bibinfo {author} {\bibfnamefont {S.}~\bibnamefont
  {Chatrchyan}} \emph {et~al.} (\bibinfo {collaboration} {CMS Collaboration}),\
  }\href {\doibase 10.1016/j.physletb.2012.08.021} {\bibfield  {journal}
  {\bibinfo  {journal} {Phys.Lett.}\ }\textbf {\bibinfo {volume} {B716}},\
  \bibinfo {pages} {30} (\bibinfo {year} {2012})},\ \Eprint
  {http://arxiv.org/abs/1207.7235} {arXiv:1207.7235 [hep-ex]} \BibitemShut
  {NoStop}%
\bibitem [{\citenamefont {Carroll}(1998)}]{Carroll}%
  \BibitemOpen
  \bibfield  {author} {\bibinfo {author} {\bibfnamefont {S.~M.}\ \bibnamefont
  {Carroll}},\ }\href {\doibase 10.1103/PhysRevLett.81.3067} {\bibfield
  {journal} {\bibinfo  {journal} {Phys.Rev.Lett.}\ }\textbf {\bibinfo {volume}
  {81}},\ \bibinfo {pages} {3067} (\bibinfo {year} {1998})},\ \Eprint
  {http://arxiv.org/abs/astro-ph/9806099} {arXiv:astro-ph/9806099 [astro-ph]}
  \BibitemShut {NoStop}%
\bibitem [{\citenamefont {Uzan}(2011)}]{uzanLR}%
  \BibitemOpen
  \bibfield  {author} {\bibinfo {author} {\bibfnamefont {J.-P.}\ \bibnamefont
  {Uzan}},\ }\href@noop {} {\bibfield  {journal} {\bibinfo  {journal} {Living
  Rev.Rel.}\ }\textbf {\bibinfo {volume} {14}},\ \bibinfo {pages} {2} (\bibinfo
  {year} {2011})},\ \Eprint {http://arxiv.org/abs/1009.5514} {arXiv:1009.5514
  [astro-ph.CO]} \BibitemShut {NoStop}%
\bibitem [{\citenamefont {Martins}(2014)}]{cjmGRG}%
  \BibitemOpen
  \bibfield  {author} {\bibinfo {author} {\bibfnamefont {C.~J. A.~P.}\
  \bibnamefont {Martins}},\ }\href {\doibase 10.1007/s10714-014-1843-7}
  {\bibfield  {journal} {\bibinfo  {journal} {Gen.Rel.Grav.}\ }\textbf
  {\bibinfo {volume} {47}},\ \bibinfo {pages} {1843} (\bibinfo {year}
  {2014})},\ \Eprint {http://arxiv.org/abs/1412.0108} {arXiv:1412.0108
  [astro-ph.CO]} \BibitemShut {NoStop}%
\bibitem [{\citenamefont {Webb}\ \emph {et~al.}(2011)\citenamefont {Webb},
  \citenamefont {King}, \citenamefont {Murphy}, \citenamefont {Flambaum},
  \citenamefont {Carswell} \emph {et~al.}}]{Webb}%
  \BibitemOpen
  \bibfield  {author} {\bibinfo {author} {\bibfnamefont {J.}~\bibnamefont
  {Webb}}, \bibinfo {author} {\bibfnamefont {J.}~\bibnamefont {King}}, \bibinfo
  {author} {\bibfnamefont {M.}~\bibnamefont {Murphy}}, \bibinfo {author}
  {\bibfnamefont {V.}~\bibnamefont {Flambaum}}, \bibinfo {author}
  {\bibfnamefont {R.}~\bibnamefont {Carswell}},  \emph {et~al.},\ }\href
  {\doibase 10.1103/PhysRevLett.107.191101} {\bibfield  {journal} {\bibinfo
  {journal} {Phys.Rev.Lett.}\ }\textbf {\bibinfo {volume} {107}},\ \bibinfo
  {pages} {191101} (\bibinfo {year} {2011})},\ \Eprint
  {http://arxiv.org/abs/1008.3907} {arXiv:1008.3907 [astro-ph.CO]} \BibitemShut
  {NoStop}%
\bibitem [{\citenamefont {Molaro}\ \emph {et~al.}(2013)\citenamefont {Molaro},
  \citenamefont {Centurion}, \citenamefont {Whitmore}, \citenamefont {Evans},
  \citenamefont {Murphy} \emph {et~al.}}]{LP1}%
  \BibitemOpen
  \bibfield  {author} {\bibinfo {author} {\bibfnamefont {P.}~\bibnamefont
  {Molaro}}, \bibinfo {author} {\bibfnamefont {M.}~\bibnamefont {Centurion}},
  \bibinfo {author} {\bibfnamefont {J.}~\bibnamefont {Whitmore}}, \bibinfo
  {author} {\bibfnamefont {T.}~\bibnamefont {Evans}}, \bibinfo {author}
  {\bibfnamefont {M.}~\bibnamefont {Murphy}},  \emph {et~al.},\ }\href
  {\doibase 10.1051/0004-6361/201321351} {\bibfield  {journal} {\bibinfo
  {journal} {Astron.Astrophys.}\ }\textbf {\bibinfo {volume} {555}},\ \bibinfo
  {pages} {A68} (\bibinfo {year} {2013})},\ \Eprint
  {http://arxiv.org/abs/1305.1884} {arXiv:1305.1884 [astro-ph.CO]} \BibitemShut
  {NoStop}%
\bibitem [{\citenamefont {{Evans}}\ \emph {et~al.}(2014)\citenamefont
  {{Evans}}, \citenamefont {{Murphy}}, \citenamefont {{Whitmore}},
  \citenamefont {{Misawa}}, \citenamefont {{Centurion}}, \citenamefont
  {{D'Odorico}}, \citenamefont {{Lopez}}, \citenamefont {{Martins}},
  \citenamefont {{Molaro}}, \citenamefont {{Petitjean}}, \citenamefont
  {{Rahmani}}, \citenamefont {{Srianand}},\ and\ \citenamefont
  {{Wendt}}}]{LP3}%
  \BibitemOpen
  \bibfield  {author} {\bibinfo {author} {\bibfnamefont {T.~M.}\ \bibnamefont
  {{Evans}}}, \bibinfo {author} {\bibfnamefont {M.~T.}\ \bibnamefont
  {{Murphy}}}, \bibinfo {author} {\bibfnamefont {J.~B.}\ \bibnamefont
  {{Whitmore}}}, \bibinfo {author} {\bibfnamefont {T.}~\bibnamefont
  {{Misawa}}}, \bibinfo {author} {\bibfnamefont {M.}~\bibnamefont
  {{Centurion}}}, \bibinfo {author} {\bibfnamefont {S.}~\bibnamefont
  {{D'Odorico}}}, \bibinfo {author} {\bibfnamefont {S.}~\bibnamefont
  {{Lopez}}}, \bibinfo {author} {\bibfnamefont {C.~J.~A.~P.}\ \bibnamefont
  {{Martins}}}, \bibinfo {author} {\bibfnamefont {P.}~\bibnamefont {{Molaro}}},
  \bibinfo {author} {\bibfnamefont {P.}~\bibnamefont {{Petitjean}}}, \bibinfo
  {author} {\bibfnamefont {H.}~\bibnamefont {{Rahmani}}}, \bibinfo {author}
  {\bibfnamefont {R.}~\bibnamefont {{Srianand}}}, \ and\ \bibinfo {author}
  {\bibfnamefont {M.}~\bibnamefont {{Wendt}}},\ }\href {\doibase
  10.1093/mnras/stu1754} {\bibfield  {journal} {\bibinfo  {journal}
  {M.N.R.A.S.}\ }\textbf {\bibinfo {volume} {445}},\ \bibinfo {pages} {128}
  (\bibinfo {year} {2014})}\BibitemShut {NoStop}%
\bibitem [{\citenamefont {Amendola}\ \emph {et~al.}(2012)\citenamefont
  {Amendola}, \citenamefont {Leite}, \citenamefont {Martins}, \citenamefont
  {Nunes}, \citenamefont {Pedrosa} \emph {et~al.}}]{Amendola}%
  \BibitemOpen
  \bibfield  {author} {\bibinfo {author} {\bibfnamefont {L.}~\bibnamefont
  {Amendola}}, \bibinfo {author} {\bibfnamefont {A.~C.~O.}\ \bibnamefont
  {Leite}}, \bibinfo {author} {\bibfnamefont {C.~J. A.~P.}\ \bibnamefont
  {Martins}}, \bibinfo {author} {\bibfnamefont {N.}~\bibnamefont {Nunes}},
  \bibinfo {author} {\bibfnamefont {P.~O.~J.}\ \bibnamefont {Pedrosa}},  \emph
  {et~al.},\ }\href {\doibase 10.1103/PhysRevD.86.063515} {\bibfield  {journal}
  {\bibinfo  {journal} {Phys.Rev.}\ }\textbf {\bibinfo {volume} {D86}},\
  \bibinfo {pages} {063515} (\bibinfo {year} {2012})},\ \Eprint
  {http://arxiv.org/abs/1109.6793} {arXiv:1109.6793 [astro-ph.CO]} \BibitemShut
  {NoStop}%
\bibitem [{\citenamefont {Leite}\ \emph {et~al.}(2014)\citenamefont {Leite},
  \citenamefont {Martins}, \citenamefont {Pedrosa},\ and\ \citenamefont
  {Nunes}}]{Leite}%
  \BibitemOpen
  \bibfield  {author} {\bibinfo {author} {\bibfnamefont {A.~C.~O.}\
  \bibnamefont {Leite}}, \bibinfo {author} {\bibfnamefont {C.~J. A.~P.}\
  \bibnamefont {Martins}}, \bibinfo {author} {\bibfnamefont {P.~O.~J.}\
  \bibnamefont {Pedrosa}}, \ and\ \bibinfo {author} {\bibfnamefont
  {N.}~\bibnamefont {Nunes}},\ }\href {\doibase 10.1103/PhysRevD.90.063519}
  {\bibfield  {journal} {\bibinfo  {journal} {Phys.Rev.}\ }\textbf {\bibinfo
  {volume} {D90}},\ \bibinfo {pages} {063519} (\bibinfo {year} {2014})},\
  \Eprint {http://arxiv.org/abs/1409.3963} {arXiv:1409.3963 [astro-ph.CO]}
  \BibitemShut {NoStop}%
\bibitem [{\citenamefont {Leite}\ and\ \citenamefont
  {Martins}(2015)}]{LeiteNEW}%
  \BibitemOpen
  \bibfield  {author} {\bibinfo {author} {\bibfnamefont {A.}~\bibnamefont
  {Leite}}\ and\ \bibinfo {author} {\bibfnamefont {C.}~\bibnamefont
  {Martins}},\ }\href {\doibase 10.1103/PhysRevD.91.103519} {\bibfield
  {journal} {\bibinfo  {journal} {Phys. Rev.}\ }\textbf {\bibinfo {volume}
  {D91}},\ \bibinfo {pages} {103519} (\bibinfo {year} {2015})},\ \Eprint
  {http://arxiv.org/abs/1505.05529} {arXiv:1505.05529 [astro-ph.CO]}
  \BibitemShut {NoStop}%
\bibitem [{\citenamefont {Martins}\ and\ \citenamefont {Pinho}(2015)}]{Pinho}%
  \BibitemOpen
  \bibfield  {author} {\bibinfo {author} {\bibfnamefont {C.~J. A.~P.}\
  \bibnamefont {Martins}}\ and\ \bibinfo {author} {\bibfnamefont {A.~M.~M.}\
  \bibnamefont {Pinho}},\ }\href {\doibase 10.1103/PhysRevD.91.103501}
  {\bibfield  {journal} {\bibinfo  {journal} {Phys. Rev.}\ }\textbf {\bibinfo
  {volume} {D91}},\ \bibinfo {pages} {103501} (\bibinfo {year} {2015})},\
  \Eprint {http://arxiv.org/abs/1505.02196} {arXiv:1505.02196 [astro-ph.CO]}
  \BibitemShut {NoStop}%
\bibitem [{\citenamefont {Martins}\ \emph {et~al.}(2015)\citenamefont
  {Martins}, \citenamefont {Pinho}, \citenamefont {Alves}, \citenamefont
  {Pino}, \citenamefont {Rocha},\ and\ \citenamefont {von
  Wietersheim}}]{Pinho2}%
  \BibitemOpen
  \bibfield  {author} {\bibinfo {author} {\bibfnamefont {C.~J. A.~P.}\
  \bibnamefont {Martins}}, \bibinfo {author} {\bibfnamefont {A.~M.~M.}\
  \bibnamefont {Pinho}}, \bibinfo {author} {\bibfnamefont {R.~F.~C.}\
  \bibnamefont {Alves}}, \bibinfo {author} {\bibfnamefont {M.}~\bibnamefont
  {Pino}}, \bibinfo {author} {\bibfnamefont {C.~I. S.~A.}\ \bibnamefont
  {Rocha}}, \ and\ \bibinfo {author} {\bibfnamefont {M.}~\bibnamefont {von
  Wietersheim}},\ }\href {\doibase 10.1088/1475-7516/2015/08/047} {\bibfield
  {journal} {\bibinfo  {journal} {JCAP}\ }\textbf {\bibinfo {volume} {1508}},\
  \bibinfo {pages} {047} (\bibinfo {year} {2015})},\ \Eprint
  {http://arxiv.org/abs/1508.06157} {arXiv:1508.06157 [astro-ph.CO]}
  \BibitemShut {NoStop}%
\bibitem [{\citenamefont {Dvali}\ and\ \citenamefont
  {Zaldarriaga}(2002)}]{Dvali}%
  \BibitemOpen
  \bibfield  {author} {\bibinfo {author} {\bibfnamefont {G.}~\bibnamefont
  {Dvali}}\ and\ \bibinfo {author} {\bibfnamefont {M.}~\bibnamefont
  {Zaldarriaga}},\ }\href {\doibase 10.1103/PhysRevLett.88.091303} {\bibfield
  {journal} {\bibinfo  {journal} {Phys.Rev.Lett.}\ }\textbf {\bibinfo {volume}
  {88}},\ \bibinfo {pages} {091303} (\bibinfo {year} {2002})},\ \Eprint
  {http://arxiv.org/abs/hep-ph/0108217} {arXiv:hep-ph/0108217 [hep-ph]}
  \BibitemShut {NoStop}%
\bibitem [{\citenamefont {Chiba}\ and\ \citenamefont {Kohri}(2002)}]{Chiba}%
  \BibitemOpen
  \bibfield  {author} {\bibinfo {author} {\bibfnamefont {T.}~\bibnamefont
  {Chiba}}\ and\ \bibinfo {author} {\bibfnamefont {K.}~\bibnamefont {Kohri}},\
  }\href {\doibase 10.1143/PTP.107.631} {\bibfield  {journal} {\bibinfo
  {journal} {Prog. Theor. Phys.}\ }\textbf {\bibinfo {volume} {107}},\ \bibinfo
  {pages} {631} (\bibinfo {year} {2002})},\ \Eprint
  {http://arxiv.org/abs/hep-ph/0111086} {arXiv:hep-ph/0111086 [hep-ph]}
  \BibitemShut {NoStop}%
\bibitem [{\citenamefont {Caldwell}\ and\ \citenamefont {Linder}(2005)}]{FrTh}%
  \BibitemOpen
  \bibfield  {author} {\bibinfo {author} {\bibfnamefont {R.~R.}\ \bibnamefont
  {Caldwell}}\ and\ \bibinfo {author} {\bibfnamefont {E.~V.}\ \bibnamefont
  {Linder}},\ }\href {\doibase 10.1103/PhysRevLett.95.141301} {\bibfield
  {journal} {\bibinfo  {journal} {Phys. Rev. Lett.}\ }\textbf {\bibinfo
  {volume} {95}},\ \bibinfo {pages} {141301} (\bibinfo {year} {2005})},\
  \Eprint {http://arxiv.org/abs/astro-ph/0505494} {arXiv:astro-ph/0505494
  [astro-ph]} \BibitemShut {NoStop}%
\bibitem [{\citenamefont {Chevallier}\ and\ \citenamefont
  {Polarski}(2001)}]{CPL1}%
  \BibitemOpen
  \bibfield  {author} {\bibinfo {author} {\bibfnamefont {M.}~\bibnamefont
  {Chevallier}}\ and\ \bibinfo {author} {\bibfnamefont {D.}~\bibnamefont
  {Polarski}},\ }\href {\doibase 10.1142/S0218271801000822} {\bibfield
  {journal} {\bibinfo  {journal} {Int. J. Mod. Phys.}\ }\textbf {\bibinfo
  {volume} {D10}},\ \bibinfo {pages} {213} (\bibinfo {year} {2001})},\ \Eprint
  {http://arxiv.org/abs/gr-qc/0009008} {arXiv:gr-qc/0009008 [gr-qc]}
  \BibitemShut {NoStop}%
\bibitem [{\citenamefont {Linder}(2003)}]{CPL2}%
  \BibitemOpen
  \bibfield  {author} {\bibinfo {author} {\bibfnamefont {E.~V.}\ \bibnamefont
  {Linder}},\ }\href {\doibase 10.1103/PhysRevLett.90.091301} {\bibfield
  {journal} {\bibinfo  {journal} {Phys. Rev. Lett.}\ }\textbf {\bibinfo
  {volume} {90}},\ \bibinfo {pages} {091301} (\bibinfo {year} {2003})},\
  \Eprint {http://arxiv.org/abs/astro-ph/0208512} {arXiv:astro-ph/0208512
  [astro-ph]} \BibitemShut {NoStop}%
\bibitem [{\citenamefont {Doran}\ and\ \citenamefont {Robbers}(2006)}]{EDE}%
  \BibitemOpen
  \bibfield  {author} {\bibinfo {author} {\bibfnamefont {M.}~\bibnamefont
  {Doran}}\ and\ \bibinfo {author} {\bibfnamefont {G.}~\bibnamefont
  {Robbers}},\ }\href {\doibase 10.1088/1475-7516/2006/06/026} {\bibfield
  {journal} {\bibinfo  {journal} {JCAP}\ }\textbf {\bibinfo {volume} {0606}},\
  \bibinfo {pages} {026} (\bibinfo {year} {2006})},\ \Eprint
  {http://arxiv.org/abs/astro-ph/0601544} {arXiv:astro-ph/0601544 [astro-ph]}
  \BibitemShut {NoStop}%
\bibitem [{\citenamefont {Mukhanov}(2013)}]{MKH}%
  \BibitemOpen
  \bibfield  {author} {\bibinfo {author} {\bibfnamefont {V.}~\bibnamefont
  {Mukhanov}},\ }\href {\doibase 10.1140/epjc/s10052-013-2486-7} {\bibfield
  {journal} {\bibinfo  {journal} {Eur. Phys. J.}\ }\textbf {\bibinfo {volume}
  {C73}},\ \bibinfo {pages} {2486} (\bibinfo {year} {2013})},\ \Eprint
  {http://arxiv.org/abs/1303.3925} {arXiv:1303.3925 [astro-ph.CO]} \BibitemShut
  {NoStop}%
\bibitem [{\citenamefont {Calabrese}\ \emph {et~al.}(2011)\citenamefont
  {Calabrese}, \citenamefont {Menegoni}, \citenamefont {Martins}, \citenamefont
  {Melchiorri},\ and\ \citenamefont {Rocha}}]{Erminia1}%
  \BibitemOpen
  \bibfield  {author} {\bibinfo {author} {\bibfnamefont {E.}~\bibnamefont
  {Calabrese}}, \bibinfo {author} {\bibfnamefont {E.}~\bibnamefont {Menegoni}},
  \bibinfo {author} {\bibfnamefont {C.~J. A.~P.}\ \bibnamefont {Martins}},
  \bibinfo {author} {\bibfnamefont {A.}~\bibnamefont {Melchiorri}}, \ and\
  \bibinfo {author} {\bibfnamefont {G.}~\bibnamefont {Rocha}},\ }\href
  {\doibase 10.1103/PhysRevD.84.023518} {\bibfield  {journal} {\bibinfo
  {journal} {Phys.Rev.}\ }\textbf {\bibinfo {volume} {D84}},\ \bibinfo {pages}
  {023518} (\bibinfo {year} {2011})},\ \Eprint {http://arxiv.org/abs/1104.0760}
  {arXiv:1104.0760 [astro-ph.CO]} \BibitemShut {NoStop}%
\bibitem [{\citenamefont {Nunes}\ and\ \citenamefont {Lidsey}(2004)}]{Nunes}%
  \BibitemOpen
  \bibfield  {author} {\bibinfo {author} {\bibfnamefont {N.~J.}\ \bibnamefont
  {Nunes}}\ and\ \bibinfo {author} {\bibfnamefont {J.~E.}\ \bibnamefont
  {Lidsey}},\ }\href {\doibase 10.1103/PhysRevD.69.123511} {\bibfield
  {journal} {\bibinfo  {journal} {Phys. Rev.}\ }\textbf {\bibinfo {volume}
  {D69}},\ \bibinfo {pages} {123511} (\bibinfo {year} {2004})},\ \Eprint
  {http://arxiv.org/abs/astro-ph/0310882} {arXiv:astro-ph/0310882 [astro-ph]}
  \BibitemShut {NoStop}%
\bibitem [{\citenamefont {Vielzeuf}\ and\ \citenamefont
  {Martins}(2014)}]{Phantom}%
  \BibitemOpen
  \bibfield  {author} {\bibinfo {author} {\bibfnamefont {P.~E.}\ \bibnamefont
  {Vielzeuf}}\ and\ \bibinfo {author} {\bibfnamefont {C.~J. A.~P.}\
  \bibnamefont {Martins}},\ }\href@noop {} {\bibfield  {journal} {\bibinfo
  {journal} {Mem.Soc.Ast.It.}\ }\textbf {\bibinfo {volume} {85}},\ \bibinfo
  {pages} {155} (\bibinfo {year} {2014})},\ \Eprint
  {http://arxiv.org/abs/1309.7771} {arXiv:1309.7771 [astro-ph.CO]} \BibitemShut
  {NoStop}%
\bibitem [{\citenamefont {Damour}\ and\ \citenamefont
  {Donoghue}(2010)}]{Damour}%
  \BibitemOpen
  \bibfield  {author} {\bibinfo {author} {\bibfnamefont {T.}~\bibnamefont
  {Damour}}\ and\ \bibinfo {author} {\bibfnamefont {J.~F.}\ \bibnamefont
  {Donoghue}},\ }\href {\doibase 10.1088/0264-9381/27/20/202001} {\bibfield
  {journal} {\bibinfo  {journal} {Class. Quant. Grav.}\ }\textbf {\bibinfo
  {volume} {27}},\ \bibinfo {pages} {202001} (\bibinfo {year} {2010})},\
  \Eprint {http://arxiv.org/abs/1007.2790} {arXiv:1007.2790 [gr-qc]}
  \BibitemShut {NoStop}%
\bibitem [{\citenamefont {Suzuki}\ \emph {et~al.}(2012)\citenamefont {Suzuki},
  \citenamefont {Rubin}, \citenamefont {Lidman}, \citenamefont {Aldering},
  \citenamefont {Amanullah} \emph {et~al.}}]{Union}%
  \BibitemOpen
  \bibfield  {author} {\bibinfo {author} {\bibfnamefont {N.}~\bibnamefont
  {Suzuki}}, \bibinfo {author} {\bibfnamefont {D.}~\bibnamefont {Rubin}},
  \bibinfo {author} {\bibfnamefont {C.}~\bibnamefont {Lidman}}, \bibinfo
  {author} {\bibfnamefont {G.}~\bibnamefont {Aldering}}, \bibinfo {author}
  {\bibfnamefont {R.}~\bibnamefont {Amanullah}},  \emph {et~al.},\ }\href
  {\doibase 10.1088/0004-637X/746/1/85} {\bibfield  {journal} {\bibinfo
  {journal} {Astrophys.J.}\ }\textbf {\bibinfo {volume} {746}},\ \bibinfo
  {pages} {85} (\bibinfo {year} {2012})},\ \Eprint
  {http://arxiv.org/abs/1105.3470} {arXiv:1105.3470 [astro-ph.CO]} \BibitemShut
  {NoStop}%
\bibitem [{\citenamefont {Farooq}\ and\ \citenamefont {Ratra}(2013)}]{Farooq}%
  \BibitemOpen
  \bibfield  {author} {\bibinfo {author} {\bibfnamefont {O.}~\bibnamefont
  {Farooq}}\ and\ \bibinfo {author} {\bibfnamefont {B.}~\bibnamefont {Ratra}},\
  }\href {\doibase 10.1088/2041-8205/766/1/L7} {\bibfield  {journal} {\bibinfo
  {journal} {Astrophys.J.}\ }\textbf {\bibinfo {volume} {766}},\ \bibinfo
  {pages} {L7} (\bibinfo {year} {2013})},\ \Eprint
  {http://arxiv.org/abs/1301.5243} {arXiv:1301.5243 [astro-ph.CO]} \BibitemShut
  {NoStop}%
\bibitem [{\citenamefont {{Calabrese}}\ \emph {et~al.}(2014)\citenamefont
  {{Calabrese}}, \citenamefont {{Martinelli}}, \citenamefont {{Pandolfi}},
  \citenamefont {{Cardone}}, \citenamefont {{Martins}}, \citenamefont
  {{Spiro}},\ and\ \citenamefont {{Vielzeuf}}}]{Erminia2}%
  \BibitemOpen
  \bibfield  {author} {\bibinfo {author} {\bibfnamefont {E.}~\bibnamefont
  {{Calabrese}}}, \bibinfo {author} {\bibfnamefont {M.}~\bibnamefont
  {{Martinelli}}}, \bibinfo {author} {\bibfnamefont {S.}~\bibnamefont
  {{Pandolfi}}}, \bibinfo {author} {\bibfnamefont {V.~F.}\ \bibnamefont
  {{Cardone}}}, \bibinfo {author} {\bibfnamefont {C.~J.~A.~P.}\ \bibnamefont
  {{Martins}}}, \bibinfo {author} {\bibfnamefont {S.}~\bibnamefont {{Spiro}}},
  \ and\ \bibinfo {author} {\bibfnamefont {P.~E.}\ \bibnamefont {{Vielzeuf}}},\
  }\href {\doibase 10.1103/PhysRevD.89.083509} {\bibfield  {journal} {\bibinfo
  {journal} {Phys.Rev.}\ ,\ \bibinfo {eid} {083509}} (\bibinfo {year}
  {2014})},\ \Eprint {http://arxiv.org/abs/1311.5841} {arXiv:1311.5841
  [astro-ph.CO]} \BibitemShut {NoStop}%
\bibitem [{\citenamefont {Rosenband}\ \emph {et~al.}(2008)\citenamefont
  {Rosenband}, \citenamefont {Hume}, \citenamefont {Schmidt}, \citenamefont
  {Chou}, \citenamefont {Brusch}, \citenamefont {Lorini}, \citenamefont
  {Oskay}, \citenamefont {Drullinger}, \citenamefont {Fortier}, \citenamefont
  {Stalnaker}, \citenamefont {Diddams}, \citenamefont {Swann}, \citenamefont
  {Newbury}, \citenamefont {Itano}, \citenamefont {Wineland},\ and\
  \citenamefont {Bergquist}}]{Rosenband}%
  \BibitemOpen
  \bibfield  {author} {\bibinfo {author} {\bibfnamefont {T.}~\bibnamefont
  {Rosenband}}, \bibinfo {author} {\bibfnamefont {D.}~\bibnamefont {Hume}},
  \bibinfo {author} {\bibfnamefont {P.}~\bibnamefont {Schmidt}}, \bibinfo
  {author} {\bibfnamefont {C.}~\bibnamefont {Chou}}, \bibinfo {author}
  {\bibfnamefont {A.}~\bibnamefont {Brusch}}, \bibinfo {author} {\bibfnamefont
  {L.}~\bibnamefont {Lorini}}, \bibinfo {author} {\bibfnamefont
  {W.}~\bibnamefont {Oskay}}, \bibinfo {author} {\bibfnamefont
  {R.}~\bibnamefont {Drullinger}}, \bibinfo {author} {\bibfnamefont
  {T.}~\bibnamefont {Fortier}}, \bibinfo {author} {\bibfnamefont
  {J.}~\bibnamefont {Stalnaker}}, \bibinfo {author} {\bibfnamefont
  {S.}~\bibnamefont {Diddams}}, \bibinfo {author} {\bibfnamefont
  {W.}~\bibnamefont {Swann}}, \bibinfo {author} {\bibfnamefont
  {N.}~\bibnamefont {Newbury}}, \bibinfo {author} {\bibfnamefont
  {W.}~\bibnamefont {Itano}}, \bibinfo {author} {\bibfnamefont
  {D.}~\bibnamefont {Wineland}}, \ and\ \bibinfo {author} {\bibfnamefont
  {J.}~\bibnamefont {Bergquist}},\ }\href {\doibase 10.1126/science.1154622}
  {\bibfield  {journal} {\bibinfo  {journal} {Science}\ }\textbf {\bibinfo
  {volume} {319}},\ \bibinfo {pages} {1808} (\bibinfo {year}
  {2008})}\BibitemShut {NoStop}%
\bibitem [{\citenamefont {Luo}\ \emph {et~al.}(2011)\citenamefont {Luo},
  \citenamefont {Olive},\ and\ \citenamefont {Uzan}}]{Luo}%
  \BibitemOpen
  \bibfield  {author} {\bibinfo {author} {\bibfnamefont {F.}~\bibnamefont
  {Luo}}, \bibinfo {author} {\bibfnamefont {K.~A.}\ \bibnamefont {Olive}}, \
  and\ \bibinfo {author} {\bibfnamefont {J.-P.}\ \bibnamefont {Uzan}},\ }\href
  {\doibase 10.1103/PhysRevD.84.096004} {\bibfield  {journal} {\bibinfo
  {journal} {Phys. Rev.}\ }\textbf {\bibinfo {volume} {D84}},\ \bibinfo {pages}
  {096004} (\bibinfo {year} {2011})},\ \Eprint {http://arxiv.org/abs/1107.4154}
  {arXiv:1107.4154 [hep-ph]} \BibitemShut {NoStop}%
\bibitem [{\citenamefont {Ferreira}\ \emph {et~al.}(2014)\citenamefont
  {Ferreira}, \citenamefont {Frigola}, \citenamefont {Martins}, \citenamefont
  {Monteiro},\ and\ \citenamefont {Solà}}]{Ferreira}%
  \BibitemOpen
  \bibfield  {author} {\bibinfo {author} {\bibfnamefont {M.~C.}\ \bibnamefont
  {Ferreira}}, \bibinfo {author} {\bibfnamefont {O.}~\bibnamefont {Frigola}},
  \bibinfo {author} {\bibfnamefont {C.~J. A.~P.}\ \bibnamefont {Martins}},
  \bibinfo {author} {\bibfnamefont {A.~M. R. V.~L.}\ \bibnamefont {Monteiro}},
  \ and\ \bibinfo {author} {\bibfnamefont {J.}~\bibnamefont {Solà}},\ }\href
  {\doibase 10.1103/PhysRevD.89.083011} {\bibfield  {journal} {\bibinfo
  {journal} {Phys. Rev.}\ }\textbf {\bibinfo {volume} {D89}},\ \bibinfo {pages}
  {083011} (\bibinfo {year} {2014})},\ \Eprint {http://arxiv.org/abs/1405.0299}
  {arXiv:1405.0299 [astro-ph.CO]} \BibitemShut {NoStop}%
\bibitem [{\citenamefont {Ferreira}\ and\ \citenamefont
  {Martins}(2015)}]{Ferreira2}%
  \BibitemOpen
  \bibfield  {author} {\bibinfo {author} {\bibfnamefont {M.}~\bibnamefont
  {Ferreira}}\ and\ \bibinfo {author} {\bibfnamefont {C.}~\bibnamefont
  {Martins}},\ }\href {\doibase 10.1103/PhysRevD.91.124032} {\bibfield
  {journal} {\bibinfo  {journal} {Phys. Rev.}\ }\textbf {\bibinfo {volume}
  {D91}},\ \bibinfo {pages} {124032} (\bibinfo {year} {2015})},\ \Eprint
  {http://arxiv.org/abs/1506.03550} {arXiv:1506.03550 [astro-ph.CO]}
  \BibitemShut {NoStop}%
\bibitem [{\citenamefont {Songaila}\ and\ \citenamefont
  {Cowie}(2014)}]{Songaila}%
  \BibitemOpen
  \bibfield  {author} {\bibinfo {author} {\bibfnamefont {A.}~\bibnamefont
  {Songaila}}\ and\ \bibinfo {author} {\bibfnamefont {L.}~\bibnamefont
  {Cowie}},\ }\href {\doibase 10.1088/0004-637X/793/2/103} {\bibfield
  {journal} {\bibinfo  {journal} {Astrophys.J.}\ }\textbf {\bibinfo {volume}
  {793}},\ \bibinfo {pages} {103} (\bibinfo {year} {2014})},\ \Eprint
  {http://arxiv.org/abs/1406.3628} {arXiv:1406.3628 [astro-ph.CO]} \BibitemShut
  {NoStop}%
\bibitem [{\citenamefont {Molaro}\ \emph {et~al.}(2008)\citenamefont {Molaro},
  \citenamefont {Reimers}, \citenamefont {Agafonova},\ and\ \citenamefont
  {Levshakov}}]{alphaMolaro}%
  \BibitemOpen
  \bibfield  {author} {\bibinfo {author} {\bibfnamefont {P.}~\bibnamefont
  {Molaro}}, \bibinfo {author} {\bibfnamefont {D.}~\bibnamefont {Reimers}},
  \bibinfo {author} {\bibfnamefont {I.~I.}\ \bibnamefont {Agafonova}}, \ and\
  \bibinfo {author} {\bibfnamefont {S.~A.}\ \bibnamefont {Levshakov}},\ }\href
  {\doibase 10.1140/epjst/e2008-00818-4} {\bibfield  {journal} {\bibinfo
  {journal} {Eur.Phys.J.ST}\ }\textbf {\bibinfo {volume} {163}},\ \bibinfo
  {pages} {173} (\bibinfo {year} {2008})},\ \Eprint
  {http://arxiv.org/abs/0712.4380} {arXiv:0712.4380 [astro-ph]} \BibitemShut
  {NoStop}%
\bibitem [{\citenamefont {{Chand}}\ \emph {et~al.}(2006)\citenamefont
  {{Chand}}, \citenamefont {{Srianand}}, \citenamefont {{Petitjean}},
  \citenamefont {{Aracil}}, \citenamefont {{Quast}},\ and\ \citenamefont
  {{Reimers}}}]{alphaChand}%
  \BibitemOpen
  \bibfield  {author} {\bibinfo {author} {\bibfnamefont {H.}~\bibnamefont
  {{Chand}}}, \bibinfo {author} {\bibfnamefont {R.}~\bibnamefont {{Srianand}}},
  \bibinfo {author} {\bibfnamefont {P.}~\bibnamefont {{Petitjean}}}, \bibinfo
  {author} {\bibfnamefont {B.}~\bibnamefont {{Aracil}}}, \bibinfo {author}
  {\bibfnamefont {R.}~\bibnamefont {{Quast}}}, \ and\ \bibinfo {author}
  {\bibfnamefont {D.}~\bibnamefont {{Reimers}}},\ }\href {\doibase
  10.1051/0004-6361:20054584} {\bibfield  {journal} {\bibinfo  {journal}
  {Astron.Astrophys.}\ }\textbf {\bibinfo {volume} {451}},\ \bibinfo {pages}
  {45} (\bibinfo {year} {2006})},\ \Eprint
  {http://arxiv.org/abs/astro-ph/0601194} {astro-ph/0601194} \BibitemShut
  {NoStop}%
\bibitem [{\citenamefont {{Agafonova}}\ \emph {et~al.}(2011)\citenamefont
  {{Agafonova}}, \citenamefont {{Molaro}}, \citenamefont {{Levshakov}},\ and\
  \citenamefont {{Hou}}}]{alphaAgafonova}%
  \BibitemOpen
  \bibfield  {author} {\bibinfo {author} {\bibfnamefont {I.~I.}\ \bibnamefont
  {{Agafonova}}}, \bibinfo {author} {\bibfnamefont {P.}~\bibnamefont
  {{Molaro}}}, \bibinfo {author} {\bibfnamefont {S.~A.}\ \bibnamefont
  {{Levshakov}}}, \ and\ \bibinfo {author} {\bibfnamefont {J.~L.}\ \bibnamefont
  {{Hou}}},\ }\href {\doibase 10.1051/0004-6361/201016194} {\bibfield
  {journal} {\bibinfo  {journal} {Astron.Astrophys.}\ }\textbf {\bibinfo
  {volume} {529}},\ \bibinfo {eid} {A28} (\bibinfo {year} {2011})},\ \Eprint
  {http://arxiv.org/abs/1102.2967} {arXiv:1102.2967 [astro-ph.CO]} \BibitemShut
  {NoStop}%
\bibitem [{\citenamefont {Wagner}\ \emph {et~al.}(2012)\citenamefont {Wagner},
  \citenamefont {Schlamminger}, \citenamefont {Gundlach},\ and\ \citenamefont
  {Adelberger}}]{Torsion}%
  \BibitemOpen
  \bibfield  {author} {\bibinfo {author} {\bibfnamefont {T.~A.}\ \bibnamefont
  {Wagner}}, \bibinfo {author} {\bibfnamefont {S.}~\bibnamefont
  {Schlamminger}}, \bibinfo {author} {\bibfnamefont {J.~H.}\ \bibnamefont
  {Gundlach}}, \ and\ \bibinfo {author} {\bibfnamefont {E.~G.}\ \bibnamefont
  {Adelberger}},\ }\href {\doibase 10.1088/0264-9381/29/18/184002} {\bibfield
  {journal} {\bibinfo  {journal} {Class. Quant. Grav.}\ }\textbf {\bibinfo
  {volume} {29}},\ \bibinfo {pages} {184002} (\bibinfo {year} {2012})},\
  \Eprint {http://arxiv.org/abs/1207.2442} {arXiv:1207.2442 [gr-qc]}
  \BibitemShut {NoStop}%
\bibitem [{\citenamefont {Müller}\ \emph {et~al.}(2012)\citenamefont
  {Müller}, \citenamefont {Hofmann},\ and\ \citenamefont {Biskupek}}]{Lunar}%
  \BibitemOpen
  \bibfield  {author} {\bibinfo {author} {\bibfnamefont {J.}~\bibnamefont
  {Müller}}, \bibinfo {author} {\bibfnamefont {F.}~\bibnamefont {Hofmann}}, \
  and\ \bibinfo {author} {\bibfnamefont {L.}~\bibnamefont {Biskupek}},\ }\href
  {\doibase 10.1088/0264-9381/29/18/184006} {\bibfield  {journal} {\bibinfo
  {journal} {Class. Quant. Grav.}\ }\textbf {\bibinfo {volume} {29}},\ \bibinfo
  {pages} {184006} (\bibinfo {year} {2012})}\BibitemShut {NoStop}%
\bibitem [{\citenamefont {Ade}\ \emph {et~al.}(2015)\citenamefont {Ade} \emph
  {et~al.}}]{Planck}%
  \BibitemOpen
  \bibfield  {author} {\bibinfo {author} {\bibfnamefont {P.~A.~R.}\
  \bibnamefont {Ade}} \emph {et~al.} (\bibinfo {collaboration} {Planck}),\
  }\href@noop {} {\  (\bibinfo {year} {2015})},\ \Eprint
  {http://arxiv.org/abs/1502.01589} {arXiv:1502.01589 [astro-ph.CO]}
  \BibitemShut {NoStop}%
\bibitem [{\citenamefont {Berg\'e}\ \emph {et~al.}(2015)\citenamefont
  {Berg\'e}, \citenamefont {Touboul},\ and\ \citenamefont
  {Rodrigues}}]{MICROSCOPE}%
  \BibitemOpen
  \bibfield  {author} {\bibinfo {author} {\bibfnamefont {J.}~\bibnamefont
  {Berg\'e}}, \bibinfo {author} {\bibfnamefont {P.}~\bibnamefont {Touboul}}, \
  and\ \bibinfo {author} {\bibfnamefont {M.}~\bibnamefont {Rodrigues}}
  (\bibinfo {collaboration} {MICROSCOPE}),\ }\bibfield  {booktitle} {\emph
  {\bibinfo {booktitle} {{Proceedings, 10th International LISA Symposium}}},\
  }\href {\doibase 10.1088/1742-6596/610/1/012009} {\bibfield  {journal}
  {\bibinfo  {journal} {J. Phys. Conf. Ser.}\ }\textbf {\bibinfo {volume}
  {610}},\ \bibinfo {pages} {012009} (\bibinfo {year} {2015})},\ \Eprint
  {http://arxiv.org/abs/1501.01644} {arXiv:1501.01644 [gr-qc]} \BibitemShut
  {NoStop}%
\bibitem [{\citenamefont {Overduin}\ \emph {et~al.}(2012)\citenamefont
  {Overduin}, \citenamefont {Everitt}, \citenamefont {Worden},\ and\
  \citenamefont {Mester}}]{STEP}%
  \BibitemOpen
  \bibfield  {author} {\bibinfo {author} {\bibfnamefont {J.}~\bibnamefont
  {Overduin}}, \bibinfo {author} {\bibfnamefont {F.}~\bibnamefont {Everitt}},
  \bibinfo {author} {\bibfnamefont {P.}~\bibnamefont {Worden}}, \ and\ \bibinfo
  {author} {\bibfnamefont {J.}~\bibnamefont {Mester}},\ }\href {\doibase
  10.1088/0264-9381/29/18/184012} {\bibfield  {journal} {\bibinfo  {journal}
  {Class. Quant. Grav.}\ }\textbf {\bibinfo {volume} {29}},\ \bibinfo {pages}
  {184012} (\bibinfo {year} {2012})},\ \Eprint {http://arxiv.org/abs/1401.4784}
  {arXiv:1401.4784 [gr-qc]} \BibitemShut {NoStop}%
\bibitem [{\citenamefont {Peccei}\ and\ \citenamefont {Quinn}(1977)}]{Peccei}%
  \BibitemOpen
  \bibfield  {author} {\bibinfo {author} {\bibfnamefont {R.~D.}\ \bibnamefont
  {Peccei}}\ and\ \bibinfo {author} {\bibfnamefont {H.~R.}\ \bibnamefont
  {Quinn}},\ }\href {\doibase 10.1103/PhysRevLett.38.1440} {\bibfield
  {journal} {\bibinfo  {journal} {Phys. Rev. Lett.}\ }\textbf {\bibinfo
  {volume} {38}},\ \bibinfo {pages} {1440} (\bibinfo {year}
  {1977})}\BibitemShut {NoStop}%
\end{thebibliography}%
\end{document}